\documentclass[useAMS,usenatbib]{mn2e}

\pdfoutput=1

\usepackage{txfonts}
\usepackage{wasysym}
\usepackage{graphicx}
\usepackage{multirow}
\usepackage{subfig}

\title[A wide search for obscured AGN using XMM--Newton and WISE]
{A wide search for obscured Active Galactic Nuclei using \emph{XMM--Newton} and \emph{WISE}}

\author[E. Rovilos et al.]
{E. Rovilos,$^1$
I. Georgantopoulos,$^2$
A. Akylas,$^2$
J. Aird,$^1$
D. M. Alexander,$^1$
A. Comastri,$^3$
\newauthor
A. Del Moro,$^1$
P. Gandhi,$^1$
A. Georgakakis,$^{4,2}$
C. M. Harrison,$^1$
J. R. Mullaney,$^1$\\
\\
$^1$Department of Physics, Durham University, South Road, Durham, DH1 3LE, UK\\
$^2$Institute of Astronomy \& Astrophysics, National Observatory of Athens, Palaia Penteli, 15236, Athens, Greece\\
$^3$INAF - Osservatorio Astronomico di Bologna, via Ranzani 1, 40127, Bologna, Italy\\
$^4$Max--Planck--Institut f\"{u}r extraterrestrische Physik, Giessenbachstra\ss e, 85748 Garching bei M\"{u}nchen, Germany}

\begin{document}

\date{Accepted 2013 November 13. Received 2013 November 8; in original form 2013 August 9}

\pagerange{\pageref{firstpage}--\pageref{lastpage}} \pubyear{2013}

\maketitle

\label{firstpage}

\begin{abstract}
Heavily obscured and Compton-thick AGN are missing even in the deepest X-ray surveys, and indirect methods are required to detect them. Here we use a combination of the \emph{XMM--Newton} serendipitous X-ray survey with the optical SDSS, and the infrared \emph{WISE} all-sky survey in order to check the efficiency of the low X-ray to infrared luminosity selection method in finding heavily obscured AGN. We select the sources which are detected in the hard X-ray band ($\rm 2-8\,keV$), and also have a redshift determination (photometric or spectroscopic) in the SDSS catalogue. We match this sample with the \emph{WISE} catalogue, and fit the spectral energy distributions (SEDs) of the 2\,844 sources which have three, or more, photometric data-points in the infrared. We then select the heavily obscured AGN candidates by comparing their $\rm 12\,\mu m$ AGN luminosity to the observed 2--10\,keV X-ray luminosity and the intrinsic relation between the X-ray and the mid-infrared luminosities. With this approach we find 20 candidate heavily obscured AGN and we then examine their X-ray and optical spectra. Of the 20 initial candidates, we find nine (64\%; out of the 14, for which X-ray spectra could be fit) based on the X-ray spectra, and seven (78\%; out of the nine detected spectroscopically in the SDSS) based on the [OIII] line fluxes. Combining all criteria, we determine the final number of heavily obscured AGN to be 12--19, and the number of Compton-thick AGN to be 2--5, showing that the method is reliable in finding obscured AGN, but not Compton-thick. However those numbers are smaller than what would be expected from X-ray background population synthesis models, which demonstrates how the optical--infrared selection and the scatter of the $L_{\rm x}-L_{\rm MIR}$ relation introduced by observational constraints limit the efficiency of the method. Finally, we  test popular obscured AGN selection methods based on mid-infrared colours, and find that the probability of an AGN to be selected by its mid-infrared colours increases with the X-ray luminosity. The (observed) X-ray luminosities of heavily obscured AGN are relatively low ($L_{\rm 2-10keV}<10^{44}\,{\rm erg\,s^{-1}}$), even though most of them are located in the ``QSO locus''. However, a selection scheme based on a relatively low X-ray luminosity and mid-infrared colours characteristic of QSOs would not select $\sim25\%$ of the heavily obscured AGN of our sample.
\end{abstract} 


\section{Introduction}

Super-massive black holes (SMBHs) are considered to be one of the major building blocks of the universe. Most nearby galaxies are seen to harbour a SMBH \citep[e.g.][]{Kormendy1987,Matt1996,Ishisaki1996}, including the Milky Way \citep*{Genzel2010}, and it is found that the mass of the SMBH is tightly connected to properties of the bulge of the galaxy \citep[e.g.][]{Gebhardt2000,Ferrarese2000}. The growth of a black hole to reach a mass of $\gtrsim10^{6}\,M_{\odot}$ must include a phase of rapid accretion, i.e. an active galactic nucleus \citep[AGN;][]{Rees1984}, unless it forms from an already massive primordial black hole \citep[see][]{Volonteri2012}. This has implications for the formation and growth of galaxies and other structures in the universe \citep[see also][]{Alexander2012,Fabian2012}, therefore a complete census of AGN in the universe is essential in order to study its evolution.

The most efficient way to detect an AGN is through its high-energy emission detected in the X-rays. The deepest X-ray surveys with \emph{Chandra} and \emph{XMM--Newton} \citep{Alexander2003,Brunner2008,Xue2011,Ranalli2013} have detected a large number of AGN, with a surface density tens of times higher than that found in optical surveys \citep{Bauer2004,Xue2011}. A representative sample of AGN in the universe over different scales and redshifts can be drawn by combining deep pencil-beam surveys with wider, intermediate-depth surveys \citep[e.g.][]{Cappelluti2009,Elvis2009}, and shallow large-area surveys \citep[e.g.][]{Voges2000,Watson2009}. Most of the AGN detected in the X-rays show some level of obscuration \citep{Hasinger2008}, but the hard X-rays ($\rm2-10\,keV$) can easily penetrate large columns of obscuring material in cases where the dominant X-ray absorption mechanism is photoelectric absorption, because of the strong dependance of its cross-section on the photon energy ($\sigma_{\rm PE}\propto E^{-3}$). However, when the column density of the obscuring material reaches $N_{\rm H}\rm\approx1.5\times10^{24}\,cm^{-2}$ (i.e. the inverse of the Thomson cross-section for electrons, $\rm \sigma_{T}^{-1}$) it becomes optically thick to Compton scattering, the relativistic equivalent of Thomson scattering applied at higher energies, which has a lower dependancy with energy. Such sources are called  Compton-thick (CT) AGN and even high-energy photons are obscured. If the column density is $N_{\rm H}\rm\lesssim5\times10^{24}\,cm^{-2}$ we can still detect some X-ray photons from the source, with a hard spectrum peaking at $\rm\sim10\,keV$, where the Compton and photoelectric cross-sections are equal (transmission-dominated CT AGN). If the column density is even higher, any detected X-ray emission comes from a reflected component at the back side of the obscuring torus \citep*[reflection-dominated CT AGN; e.g.][]{Matt1996b}, giving a characteristic flat X-ray spectrum, with an observed luminosity typically a few percent of the intrinsic AGN luminosity \citep[e.g.][]{Maiolino1998,Matt2000}. In some cases a soft ($\Gamma>1.8$) component scattered possibly from electrons in the narrow-line region is also detected in lower X-ray energies \citep*[$\rm\lesssim2\,keV$; see e.g.][]{Netzer1998}, which is a blend of photo-ionised lines \citep{Guainazzi2007}.

The fact that the observed X-ray emission from CT AGN is only a fraction of the intrinsic emission, even at the highest energies detected by X-ray telescopes, makes them challenging to detect in even the deepest X-ray surveys. Therefore, other techniques have been developed, that use the combination of a low detected X-ray luminosity (or even a non-detection) with secondary processes taking place in the AGN. The most widely used methods employ optical (or near-infrared) spectroscopy focusing on high-excitation spectral lines coming from the narrow-line region \citep[e.g.][]{Bassani1999,Cappi2006,Akylas2009,Gilli2010,Vignali2010,Mignoli2013}, and mid-infrared photometry tracing the reprocessed dust emission from the absorbing material \citep[e.g.][]{Daddi2007,Fiore2008,Fiore2009,Alexander2011}. The spectral line technique is observationally challenging, as it requires relative bright sources in the optical wavelengths. It has been mostly used in narrow fields utilising multi-slit spectroscopy \citep{Juneau2011}, or to a limited number of sources in wide fields \citep[e.g.][]{Cappi2006}. In this paper we will use the mid-infrared emission, which is easier to apply to wide fields.

The mid-infrared emission from the AGN is due to the obscuring dust heated by the AGN X-ray and ultra-violet emission. However dust is also abundantly found around massive O--B stars in the host galaxies, and is heated by their ultra-violet radiation, making infrared emission also a star-formation tracer \citep[e.g.][]{Calzetti2010}. In order to differentiate between the two different generators of infrared emission, we must take into account the high energy produced by the AGN, which heats the dust to higher temperatures than O--B stars and gives a characteristic power-law spectrum in the mid-infrared \citep{Neugebauer1979} and peaks at $\rm\sim10-20\,\mu m$ \citep[see][for models involving clumpy tori]{Nenkova2008,Stalevski2012}. This feature is used to select AGN based on their mid-infrared colours \citep[e.g.][]{Stern2005,Donley2012,Mateos2012,Mateos2013} or power-law shape of the SED \citep[e.g.][]{AlonsoHerrero2006,Donley2007}. The peak of the AGN-powered IR emission at $\rm\sim10-20\,\mu m$ also coincides with the minimum of the host SED at these wavelengths \citep[see e.g.][]{Chary2001}, which makes a direct mid-infrared selection possible. This has been extensively used to select obscured AGN in medium-to-deep surveys \citep[][and references therein]{Georgantopoulos2011} by their low X-ray to infrared luminosity ratio, utilising the empirical intrinsic $L_{\rm x}/L_{\rm MIR}$ relation \citep{Lutz2004,Gandhi2009,Asmus2011}.

The low $L_{\rm x}/L_{\rm MIR}$ selection technique has not been widely used in broad surveys, because of the lack of MIR observations covering a large part of the sky. Before the advent of \emph{WISE} \citep{Wright2010}, the only all-sky survey products in the mid-infrared were the \emph{AKARI} survey, and the IRAS point-source catalogue \citep{Beichman1988}, which is used by \citet*{Severgnini2012}, giving promising results. In this work we will use the recently publicly available results from the \emph{WISE} all-sky survey, in conjunction with the wide-field \emph{XMM}--SDSS catalogue \citep{Georgakakis2011} to perform a wide search for X-ray detected Compton-thick AGN. We will also use a new SED decomposition technique to isolate the mid-infrared emission from the AGN and thus minimise the host galaxy contamination. We will then test the efficiency of the low X-ray to mid-infrared luminosity method by examining the X-ray and optical spectral properties of the candidate sources, and comparing their number with what expected form X-ray background synthesis models. We adopt $H_{0}=71\,{\rm km\,s^{-1}\,Mpc^{-1}}$, $\Omega_{\rm M}=0.27$, and $\Omega_{\Lambda}=0.73$ throughout the paper.

\section{Data}

\subsection{X-ray catalogue}
\label{XrayData}

We use the X-ray catalogue compiled by \citet{Georgakakis2011}, which contains about 40\,000 X-ray point-sources over an area of 122\,deg$^2$, with a half-area detection limit of $\rm1.5\times10^{-14}\,erg\,s^{-1}\,cm^{-2}$ in the 0.5--10\,keV band and $\rm3\times10^{-14}\,erg\,s^{-1}\,cm^{-2}$ in the 2--10\,keV band. This survey uses \emph{XMM--Newton} pointings which coincide with the SDSS DR7 \citep{Abazajan2009}, and we use it in order to have optical and near-infrared photometric information, as well as a spectroscopic or photometric redshift for our candidates. The source detection has been performed by \citet{Georgakakis2011} straight from the \emph{XMM--Newton} observations without using the automated source extraction of \citet{Watson2009}. All \emph{XMM--Newton} observations performed prior to July 2009, overlapping with the SDSS have been used in the analysis, and X-ray photometry is provided in five bands, including the 0.5--2.0\,keV and 2--8\,keV, hereafter ``soft'' and ``hard'' bands, respectively, that we investigate here.

\subsection{Infrared}

For the (mid-)infrared identification of our candidates, we use the all-sky source catalogue of \emph{WISE} \citep{Wright2010}. This is a space telescope launched in December 2009, operating in the mid-infrared part of the spectrum. It has a 40\,cm primary mirror and performed an all-sky survey in the 3.4, 4.6, 12, and 22\,$\rm \mu m$ bands, reaching $5\sigma$ point source sensitivities of 0.08, 0.11, 1 and 6\,mJy, or lower, depending on the position in the sky. The FWHM of the PSFs are 6.1\arcsec, 6.4\arcsec, 6.5\arcsec, and 12.0\arcsec for the four bands respectively, which is comparable to that of \emph{XMM--Newton} ($\approx5-10\arcsec$, depending on the instrument and off-axis angle), allowing us to perform a reliable search for counterparts between the two telescopes. We use the magnitudes measured with profile-fitting photometry, and the zero points of \citet{Jarrett2011}.

\section{The sample}

The X-ray sample of \citet{Georgakakis2011} contains 39\,830 X-ray sources within the footprint of the SDSS DR7 survey. \citet{Georgakakis2011} use the likelihood ratio method\footnote{The likelihood ratio method \citep{Sutherland1992} is usually adopted in cases where a counterpart is sought in a crowded catalogue (in this case the SDSS catalogue), and it uses the surface density of objects of a given magnitude to estimate the probability that a counterpart at a certain distance is a chance match.} to find optical counterparts for the X-ray sources. At a limit of $LR=1.5$ they find a counterpart for almost half of X-ray sources (19\,431/39\,830) with an expected spurious identification rate of 7\%. The probability that an X-ray source has an optical counterpart is strongly dependent on the X-ray flux, and is typically $\approx50\%$ for sources with $f_{\rm0.5-10\,keV}\approx2\times10^{-14}\,{\rm erg\,s\,cm^{-2}}$ and $\approx90\%$ for sources with $f_{\rm0.5-10\,keV}\approx2\times10^{-13}\,{\rm erg\,s\,cm^{-2}}$ \citep{Georgakakis2011}. A redshift determination requires a spectroscopic follow-up in the optical (or the near-infrared) and a good enough quality spectrum, which is the case for the brightest optical sources (typically with $r<17.77$). We note that in addition to SDSS spectroscopy, a number of optical spectroscopic programs were used in \citet{Georgakakis2011}, so there are sources with optical spectra with magnitudes exceeding the $r=17.77$ limit, but not with uniform coverage in terms of spatial distribution, or source type. In addition to spectroscopic redshifts, a source might have a photometric redshift determination, if it is detected in enough optical and near-infrared bands. The typical detection limits for the SDSS DR7 are $r<22.2$ and $z<20.5$. Only half of the SDSS-detected X-ray sources have a redshift determination (9\,029/19\,431), 2\,172 of them spectroscopic.

In identifying heavily obscured sources by their low X-ray to infrared ratio, it is possible that the sample will be contaminated by a number of normal galaxies, i.e. X-ray sources that do not host an AGN, and their X-ray flux is attributed to star-formation. The normalisation of the X-ray to infrared relation for star-forming galaxies \citep*[e.g.][]{Ranalli2003} is 1--2 orders of magnitude lower than the X-ray to infrared ratio of typical AGN (see Sect.\ \ref{contamination}), so normal galaxies could be mistaken for highly obscured AGN. To minimise this effect, we limit our X-ray sample to those X-ray sources that are detected in the hard band (2--8\,keV), so that we are able to have an initial hint of the shape of the X-ray spectrum through the hardness ratio, without having to analyse all the spectra prior to the candidate selection. 4\,553/9\,029 X-ray sources with a redshift determination are detected in the hard band. 

\subsection{Looking for WISE counterparts}
\label{counterparts}

\begin{figure}
\subfloat{\resizebox{\hsize}{!}{\includegraphics{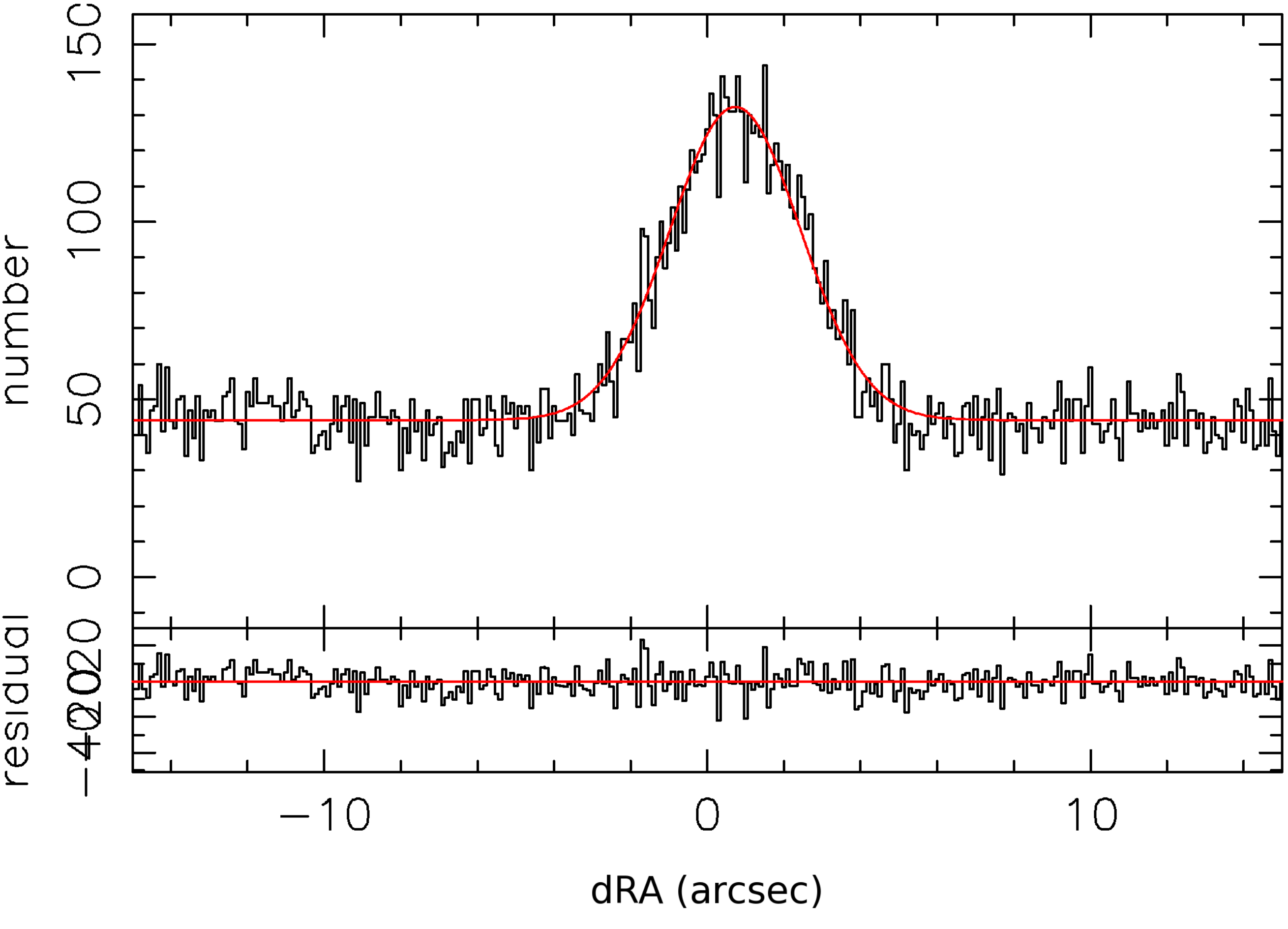}}}
\\
\subfloat{\resizebox{\hsize}{!}{\includegraphics{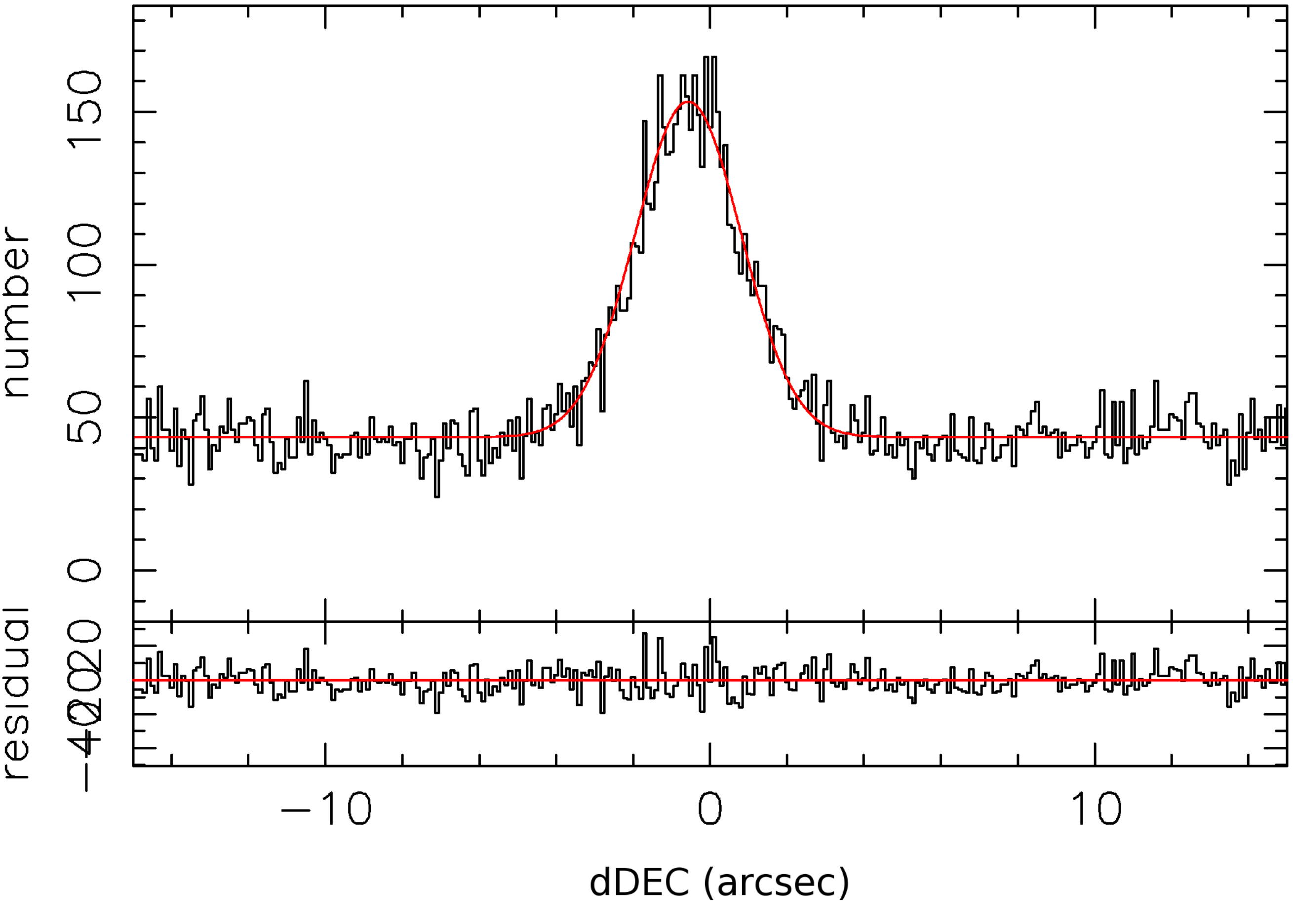}}}
  \caption{Histograms of the differences between the RA and DEC values of the \emph{XMM--Newton} and \emph{WISE} sources. Both distributions are fitted with gaussians (red lines), where the mean and standard deviation values quoted in \S\,\ref{counterparts} are derived from. We detect a $\rm\sim0.7\,arcsec$ and  $\rm\sim0.6\,arcsec$ shift in RA and DEC respectively, and we correct the matching coordinates accordingly.}
  \label{dRAdDEC}
\end{figure}

We look for counterparts to the 4\,553 X-ray sources described in the previous section in the \emph{WISE} all-sky catalogue. Because at the flux limits of both \emph{XMM--Newton} and \emph{WISE} the confusion of the sources is minimal (within 5\,arcsec of the X-ray positions there are 4\,100 \emph{WISE} counterparts with seven duplicates), we use a simple proximity criterion to select the counterparts \citep[see e.g.][]{Rovilos2009}. In order to have an estimate of the number of spurious counterparts, we initially select sources from the \emph{WISE} catalogue that are within 60\,arcsec of the X-ray positions. In Figure\,\ref{dRAdDEC} we plot the histograms of the difference in RA and Dec of the counterparts, and in red we plot Gaussians fitted to the distributions. We find a mean $\rm dRA=0.73\pm2.42\,arcsec$ and $\rm dDEC=-0.57\pm1.95\,arcsec$, which are consistent with the astrometric accuracies of the \emph{XMM--Newton} catalogue \citep[$\rm 1-2\,arcsec$; see][]{Watson2009,Georgakakis2011}. The nominal astrometric accuracy of the \emph{WISE} all-sky catalogue is $\sim0.3$\,arcsec at the faintest fluxes. We correct the positional differences between the sources of the two catalogues by the above mean values.

\begin{figure}
\resizebox{\hsize}{!}{\includegraphics{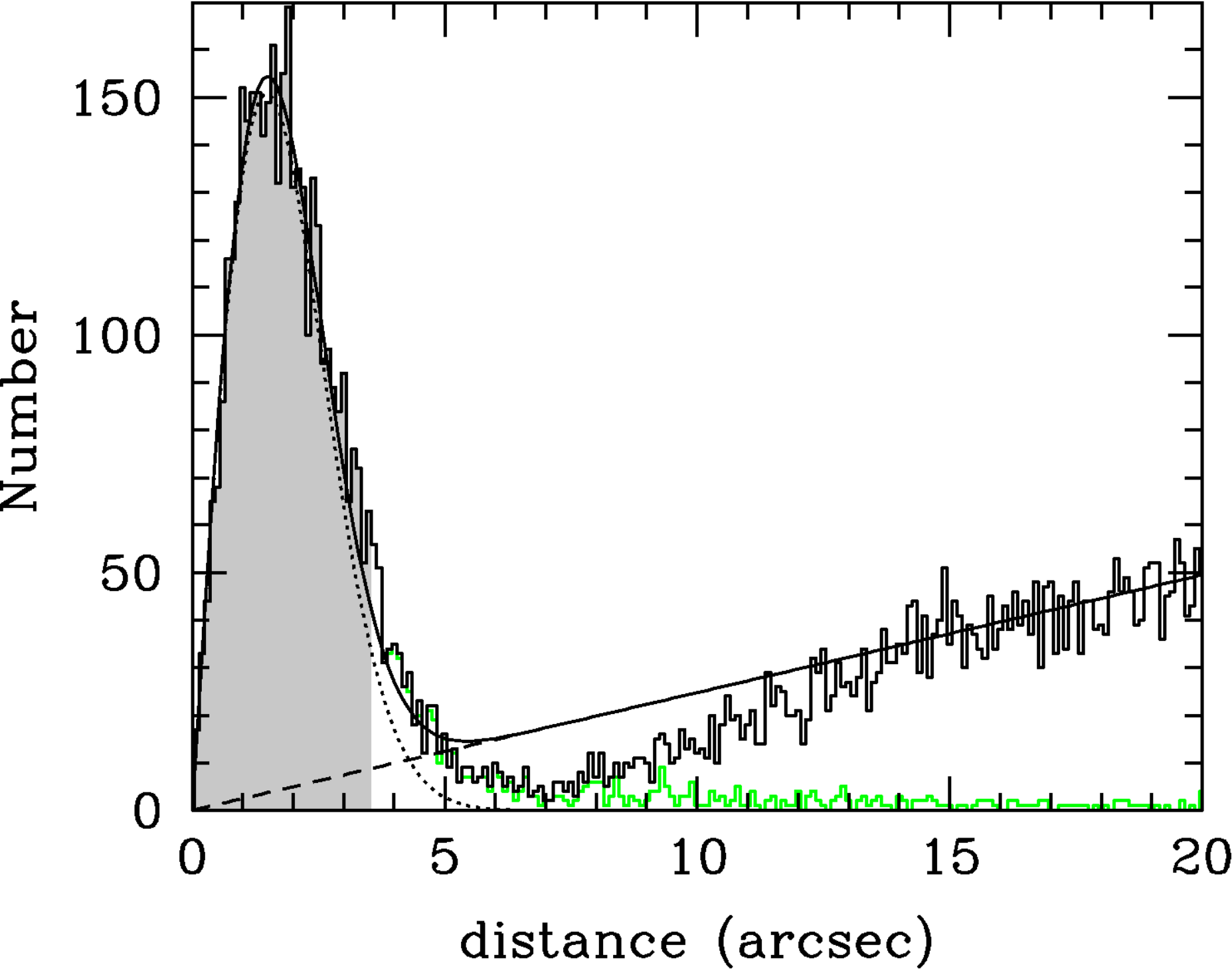}}
  \caption{Histogram of the distances between matched sources form the \emph{XMM--Newton} and \emph{WISE} catalogues. The dotted curve is a Rayleigh distribution calculated from the gaussian fits of Figure\,\ref{dRAdDEC}, and the dashed line is the expected distribution of spurious counterparts, constant in the dRA--dDEC parameter space. The solid curve is their sum, which is a good representation of the observed histogram, except for the 5--12\,arcsec region due to the \emph{WISE} PSF (see \S\,\ref{counterparts}). The shaded area represents the 3.5\,arcsec threshold used in this study, and the green line is the histogram of the distribution of unique counterparts.}
  \label{distances}
\end{figure}

Next, we estimate the number of spurious counterpart matches. Given the distributions of Figure\,\ref{dRAdDEC} and the Gaussian fits, we can assume that most of the counterparts with dRA and dDEC greater than 6\,arcsec ($\approx2\sigma$) are chance matches. In order not to include any real counterparts when assessing the spurious ratio, we measure the number of matches with distances of 20--40\,arcsec, and find 14\,861 cases. Therefore, the density of spurious counterparts in the dRA--dDEC space is 3.9\,arcsec$^{-2}$. In Figure\,\ref{distances} we plot the histogram of the distances of all the counterparts. We model this with a Rayleigh distribution with an amplitude set to be the mean of the amplitudes of the two gaussian distributions and parameter $\sigma^{2}=(\sigma_{\rm RA}^{2}+\sigma_{\rm DEC}^{2})/2$ (dotted line). We also add the expected number of spurious counterparts calculated above (dashed line). The sum of those two distributions is plotted with the solid curve in Figure\,\ref{distances}. We over-predict the number of counterparts with distances 5--12\,arcsec, and we attribute this difference to the finite PSF of the \emph{WISE} survey: if there is a \emph{WISE} source detected close to the position of the X-ray source (being the ``true'' counterpart), another detection is unlikely in its immediate vicinity (5--12\,arcsec), which would be the spurious counterpart, because of the blending of their PSFs. The two sources would become distinguishable if their distance is more than two times the FWHM of the PSF and in \emph{WISE} this is 12\,arcsec. With the green histogram in Figure\ \ref{distances} we plot the distribution of unique counterparts, choosing the nearest case, and this is almost identical with the black histogram below 5\,arcsec. The two distributions (Rayleigh of ``correct'' counterparts and linear of spurious) meet at 4.3\,arcsec, and choosing a limiting radius larger than that would give more chance matches than true counterparts. Since in this study we are searching for a rare type of object (given the high flux density limits), we are more conservative and use 3.5\,arcsec as our limiting radius, indicated by the grey area in Figure\ \ref{distances}. Within this radius we find 3\,689 (3\,685 unique) matches between the \emph{XMM--Newton} and \emph{WISE} catalogues, and the number of spurious counterparts expected within this radius is 150 (4.1\%). We do not find a \emph{WISE} counterpart for 868 sources, something that might introduce a bias in the selection of obscured AGN (see Section\,\ref{CTnumber}). However, such cases have by definition high X-ray to mid-infrared luminosity ratios and would not be selected as candidates, even if the infrared lower limits were such that they would be detected.

\section{Candidate obscured sources}

We select our sample of candidate heavily-obscured AGN based on the X-ray to mid-infrared rest-frame luminosity ratio. \citet{Gandhi2009}, exploring the nuclear X-ray ($\rm 2-10\,keV$) and mid-infrared ($\rm 12.3\,\mu m$) properties of a sample of nearby Seyferts, found a correlation between their rest-frame luminosities, when correcting the X-ray fluxes for internal absorption and using high angular resolution in the mid-infrared to resolve out the host emission \citep[see also][]{Asmus2011}. This correlation is thought to be characteristic of AGN, and any deviations from it (in the form of an infrared excess) should arise from severe obscuration of the X-ray photons. This assumption has been used in the past to select heavily obscured AGN \citep[e.g.][]{Alexander2008,Goulding2011}, and although the samples acquired are not complete, they are reliable \citep[see also][]{Georgantopoulos2011} in the sense that the majority of selected sources have indications of being heavily obscured, especially in the local universe. However, as shown by \citet{Georgakakis2010} and \citet{Asmus2011}, the host galaxy is a contaminant of the mid-infrared flux, which affects relatively low luminosity AGN; these can be mistaken for obscured AGN, whereas in reality they are ``low AGN-to-host infrared sources. To avoid such cases, we de-compose the infrared SEDs of the sources in our sample, as explained below.

\subsection{SED decomposition}
\label{SEDfit}

For the X-ray luminosities we use the $\rm 2-10\,keV$ fluxes from the catalogue of \citet{Georgakakis2011} and a photon index $\Gamma=1.4$ for the k-corrections to obtain rest-frame $\rm 2-10\,keV$ luminosities. Although the detection band is $\rm 2-8\,keV$, the fluxes are for the $\rm 2-10\,keV$ band and are calculated from the photon counts of all three detectors of \emph{XMM-Newton}, after carefully modelling and subtracting the background \citep[see][for more details]{Georgakakis2011}. To calculate the mid-infrared luminosities, we use all the near- and mid-infrared information provided by \emph{WISE}: the photometry in the four \emph{WISE} bands (3.4, 4.6, 12, and $\rm 22\,\mu m$), as well as the photometry in the three 2MASS bands ($J$, $H$, and $K$) for detected sources. The \emph{WISE} catalogue provides near-infrared ($JHK$) photometry for sources with a counterpart in the 2MASS point-source catalogue, based on the best matching 2MASS source. However, some of the low-redshift sources are extended and their near-infrared counterparts are in the 2MASS extended-source catalogue, which is not taken into account. Therefore we look for counterparts of the \emph{XMM}-\emph{WISE} sources in the 2MASS extended source catalogue and find 321 counterparts within 3\,arcsec of the \emph{WISE} positions. For those cases we correct the near-infrared photometry.

In this study we are interested in the mid-infrared luminosity from the AGN and the host galaxy is a potential contaminant that cannot be resolved by \emph{WISE}, we de-compose the infrared SED into an AGN and a galaxy component. We use a custom-built maximum likelihood method to find an optimum combination of a semi-empirical galaxy template from \citet{Chary2001}, with an AGN template of \citet{Silva2004}, and measure the $12\,\mu m$ monochromatic luminosity ($\rm \nu L_{\nu}(12\,\mu m)$) from the AGN template. We do this to sources with a photometric detection in at least three of the $J$, $H$, $K$, $\rm3.4\,\mu m$, $\rm4.6\,\mu m$, $\rm12\,\mu m$, $\rm22\,\mu m$ bands, since we use the combination of two templates, and this selection limits the number of sources from 3\,685 to 2\,844. We combine the photometric errors given in the \emph{WISE} (and/or the 2MASS extended) catalogue with a 10\%-level error of the photometric value in quadrature, to account for the intrinsic error on the SED templates used. The different criteria that were used to select these 2\,844 sources whose SEDs are fitted from the 39\,830 X-ray sources in the \emph{XMM}--SDSS catalogue are summarised in Table\,\ref{numbers}.

\begin{figure*}
\subfloat{\resizebox{87mm}{!}{\includegraphics{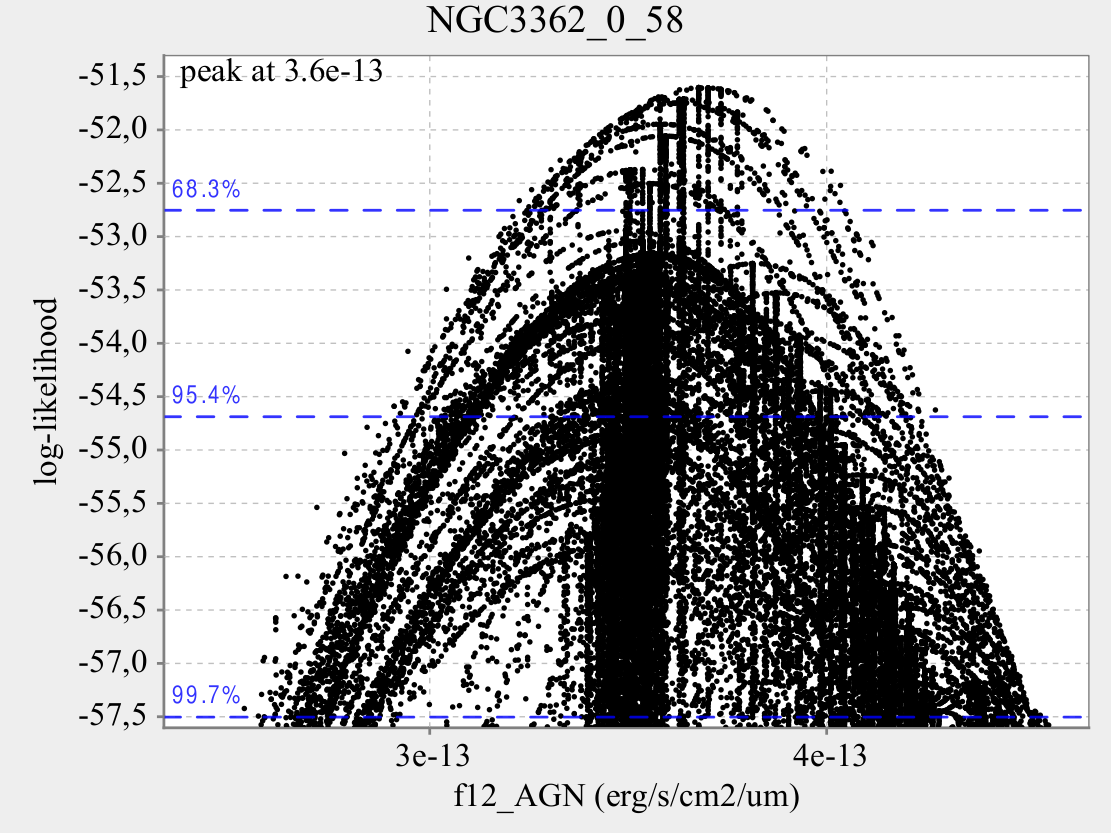}}}
\hspace{1mm}
\subfloat{\resizebox{87mm}{!}{\includegraphics{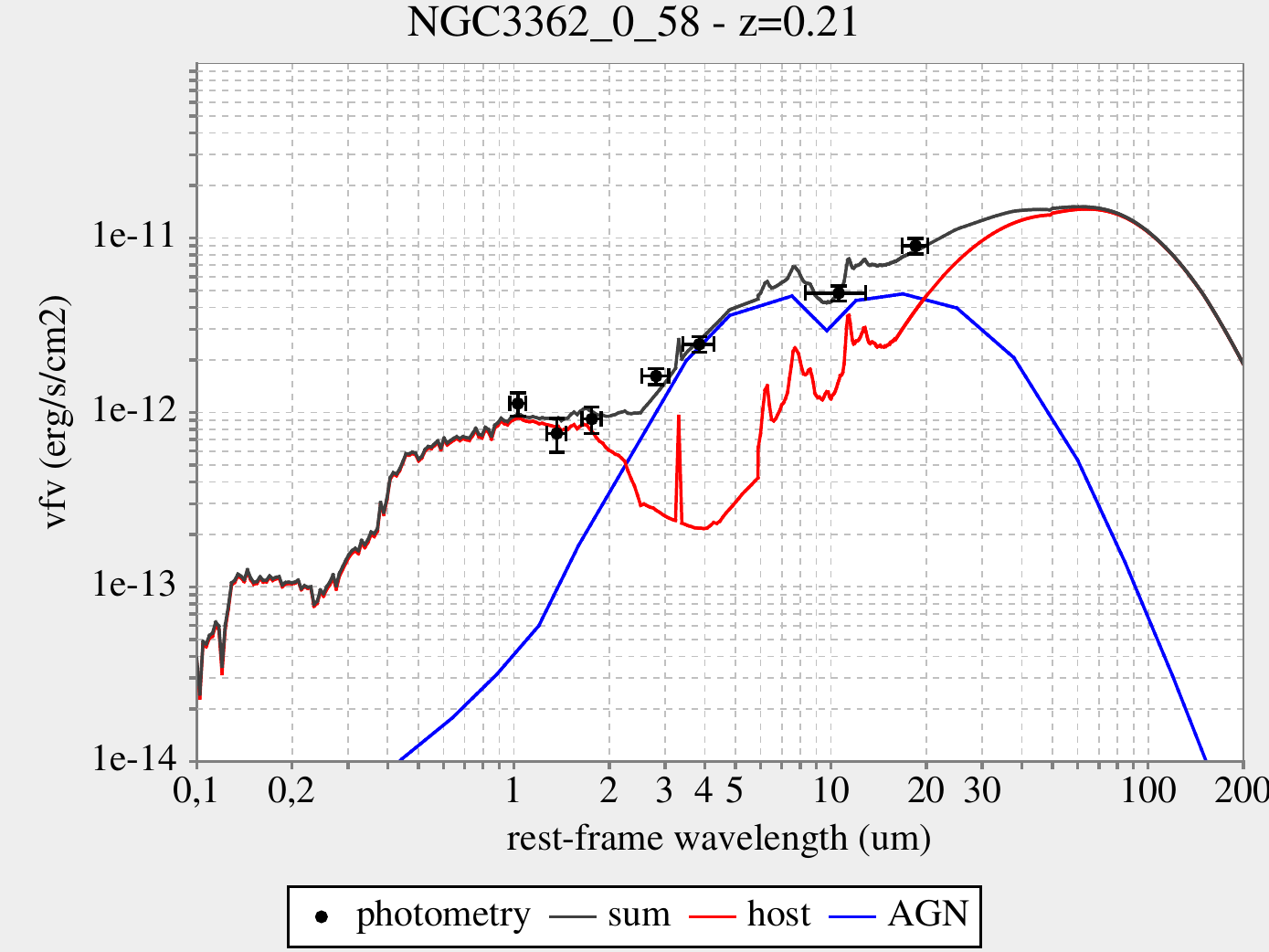}}}
\\
\subfloat{\resizebox{87mm}{!}{\includegraphics{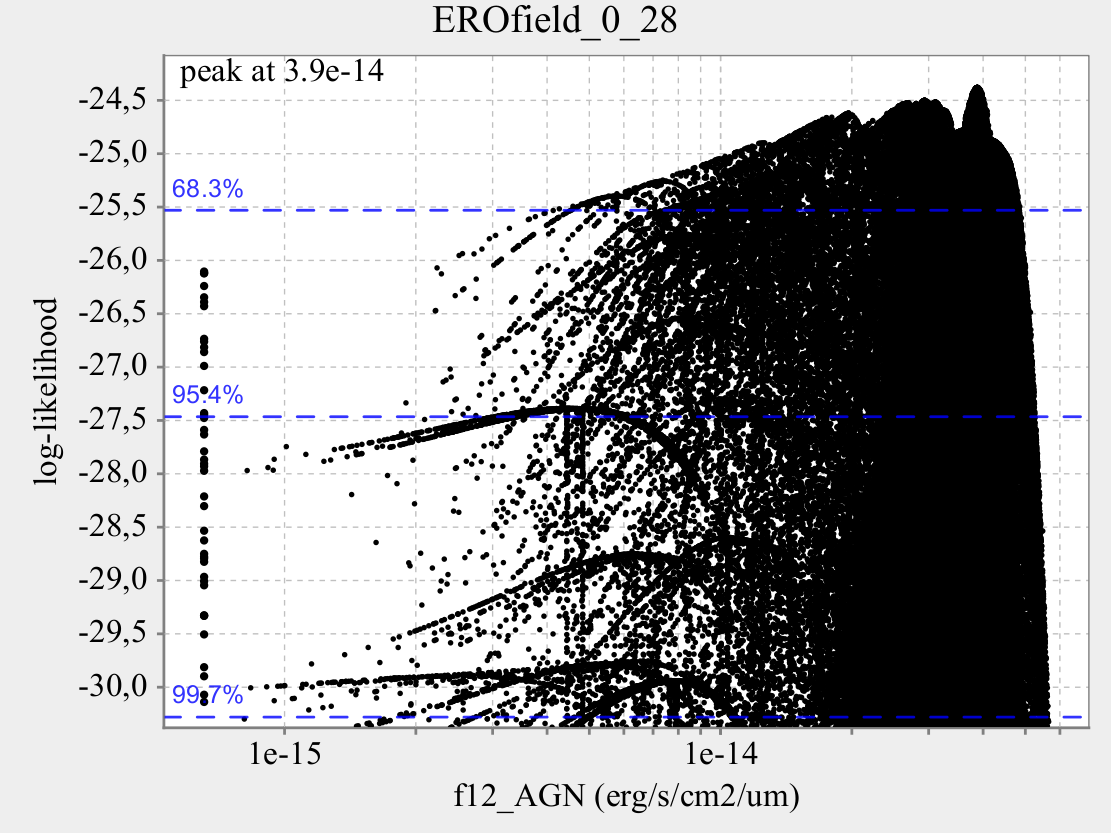}}}
\hspace{1mm}
\subfloat{\resizebox{87mm}{!}{\includegraphics{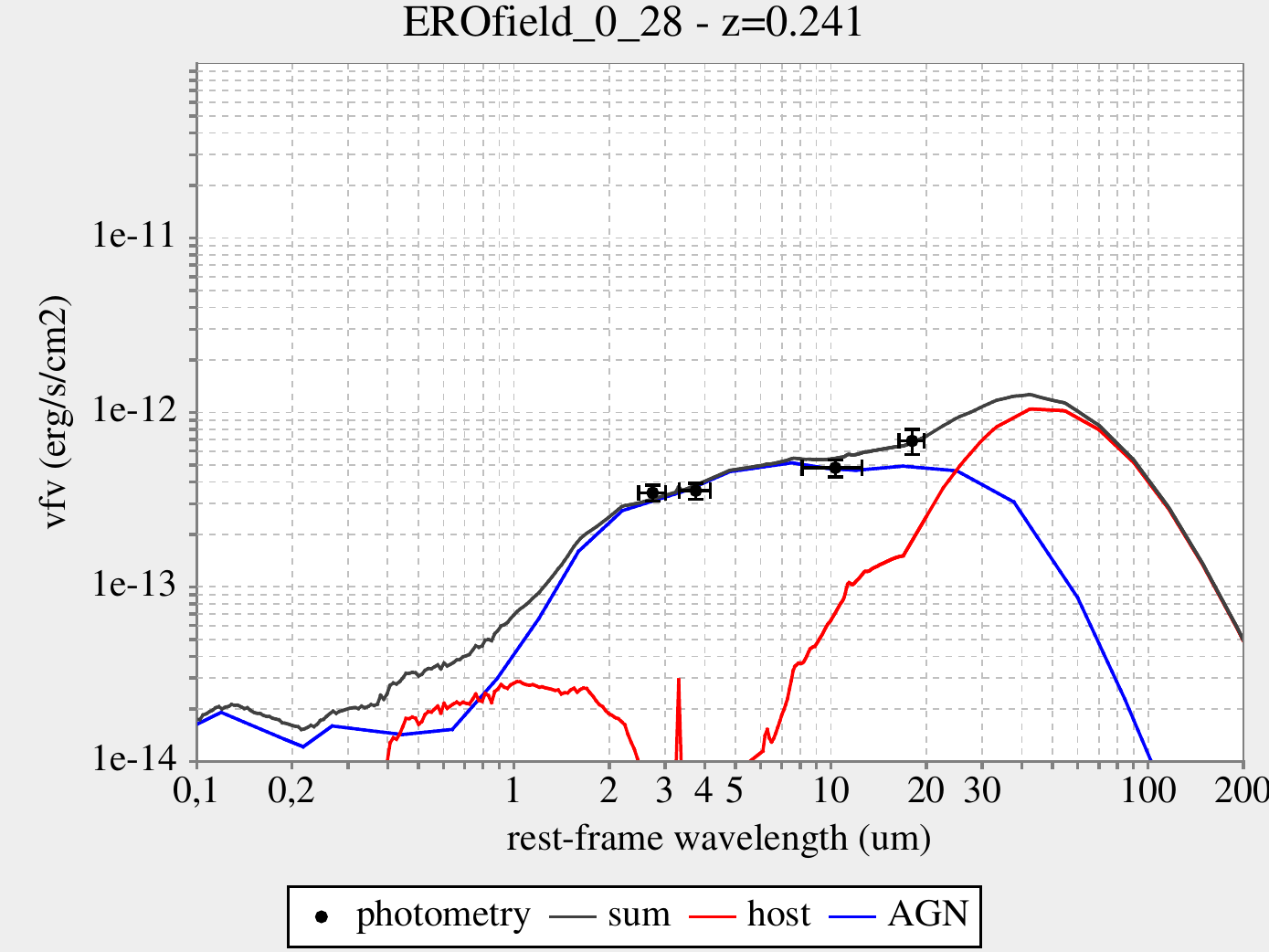}}}
  \caption{Examples of a well-defined value of monochromatic AGN $\rm 12\,\mu m$ luminosity (top panel), and one where only an upper limit can be defined, so that an AGN component is not required for the SED fitting (bottom panel). In the left panels we plot the AGN monochromatic flux density against the log-likelihoods of all the trial fits involved in finding the best-fitting solutions plotted in the right panels. On the right panels the red lines are the host templates that give the best overall fit, the blue lines the AGN templates, and the grey lines their combinations.}
  \label{SED_figures}
\end{figure*}

When trying to decompose the SEDs using multiple components and a limited number of data-points, we are expecting degeneracies between the different fitted components. Therefore, in order to have an estimate of the uncertainty of the $\rm \nu L_{\nu}(12\,\mu m)$ value calculated, for every trial fit we plot the $12\,\mu m$ flux of the AGN template against the (log) likelihood referring to it in the left panels of Figure\,\ref{SED_figures}. An example of reliable and unreliable estimates of $\rm \nu L_{\nu}(12\,\mu m)$, as well as the SED combination with the highest likelihood, are shown in Figure\,\ref{SED_figures}: the right panels show the composite best-fitting SED with the grey line, using the combination of the galaxy (red) and AGN (blue) templates that give the maximum likelihood value. In this case we do not use any priors in the maximum likelihood estimation, so the difference in the natural logarithms of the likelihoods is equivalent to the difference in $\chi^{2}$ of the fits. With the dashed lines on the left plots we indicate the $\chi^{2}$ differences from the best fit corresponding to 68.3\%, 95.4\%, and 99.7\% (or 1, 2, and 3\,$\sigma$) confidence levels. In order to check whether the AGN template is indeed needed, we plot the likelihood values of a single-template fit using only the host template in the far-left column of the likelihood plots. For the cases shown in Figure\,\ref{SED_figures} this is visible only in the lower panel, where the likelihood values are comparable to the ones of the fits involving two templates. In this case it indicates that a solution with no AGN template is almost as likely as the best solution involving the combination of two templates, therefore the AGN template is not statistically important assuming a $2\sigma$ confidence level; its significance is slightly higher than $1\sigma$ according to Figure\,\ref{SED_figures}. The SED decomposition procedure is explained in more detail in Appendix\,\ref{decomposition_details}.

\begin{table}
\caption{Selections made to the initial X-ray source sample}
\label{numbers}
\centering
\begin{tabular}{lc}
\hline\hline
Selection & Number of residual sources \\
\hline
Initial & 39\,830 \\
SDSS counterpart & 19\,431 \\
Redshift determination & 9\,029 \\
Hard X-ray detection & 4\,553 \\
\emph{WISE} counterpart & 3\,685 \\
SED fit ($\geq3$ \emph{WISE}--2MASS bands) & 2\,844 \\
\hline
\end{tabular}
\end{table}

\subsection{Identifying heavily obscured AGN candidates}
\label{CTselection}

\begin{figure}
\resizebox{\hsize}{!}{\includegraphics{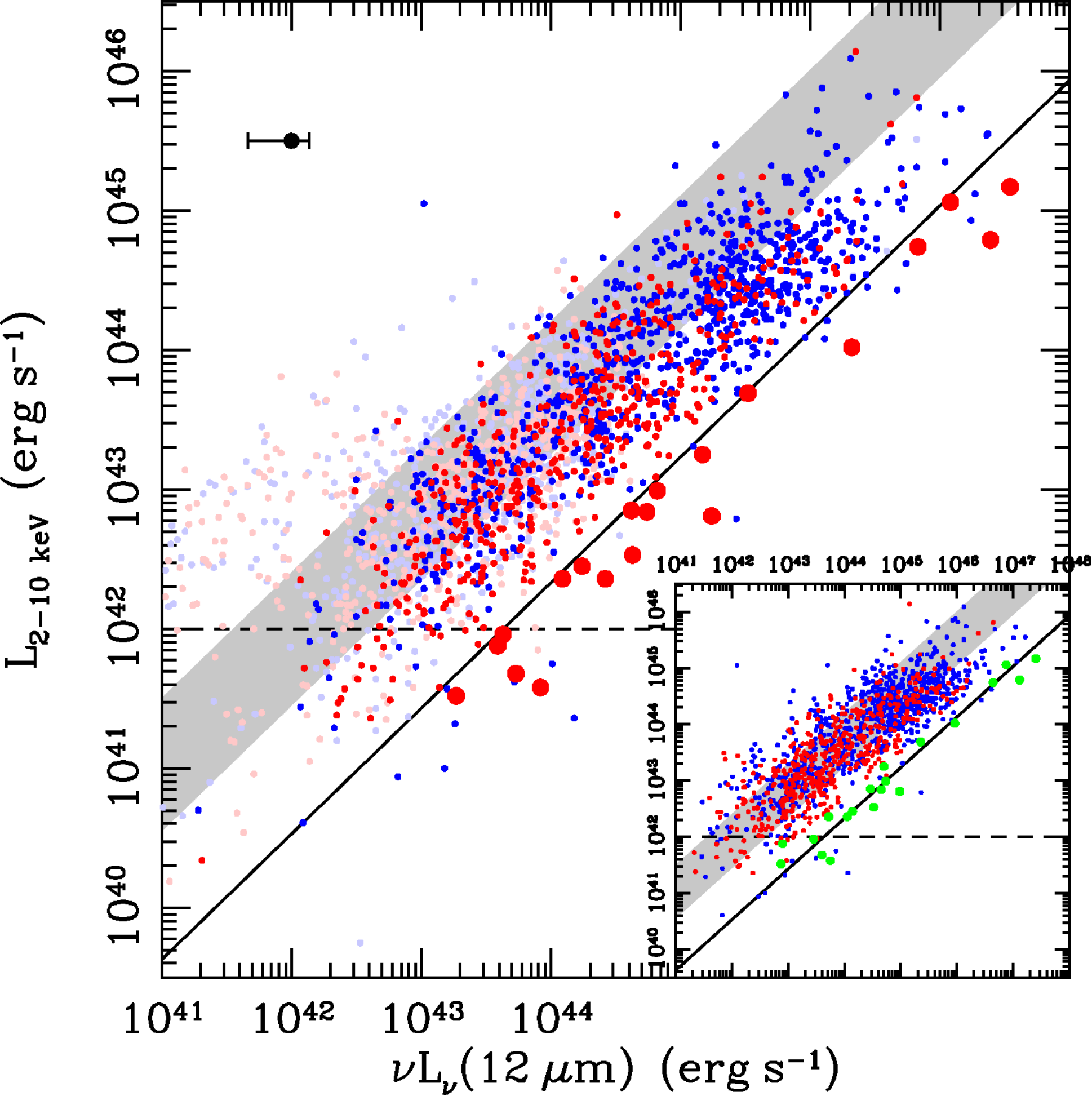}}
  \caption{Hard X-ray ($\rm 2-10\,keV$) luminosities plotted against the $\rm 12\,\mu m$ monochromatic luminosities of the AGN component for soft (blue; $HR<-0.35$) and hard (red; $HR\geq -0.35$) X-ray sources. In light blue and red symbols are plotted cases where the significance of the torus component is below $2\sigma$ (95\%). The black point on the top-left of the diagram shows the median 95\% uncertainty of the intrinsic $\rm 12\,\mu m$ luminosity as a representative error-bar. The expected relation from \citet{Gandhi2009} ($\pm\log(3)$) is shown by the grey area. The selection line (solid line) for our heavily obscured candidates represents a 4\% reflection component from the obscured AGN, and the candidates are represented with the large red symbols. The dashed line represents the limit of $\rm L_{x}=10^{42}\,erg\,s^{-1}$, below that the host galaxy contamination in the X-rays cannot be considered negligible. In the inset image we plot the same values but with the $2\sigma$ lower limit of the mid-infrared luminosity in the x-axis. The candidates are plotted with green colour, and half of them are below the black solid line.}
  \label{lxl12}
\end{figure}

In Figure\,\ref{lxl12} we plot the $\rm 2-10\,keV$ X-ray luminosity against the $\rm 12\,\mu m$ monochromatic luminosity of the AGN component, The grey area represents the relation expected from \citet{Gandhi2009} ($\pm\log(3)$, corresponding to $\rm\sigma\approx0.5\,dex$). The hardness provides an initial indication of the obscuration of the sources, and we plot the soft and hard sources in blue and red colours respectively, taking a hardness ratio ($HR=\frac{H-S}{H+S}$, where $H$ and $S$ are the count rates in the hard and soft bands, respectively) threshold of $HR=-0.35$, which corresponds to $\Gamma=1.4$ \citep[see also][]{Mainieri2007}. Using this threshold 38\% of the fitted sources (1\,089/2\,844) are obscured. In light blue and red symbols we plot sources where the infrared AGN component is not detected with a significance above $2\sigma$, and in the black data-point on the top-left of the image we indicate the median 95\% uncertainty of the intrinsic $\rm 12\,\mu m$ luminosity. The error on the X-ray luminosity coming from the count rate is too small to be plotted in this diagram, however there is an uncertainty on the X-ray flux raising from the shape of the X-ray spectrum, which can be substantial, but a detailed a-priori knowledge is impossible. We correct for this factor for the sources whose spectra we fit, and we note here that it can be a source of the scatter we observe in Figure\,\ref{lxl12}, especially since we are plotting observed values, not corrected for absorption. We observe a shift of 0.08\,dex between the mean values of $L_{\rm2-10\,keV}/\nu L_{\rm\nu(12\,\mu m)}$ of X-ray obscured and unobscured sources, but given that the scatter in both cases is 0.5\,dex, we do not consider the shift important. We observe an overall average shift of a factor of $\sim 2$ between the position of the data-points and the \citet{Gandhi2009} relation, which could be attributed to an absorbing column of $\rm\sim10^{23}\,cm^{-2}$, or some residual contribution of the host galaxy to the derived AGN infrared luminosity. There is evidence however that the X-ray to mid-infrared relation at higher redshifts deviates from the local relation of \citet{Gandhi2009}, but we should consider that the points plotted here correspond to cases where a contribution to the mid-infrared flux from the AGN is detected by the SED fitting, and this is the case for 2\,617 out of the 2\,844 sources fitted, half of them (1\,313) being upper limits within $2\sigma$. The average $L_{\rm2-10\,keV}/\nu L_{\rm\nu(12\,\mu m)}$ ratios therefore are biased towards lower values, and we cannot draw safe conclusions on any significant deviation from the \citet{Gandhi2009} relation.

In order to find the most heavily obscured X-ray sources, we search for sources that significantly deviate from the bulk of the X-ray -- mid-infrared correlation. In heavily obscured sources, only a fraction of the direct X-ray emission from the AGN is detectable, and especially in Compton-thick sources, all we detect are the X-ray photons reflected at the back side of the torus, or scattered by it; the intensity of this reflection/scattered component is typically a few percent of the X-ray energy output of the AGN at 2--8\,keV energies \citep[see][]{Matt2000,Risaliti2004}. The solid line in Figure\,\ref{lxl12} represents the X-ray -- mid-infrared correlation shifted by a factor of 25 in the X-rays, and we will use this to select the heavily obscured candidates for this paper. There are 42 sources (out of the 2\,844 whose infrared SED is fitted) lying below the solid line of Figure\,\ref{lxl12}.

If we consider local Compton-thick AGN where the X-ray luminosity is dominated by the nucleus, they are in general hard X-ray sources. Here we give the examples of Mrk\,3 \citep{Griffiths1998}, NGC\,4945 \citep{Yaqoob2012}, NGC\,7582 \citep{Schachter1998}, NGC\,6240 \citep{Iwasawa1998,Komossa2003}, NGC\,424 \citep{Marinucci2011}, and ESO\,565--G019 \citep{Gandhi2013}. Their observed X-ray emission comes predominantly from a (flat) reflection component and a photo-ionised scattered component, which in general has a soft spectrum. However in most cases the reflected component seems to dominate. In some cases a star-formation X-ray component is also present \citep*[see][]{LaMassa2012}. However, the hardness ratios of these CT AGN in the \emph{XMM--Newton} bands used in this paper would all be $>+0.4$. See \citet{Comastri2004,DellaCeca2008} for a more complete list of nearby Compton-thick X-ray AGN. A notable exception in NGC\,1068, which has a steep X-ray spectrum below $\sim2$\,keV and a flat spectrum at higher energies \citep{Elvis1988}, with the lower energy components being dominant, so that its hardness ratio in the \emph{XMM--Newton} bands used here is $HR\approx-0.45$. As revealed by high resolution (grating) X-ray spectroscopy, the soft X-ray component of NGC\,1068 is a blend of recombination lines coming from photo-ionised regions \citep{Kinkhabwala2002,Brinkman2002}, hence intrinsic in the nuclear region, which dominates in the X-rays, despite the fact that the nuclear region of NGC\,1068 is a vigorously star-forming region \citep*[e.g.][]{Thronson1989,Davies1998}.

Since the reflection component of the AGN is usually flat with $\Gamma\sim1$ \citep[see also][]{George1991}, we exclude sources with $HR<-0.35$, which corresponds to $\Gamma>1.4$, reducing the sample to 22 sources. Our final sample of heavily obscured candidates contains 20 sources, which are plotted with large red symbols in Figure\,\ref{lxl12}, and lie below the black solid line. For two of the 22 initial candidates the torus contribution to the infrared SED is not significant at the $2\sigma$ level, according to the SED decomposition described in the previous section. In the inset plot of Figure\,\ref{lxl12} we plot the 2$\sigma$ lower limits of the torus $\rm 12\,\mu m$ luminosities in the x-axis. The 20 candidate sources are plotted in green colour, and we can see that 10 of them are still below the black solid line. These sources are under-luminous in the X-rays with respect to their mid-infrared luminosities even when their mid-infrared lower limits are considered, and form the most reliable half of the candidate sample.

For the 20 candidate heavily obscured sources we search for the correct counterparts in the 2MASS and SDSS catalogues and also for a far-infrared detection in the \emph{IRAS} catalogue, and perform the SED fitting again, this time also including the optical ($ugriz$) data-points and a separate stellar component. For the latter, we use the stellar population synthesis models of \citet{Bruzual2003}, reddened with the reddening law described in \citet{Calzetti2000}. We use solar metallicity and a varying age and star-formation history to find the optimum template for each source, and using the best-fitting template gives an estimate of the stellar mass of the source. However, some sources in our sample have a point-like morphology, which indicates that the AGN dominates over the optical flux. For such cases, we use AGN templates from the SWIRE template library \citep{Polletta2007}, which have a prominent blue component, and introduce a bayesian prior for the maximum likelihood fit. Indeed in the best-fitting solutions, we see that the AGN template dominates the SDSS bands. For the far-infrared (star formation) part of the SED we use the templates of \citet{Mullaney2011}, which have a better representation of the PAH features than those of \citet{Chary2001}. The new SED fitting confirms that the torus component is statistically important for all 20 sources and that they are under-luminous in the X-rays compared to what predicted from their best-fit torus mid-infrared luminosities and the relation of \citet{Gandhi2009}. We also fit a random sub-sample (450) of the overall sample of the 2\,844 using the three-template approach including the optical bands, and find that their AGN mid-infrared luminosities are similar to that measured using the two-component approach; 85\% are within the 95\% $\nu L_{\rm\nu(12\,\mu m)}$ uncertainty plotted in Figure\,\ref{lxl12}. This happens because the extra (optical) data are fit using the extra (stellar or blue bump) component, without having a major effect on the infrared fit.

\begin{figure}
\resizebox{\hsize}{!}{\includegraphics{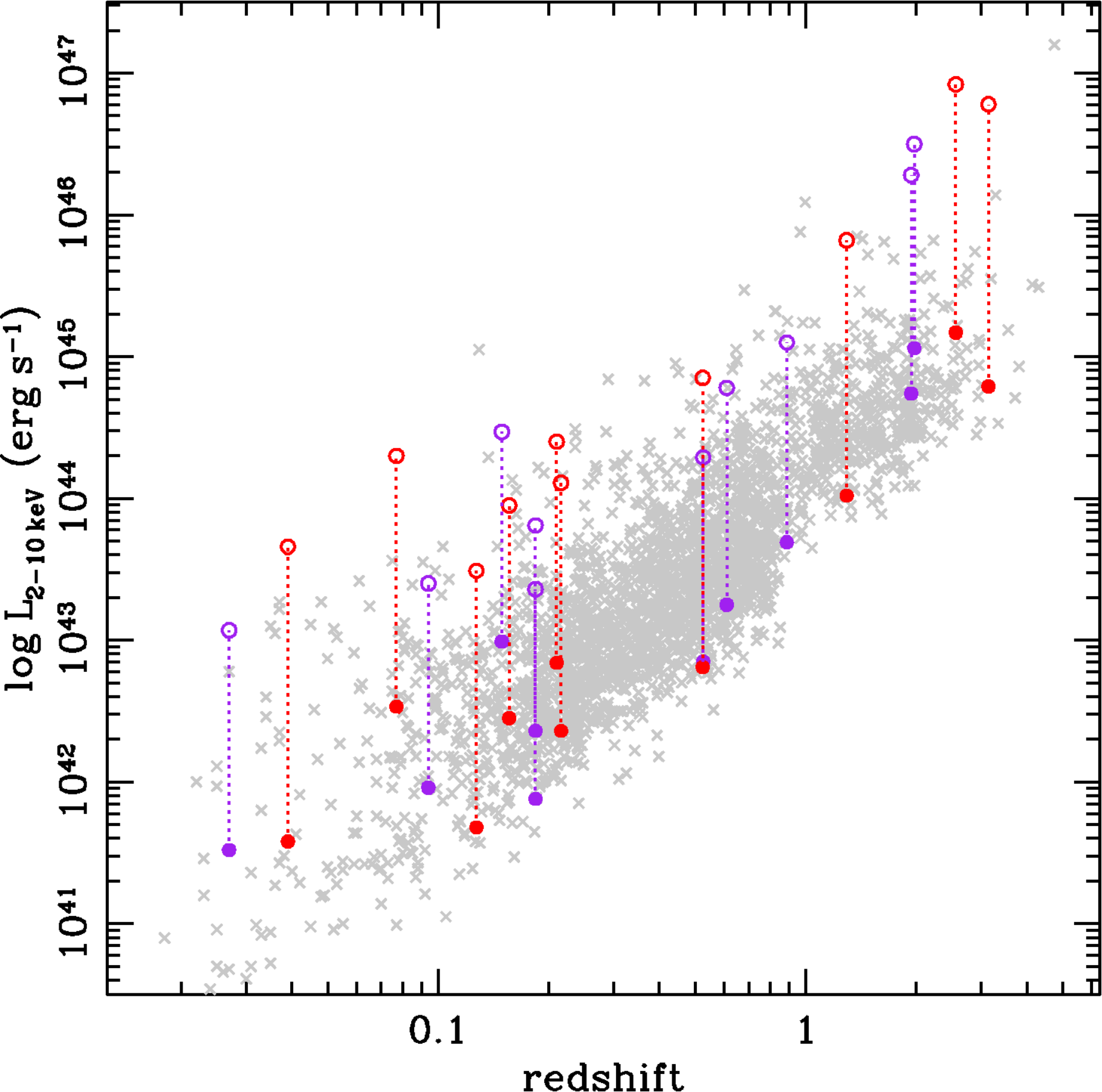}}
  \caption{The redshift distribution of the heavily obscured candidate sources plotted with red and purple points, the red points marking the most reliable outliers. In solid circles we plot the observed X-ray luminosity and in open circles the intrinsic X-ray luminosity inferred from the mid-infrared luminosity of each source. In grey crosses are plotted all X-ray sources for which an infrared SED is fit.}
  \label{Lx-z}
\end{figure}

The basic properties of the heavily obscured AGN candidates are shown in Table\,\ref{candidates}, where in the last column we indicate the number of infrared datapoints used. We note that most sources have seven points, but there are six sources for which only four infrared datapoints can reveal the presence of an AGN. This shows the power of the mid-infrared band in selecting AGN, and happens because of the seemingly different shape of the hot dust SED to that of a typical host at those wavelengths. The redshift distribution is shown in Figure\,\ref{Lx-z} with red and purple points, the red points marking the most reliable outliers. With filled circles we mark the observed X-ray luminosities, and with open circles the intrinsic luminosity derived from the infrared luminosity and the relation of \citet{Gandhi2009}. The grey crosses represent all the X-ray sources for which an infrared SED is fit; we note that the intrinsic luminosities are significantly higher than the mean luminosities of the X-ray sample, and this is what expected for a sample of heavily obscured AGN, yet still observed in the X-rays.

\begin{table*}
\caption{Basic properties of the Compton-thick candidate sources}
\label{candidates}
\centering
\begin{tabular}{ccccccccccl}
\hline\hline
Number & Name & $\log\nu L_{\rm\nu(12\,\mu m)}^{\rm AGN}$ & $\log\nu L_{\rm\nu(12\,\mu m)}^{\rm AGN,lim}$ & $\log L_{\rm 2-10\,keV}$ & $\log L_{\rm 2-10\,keV}^{\rm MIR}$ & $z$ & $[3.4]-[4.6]$ & AGN & $HR$ & N phot \\
(1) & (2) & (3) & (4) & (5) & (6) & (7) & (8) & (9) & (10) & (11) \\
\hline
 1 & J073502.30+265911.6 & 47.08 & 46.87 & 45.06 & 46.50 & 1.973 & 1.251 &  97\% & -0.24 & (7) \\
 2 & J075820.97+392336.0 & 44.42 & 44.05 & 42.36 & 44.11 & 0.216 & 1.580 &  43\% & -0.13 & (7) $\checked$ \\
 3 & J082501.49+300257.3 & 45.52 & 45.35 & 43.69 & 45.10 & 0.888 & 1.530 &  56\% & +0.20 & (4) \\
 4 & J090959.59+542340.5 & 44.62 & 44.46 & 42.85 & 44.29 & 0.526 & 1.173 &  42\% & +0.04 & (4) \\
 5 & J091848.61+211717.1 & 44.82 & 44.73 & 42.99 & 44.47 & 0.149 & 1.034 &  99\% & -0.22 & (7) \\
 6 & J093551.60+612111.8 & 43.92 & 43.75 & 41.58 & 43.66 & 0.039 & 1.688 &  40\% & -0.15 & (7) $\checked$ \\
 7 & J093857.01+412821.1 & 46.83 & 46.64 & 44.74 & 46.28 & 1.935 & 1.310 &  97\% & -0.20 & (7) \\
 8 & J094021.12+033144.8 & 46.32 & 45.96 & 44.02 & 45.82 & 1.292 & 1.426 &  80\% & +0.29 & (7) $\checked$ \\
 9 & J104426.70+063753.9 & 44.74 & 44.65 & 42.84 & 44.40 & 0.210 & 1.437 &  63\% & +0.58 & (7) $\checked$ \\
10 & J111847.01+075419.6 & 43.73 & 43.60 & 41.68 & 43.49 & 0.127 & 1.258 &  29\% & -0.30 & (7) $\checked$ \\
11 & J112611.63+425246.5 & 44.24 & 44.14 & 42.45 & 43.95 & 0.156 & 1.087 &  57\% & +0.69 & (7) $\checked$ \\
12 & J113240.25+525701.3 & 43.27 & 42.87 & 41.52 & 43.07 & 0.027 & 0.869 &  36\% & -0.17 & (7) \\
13 & J121839.40+470627.7 & 43.63 & 43.45 & 41.96 & 43.40 & 0.094 & 1.144 &  39\% & -0.32 & (7) \\
14 & J124410.21+164748.2 & 45.17 & 44.70 & 43.25 & 44.78 & 0.609 & 1.462 &  98\% & +0.51 & (4) \\
15 & J132415.92+655337.8 & 44.09 & 43.72 & 42.36 & 43.81 & 0.184 & 1.317 &  96\% & +0.46 & (6) \\
16 & J132827.08+581836.9 & 47.39 & 47.11 & 44.79 & 46.78 & 3.139 & 1.028 & 100\% & -0.20 & (4) $\checked$ \\
17 & J133332.07+503519.7 & 45.24 & 44.98 & 42.81 & 44.85 & 0.524 & 1.604 &  66\% & -0.08 & (4) $\checked$ \\
18 & J133756.94+043325.8 & 43.59 & 42.90 & 41.88 & 43.36 & 0.184 & 1.283 &  87\% & +0.43 & (4) \\
19 & J140700.40+282714.7 & 44.63 & 44.52 & 42.53 & 44.30 & 0.077 & 1.008 &  70\% & -0.17 & (7) $\checked$ \\
20 & J141546.24+112943.5 & 47.54 & 47.40 & 45.17 & 46.92 & 2.560 & 1.333 &  76\% & +0.19 & (7) $\checked$ \\
\hline
\end{tabular}
\begin{list}{}{}
\item The columns are: (1) Number;
                       (2) Source Name;
                       (3) $\rm 12\,\mu m$ luminosity of the AGN, based on SED fitting, in $\rm erg\,s^{-1}$;
                       (4) $\rm 12\,\mu m$ luminosity lower limit of the AGN, based on SED fitting, in $\rm erg\,s^{-1}$;
                       (5) Observed X-ray luminosity, in $\rm erg\,s^{-1}$;
                       (6) Expected intrinsic X-ray luminosity based on the $\rm12\,\mu m$ luminosity and the relation of \citet{Gandhi2009}, in $\rm erg\,s^{-1}$;
                       (7) Redshift;
                       (8) Vega mid-infrared colour;
                       (9) Fraction of the AGN component to the $\rm12\,\mu m$ flux, based on SED fitting;
                       (10) Hardness ratio between the $0.5-2$\,keV and $2-10$\,keV bands;
                       (11) In brackets is the number of infrared data-points used for the SED decomposition. A `$\checked$' symbol means that the source is in the most reliable half of the candidate sample.
\end{list}
\end{table*}

\section{Sample properties} 

In this section we investigate the multi-wavelength properties of the 20 candidates, in the X-rays using the data described in \S\,\ref{XrayData}, and in the optical using the SDSS parameters.

\subsection{X-ray spectra}

We investigate the X-ray properties of the sources in our sample by performing spectral fittings with the {\sc XSPEC v.12.8} software package \citep{Arnaud1996}. The goal is to identify heavily obscured AGN via X-ray spectral analysis. The X-ray data have been obtained with the EPIC \citep[European Photon Imaging Cameras;][]{Struder2001,Turner2001} on board \emph{XMM--Newton}. The \emph{XMM--Newton} observations' details corresponding to the heavily obscured candidate sources are reported in Table\,\ref{log}. The data have been analysed using the Scientific Analysis Software ({\sl SAS v.7.1}). We produce event files for the pn-CCD and the MOS-1 and MOS-2 (Metal Oxide Semiconductor) observations using the {\sl EPCHAIN} and {\sl EMCHAIN} tasks of {\sl SAS} respectively. The event files are screened for high particle background periods. In our analysis we deal only with events corresponding to patterns 0--4 for the pn and 0--12 for the MOS instruments. Spectra for sources with more than 100 combined counts are extracted from circular regions with radius of 20\,arcsec. This area encircles at least 70 per cent of the source X-ray photons at off-axis angles less than 10\,arcmin. A ten times larger, source-free area is used for the background spectra. The response and ancillary files are also produced using {\sl SAS} tasks {\sl RMFGEN} and {\sl ARFGEN} respectively. We employ C-statistics \citep{Cash1979}, which had been specifically developed to extract spectral information from data of low signal-to-noise ratio. This statistic works on un-binned data, allowing us, in principle, to use the full spectral resolution of the instruments without degrading it by binning. We fit the PN and the MOS data simultaneously in the 0.5--8\,keV range. We assume a standard power-law model with two absorption components plus a Gaussian line to account for the FeK$\alpha$ line ({\sc wa*zwa*(po+zga)} in {\sc XSPEC} notation). The first absorption component models the Galactic absorption. Its fixed values  are obtained from \citet{Dickey1990} and are listed in Table\,\ref{log}. The second absorption component represents the AGN intrinsic absorption and it is left as a free parameter during the model fitting procedure. The rest-frame energy of the FeK$\alpha$ line is fixed to 6.4 keV. Note that in the case of source \#\,4 the FeK$\alpha$ line may have a different energy. In this case only a photometric redshift is available, z=0.53, and the PN detector shows a line, that if associated with FeK$\alpha$ would suggest a redshift of z=0.42. The EW of the line is $ 0.7^{+0.60}_{-0.65}$ keV. The best fit parameters for all sources with more than 100 net combined counts (PN+MOS) are reported in Table \ref{PLFits}, for the rest a reliable fit cannot be made. The errors quoted correspond to the 90 per cent confidence level for the parameter of interest.

\begin{table*}
\centering
\caption{Log of the \emph{XMM--Newton} observations}
\label{log}
\begin{tabular}{cccccccccc}
\hline\hline
Number & obsID      & Name                & Field       & $z$   & $N_{\rm H}$ & exp pn & exp MOS & cts pn & cts MOS \\
(1)    & (2)        & (3)                 & (4)         & (5)   & (6)         & (7)    & (8)     & (9)    & (10)    \\
\hline
 1     & 0503630101 & J073502.30+265911.6 & 2MASXJ074   & 1.973 & 4.9         &  2.0   &  2.4    &  229   &  177    \\
 2     & 0406740101 & J075820.97+392336.0 & FBQSJ0758   & 0.216 & 5.0         &  1.1   &  1.4    &   29   &   29    \\
 3     & 0504102001 & J082501.49+300257.3 & SDSS0824    & 0.89  & 3.6         &  1.9   &  2.2    &   52   &   45    \\
 4     & 0200960101 & J090959.59+542340.5 & XYUMA       & 0.53  & 2.0         &  7.0   &  8.3    &  253   &  140    \\
 5     & 0303360101 & J091848.61+211717.1 & 2MASSI091   & 0.149 & 4.2         &  1.8   &  2.1    & 7841   & 6045    \\
 6     & 0085640201 & J093551.60+612111.8 & UGC05101    & 0.039 & 2.7         &  2.7   &  3.4    &  611   &  560    \\
 7     & 0504621001 & J093857.01+412821.1 & J093857.0   & 1.935 & 1.5         &  1.4   &  1.9    &   79   &  120    \\
 8     & 0306050201 & J094021.12+033144.8 & Mrk1419     & 1.292 & 3.6         &  2.2   &  2.6    &   36   &   37    \\
 9     & 0405240901 & J104426.70+063753.9 & NGC3362     & 0.210 & 2.8         &  2.6   &  3.1    &  187   &  102    \\
10     & 0203560201 & J111847.01+075419.6 & PG1115      & 0.127 & 3.6         &  7.0   &  8.0    &   65   &   58    \\
11     & 0110660401 & J112611.63+425246.5 & HVCComple   & 0.156 & 2.0         &  0.8   &  1.3    &   24   &   65    \\
12     & 0200430501 & J113240.25+525701.3 & UGC6527     & 0.027 & 3.6         &  9.7   & 12.0    &  475   &  329    \\
13     & 0203270201 & J121839.40+470627.7 & RXJ121803   & 0.094 & 1.2         &  4.1   &  4.8    &  221   &  250    \\
14     & 0302581501 & J124410.21+164748.2 & MS1241.5    & 0.61  & 1.8         &  2.0   &  2.9    &   31   &   29    \\
15     & 0206180201 & J132415.92+655337.8 & WARPJ1325   & 0.18  & 2.0         &  3.5   &  3.5    &   93   &   55    \\
16     & 0405690101 & J132827.08+581836.9 & NGC5204     & 3.139 & 1.7         &  -     &  5.9    &   -    &   10    \\
17     & 0142860201 & J133332.07+503519.7 & RXJ1334.3   & 0.52  & 1.0         &  5.0   &  5.6    &   93   &   27    \\
18     & 0152940101 & J133756.94+043325.8 & NGC5252     & 0.184 & 2.0         &  5.5   &  6.2    &  183   &   46    \\
19     & 0140960101 & J140700.40+282714.7 & Mrk668      & 0.077 & 1.4         &  1.9   &  2.2    &  961   &  763    \\
20     & 0112250301 & J141546.24+112943.5 & H1413       & 2.560 & 1.8         &  2.0   &  2.6    &  189   &  120    \\
\hline\hline
\end{tabular}
\begin{list}{}{}
\item The columns are: (1) Number;
                       (2) Observation-ID;
                       (3) Source Name;
                       (4) Field Name;
                       (5) Redshift: two and three decimal digits denote photometric and spectroscopic redshifts respectively;
                       (6) Galactic column density in units $\rm 10^{20}\,cm^{-2}$;
                       (7) pn exposure in units $\rm 10^{4}\,s$;
                       (8) MOS exposure in units of $\rm 10^{4}\,s$;
                       (9) pn net counts;
                       (10) Sum of MOS-1 and MOS-2 net counts;
\end{list}
\end{table*}

\begin{table*}
\centering
\caption{The X-ray spectral fits using a power-law component and a Gaussian FeK$\alpha$ line}
\label{PLFits}
\begin{tabular}{ccccccccc}
\hline\hline
Number      & obsID      & $N_{\rm H}$           & $\Gamma$                & EW                           & c-stat    & $f_{\rm 2-10\,keV}$ & $L_{\rm 2-10\,keV}$ & Notes   \\
 (1)        & (2)        & (3)                   & (4)                     & (5)                          & (6)       & (7)                 & (8)                 &         \\
\hline
 1$\dagger$ & 0503630101 & $<1.9$                & $1.91^{+0.46}_{-0.34}$  & $0.77^{+0.54}_{-0.47}$       & 2070/2498 & $7.2\times10^{-14}$ & $1.8\times10^{45}$  & $\times$ \\
 4$\dagger$ & 0200960101 & $<1.7$                & $-0.32^{+0.47}_{-0.46}$ & $<0.63\ddag$                 & 2719/2498 & $8.1\times10^{-14}$ & $3.4\times10^{43}$  & $\times$ \\
 5          & 0303360101 & $0.2^{+0.02}_{-0.01}$ & $1.87^{+0.04}_{-0.04}$  & $<0.13$                      & 2756/2498 & $1.4\times10^{-12}$ & $8.1\times10^{43}$  & $\times$ \\
 6$\dagger$ & 0085640201 & $<0.04$               & $1.10^{+0.12}_{-0.09}$  & $0.86^{+0.20}_{-0.10}$       & 2244/2498 & $1.4\times10^{-13}$ & $4.6\times10^{41}$  &          \\
 7          & 0504621001 & $<5.0$                & $2.90^{+2.1}_{-1.1}$    & $<2.5$                       & 2651/2498 & $8.0\times10^{-15}$ & $4.6\times10^{44}$  &          \\
 9$\dagger$ & 0405240901 & $<0.01$               & 1.8                     & $12.5_{-3.5}^{+3.0}$         & 1529/2499 & $1.0\times10^{-14}$ & $1.2\times10^{42}$  &          \\
10          & 0203560201 & $<0.05$               & 1.8                     & $<5.9$                       & 2039/2500 & $7.5\times10^{-15}$ & $3.1\times10^{41}$  &          \\
12$\dagger$ & 0200430501 & $<0.01$               & $1.07^{+0.15}_{-0.13}$  & $2.0^{+0.7}_{-0.3}$          & 2236/2498 & $3.7\times10^{-13}$ & $6.0\times10^{41}$  & $\times$ \\
13$\dagger$ & 0203270201 & $<0.01$               & 1.8                     & $3.^{+2.2}_{-1.0}$           & 2261/2499 & $2.2\times10^{-14}$ & $4.9\times10^{41}$  &          \\
15          & 0206180201 & $3.0^{+1.7}_{-1.4}$   & 1.8                     & $<0.62$                      &  973/1621 & $2.0\times10^{-14}$ & $1.7\times10^{42}$  &          \\
17          & 0142860201 & $<1$                  & 1.8                     & $<2.0$                       & 1681/2090 & $7.5\times10^{-15}$ & $7.4\times10^{42}$  &          \\
18$\dagger$ & 0152940101 & $<81$                 & $-0.72^{+1.20}_{-1.25}$ & $<0.88$                      & 2724/2498 & $4.4\times10^{-14}$ & $2.7\times10^{42}$  & $\times$ \\
19$\dagger$ & 0140960101 & $<0.01$               & $1.17^{+0.07}_{-0.07}$  & $0.91^{-0.21}_{+0.27}$       & 2631/2498 & $3.0\times10^{-13}$ & $4.1\times10^{42}$  &          \\
20$\dagger$ & 0112250301 & $37^{+19}_{-11}$      & $2.22^{+0.62}_{-0.55}$  & $<0.10$                      & 1441/2498 & $4.1\times10^{-14}$ & $8.7\times10^{44}$  &          \\
\hline\hline
\end{tabular}
\begin{list}{}{}
\item The columns are: (1) Source number;
                       (2) obsID;
                       (3) Intrinsic hydrogen column density in units of $\rm10^{22}\,cm^{-2}$;
                       (4) Photon index;
                       (5) equivalent width of the FeK$\alpha$ line in unit of keV;
                       (6) c-statistic/degrees of freedom; 
                       (7) Flux (2--10\,keV) in units of $\rm erg\,cm^{-2}\,s^{-1}$	 
                       (8) Observed luminosity (2--10\,keV) in units of $\rm erg\,s^{-1}$. Blank entries mean that no spectral fit could be performed owing to the limited photon statistics. Values with no errors were fixed to the quoted value.
                       (9) A $\times$ symbol means that the source in not X-ray under-luminous if the X-ray luminosity is measured from the spectrum
\item $\dagger$: Possible presence of a strong FeK$\alpha$ line, a flat, or two component spectrum. More detailed fits of these sources are given in Tables\,\ref{double} and \ref{reflection}.
\item \ddag: PN shows a line at an energy of $\rm 4.5^{+0.13}_{-0.2}\,keV$ with an EW of $\rm 0.70^{+0.6}_{-0.65}\,keV$; if confirmed and this line is associated with FeK$\alpha$, this would suggest a redshift of $z=0.42$.
\end{list}
\end{table*}

The X-ray spectra provide a more accurate measurement of the $\rm 2-10\,keV$ flux, and we note that for some cases the value measured from the X-ray spectrum is seemingly different than the one measured from the counts in \citet{Georgakakis2011}. This happens because of the difference in the spectral slopes of some sources from the $\Gamma=1.4$ value used for all sources in \citet{Georgakakis2011}, a value taken to match the X-ray background. We use the new $L_{\rm 2-10\,keV}$ to check whether the 20 candidates are still X-ray under-luminous: we find that five of them would not be included in our candidate sample using their updated X-ray luminosities, none of which are among the most reliable candidates. We mark those sources with a cross in Table\,\ref{PLFits}. We will use the X-ray flux and luminosities from the spectral analysis hereafter.

\subsubsection{X-ray obscured AGN}

\begin{figure*}
\begin{center}
\includegraphics[width=5.8cm]{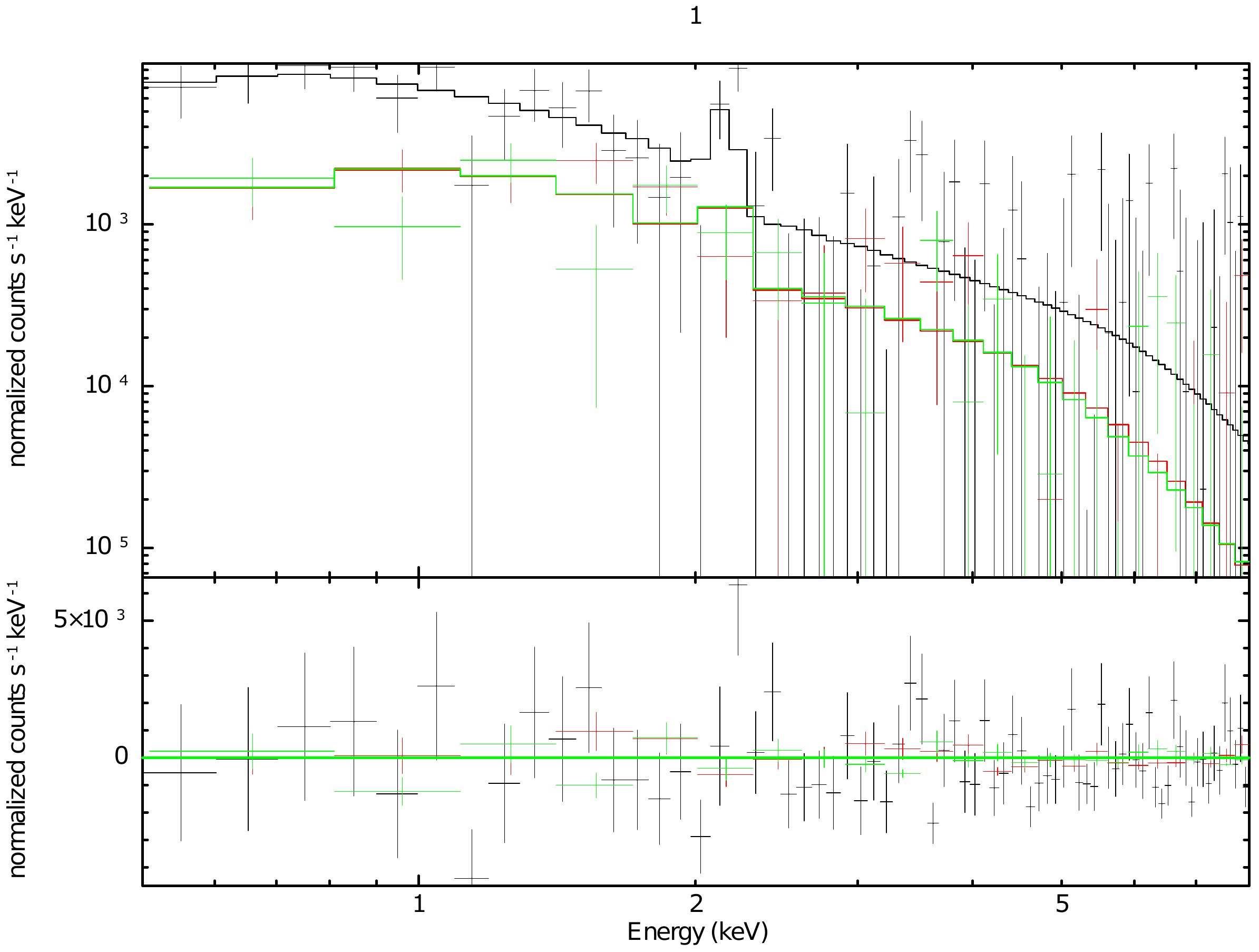} 
\includegraphics[width=5.8cm]{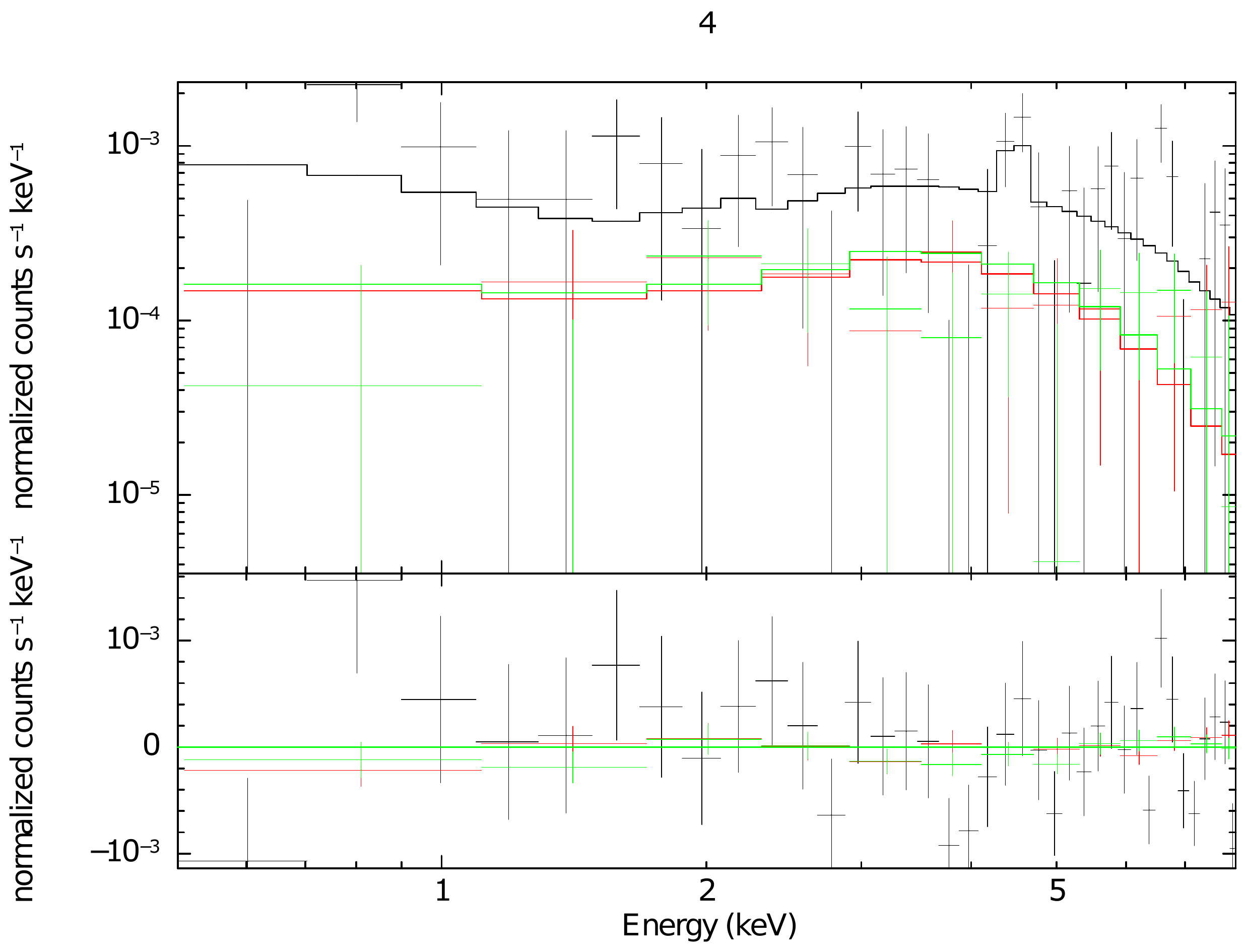} 
\includegraphics[width=5.8cm]{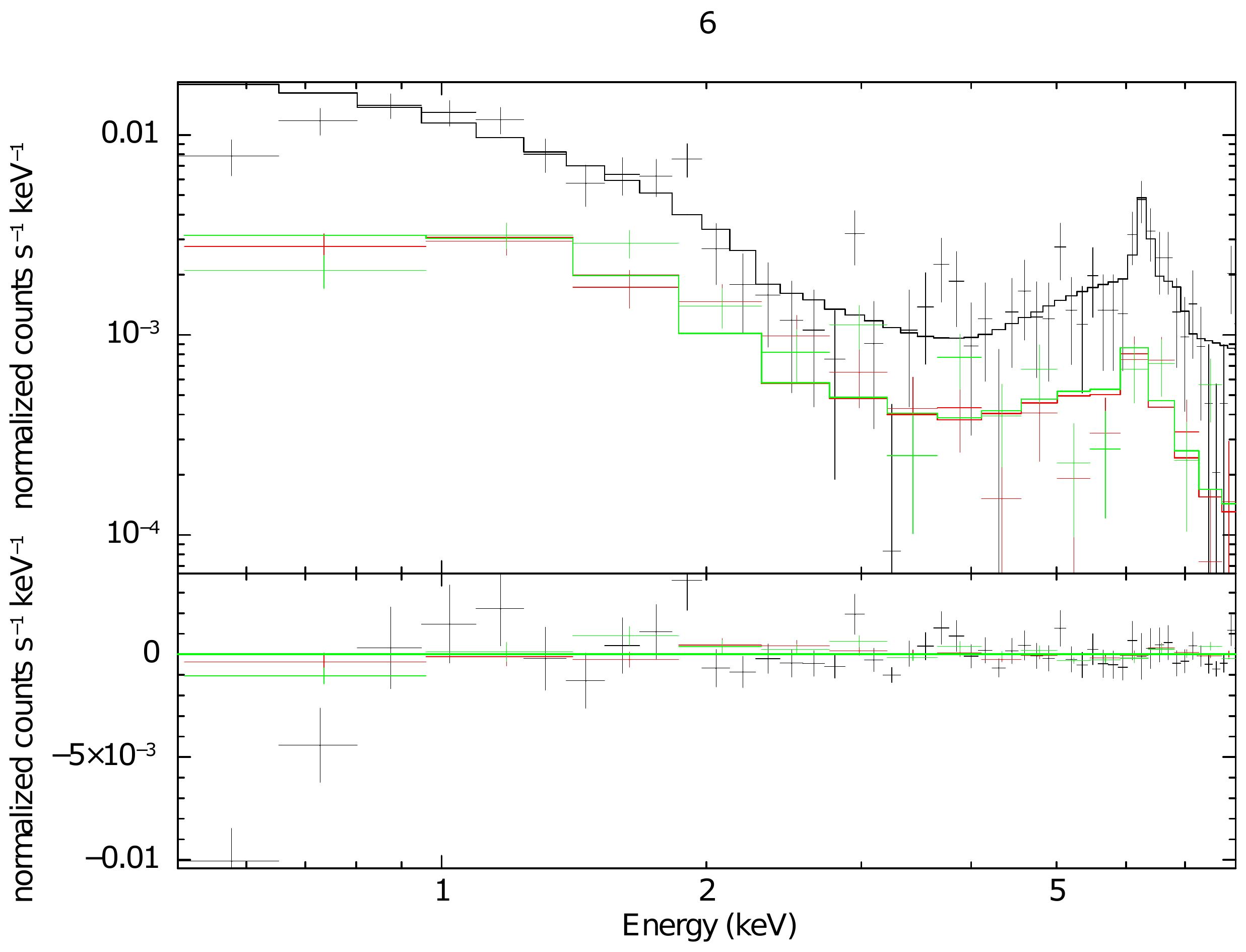} 
\includegraphics[width=5.8cm]{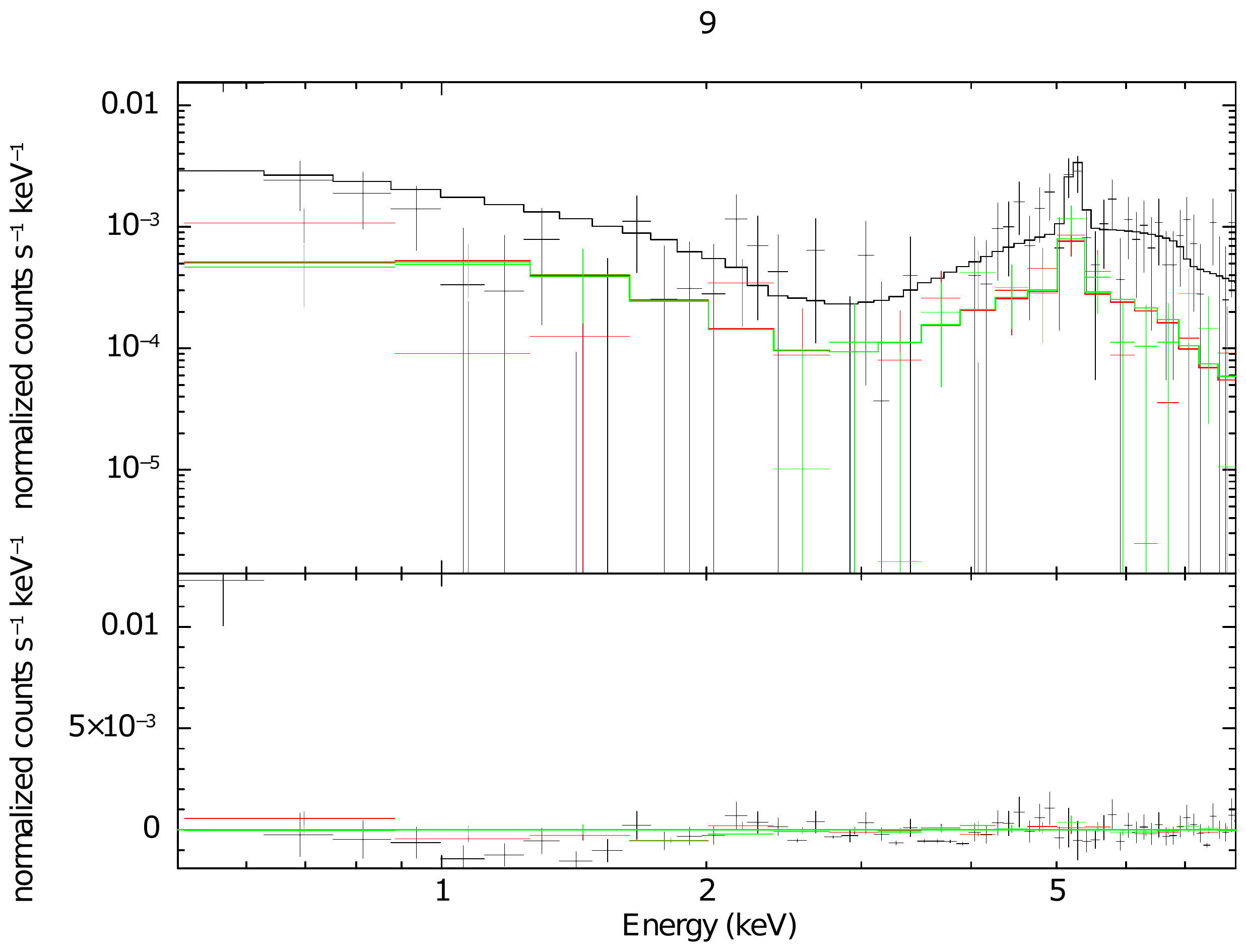} 
\includegraphics[width=5.8cm]{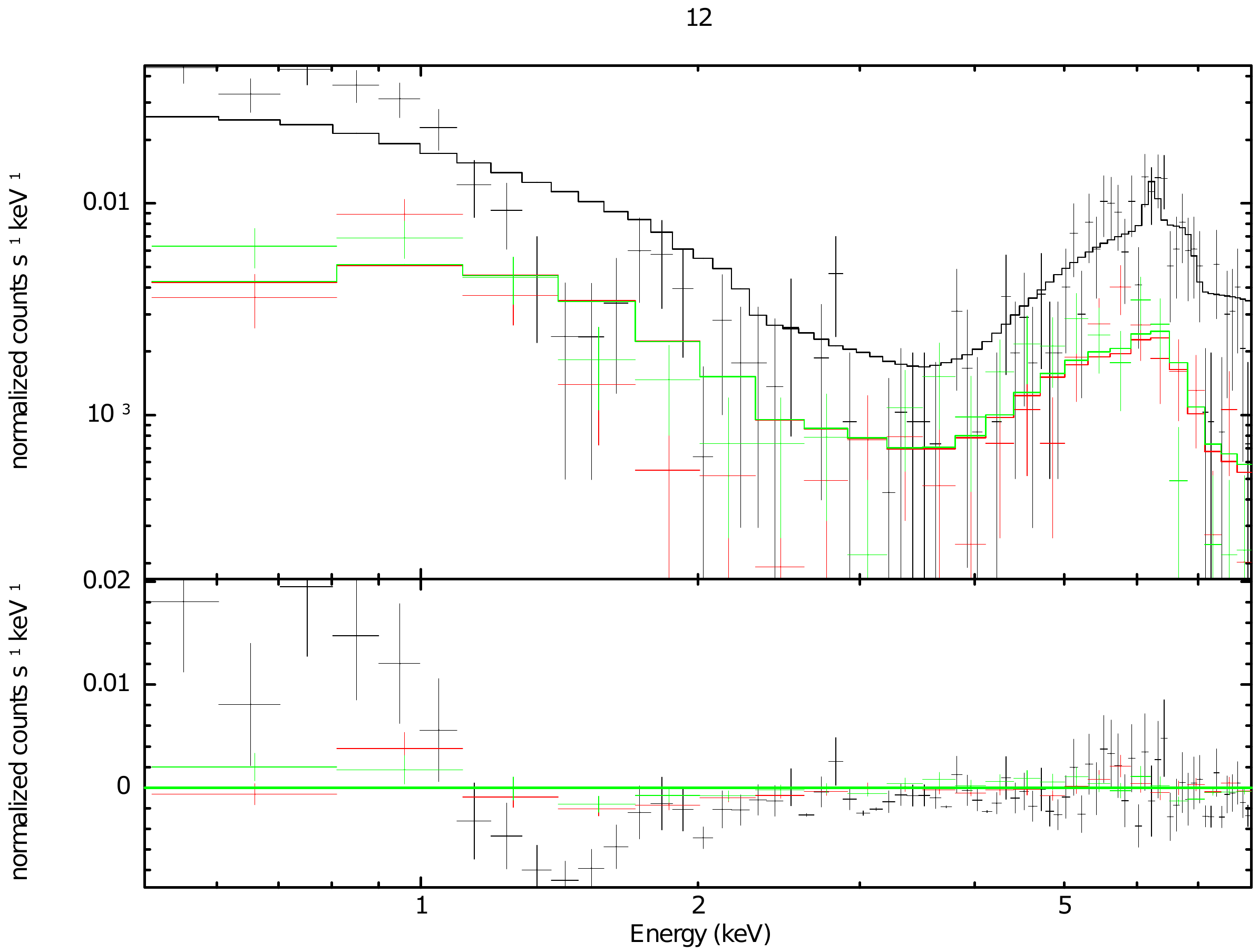} 
\includegraphics[width=5.8cm]{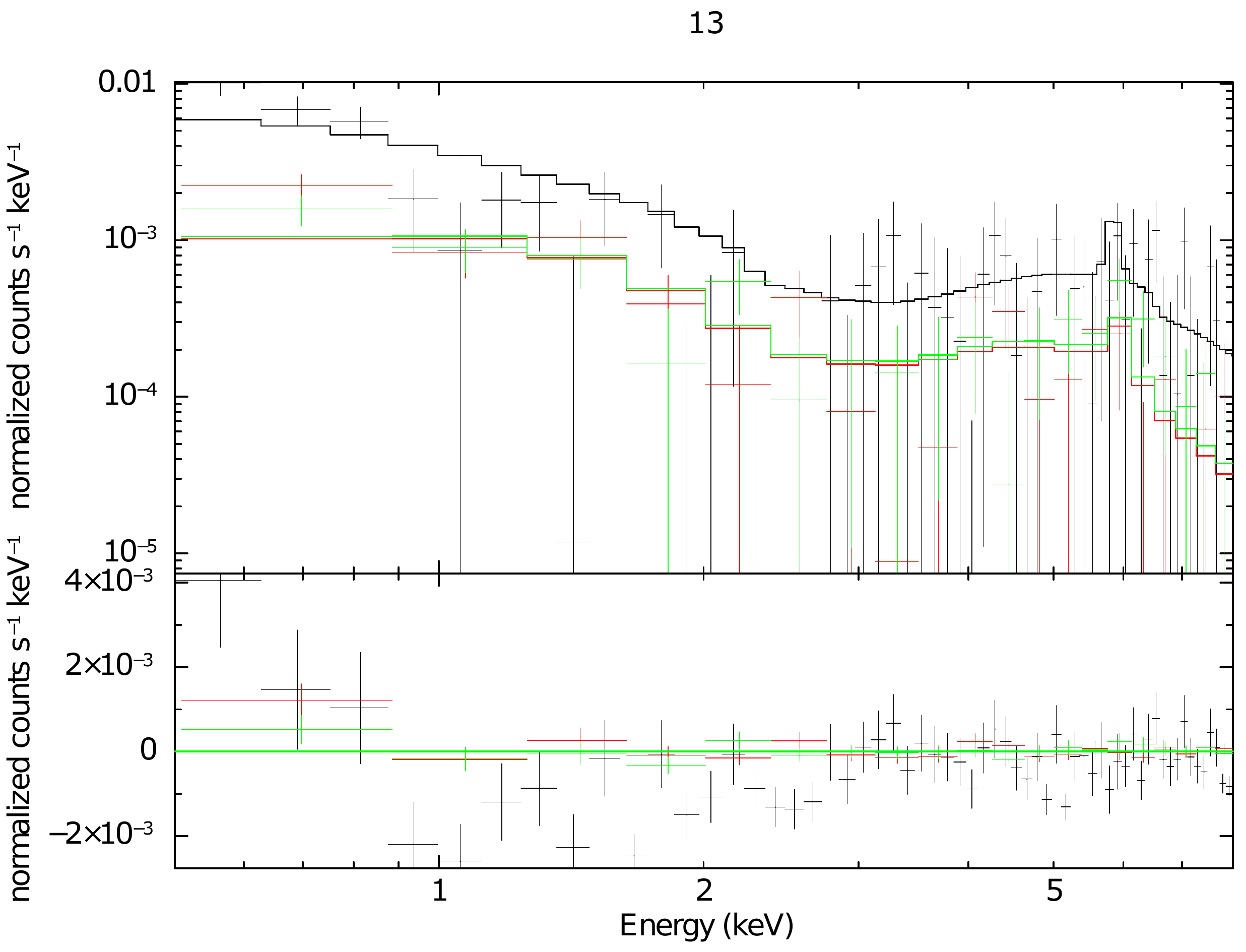} 
\includegraphics[width=5.8cm]{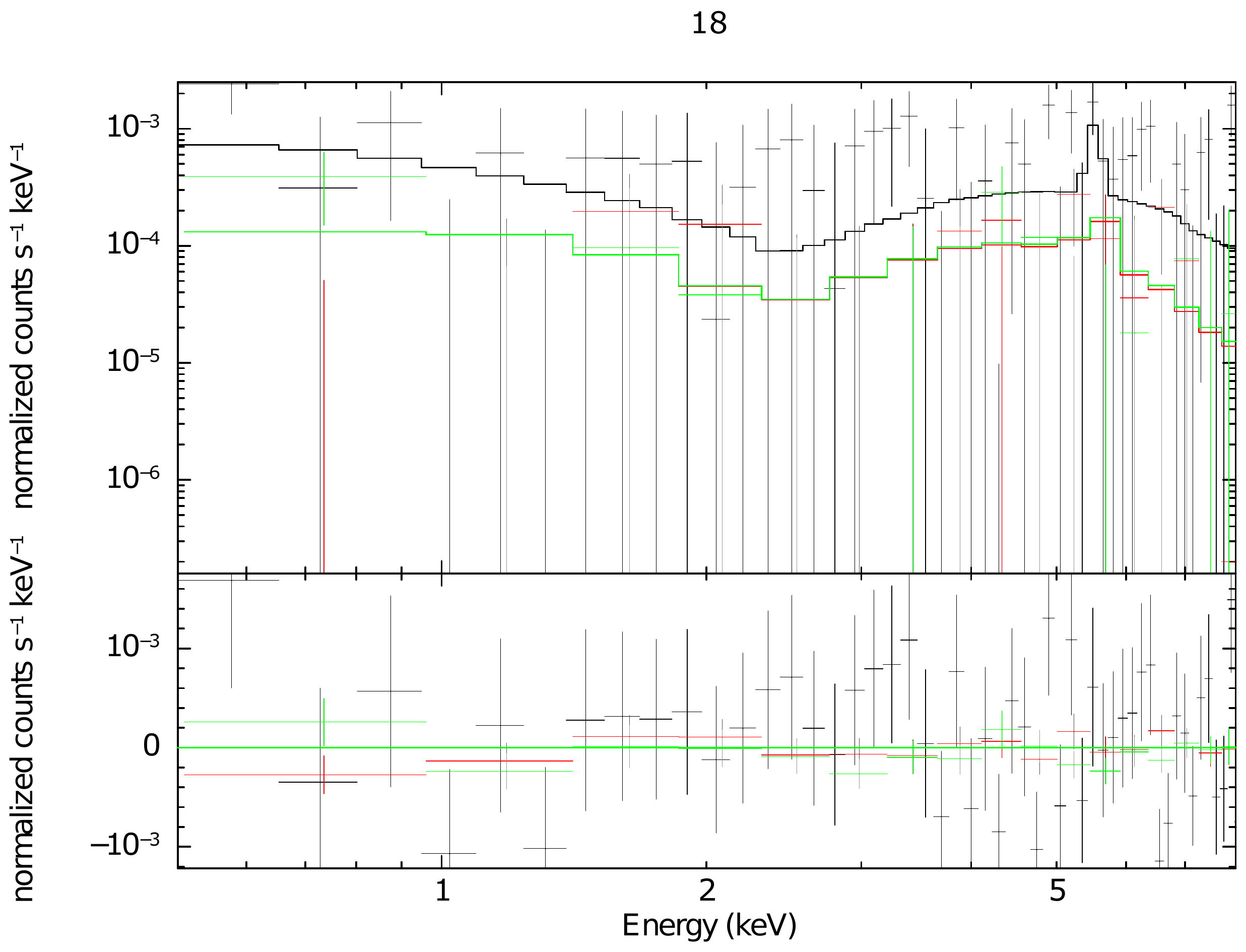} 
\includegraphics[width=5.8cm]{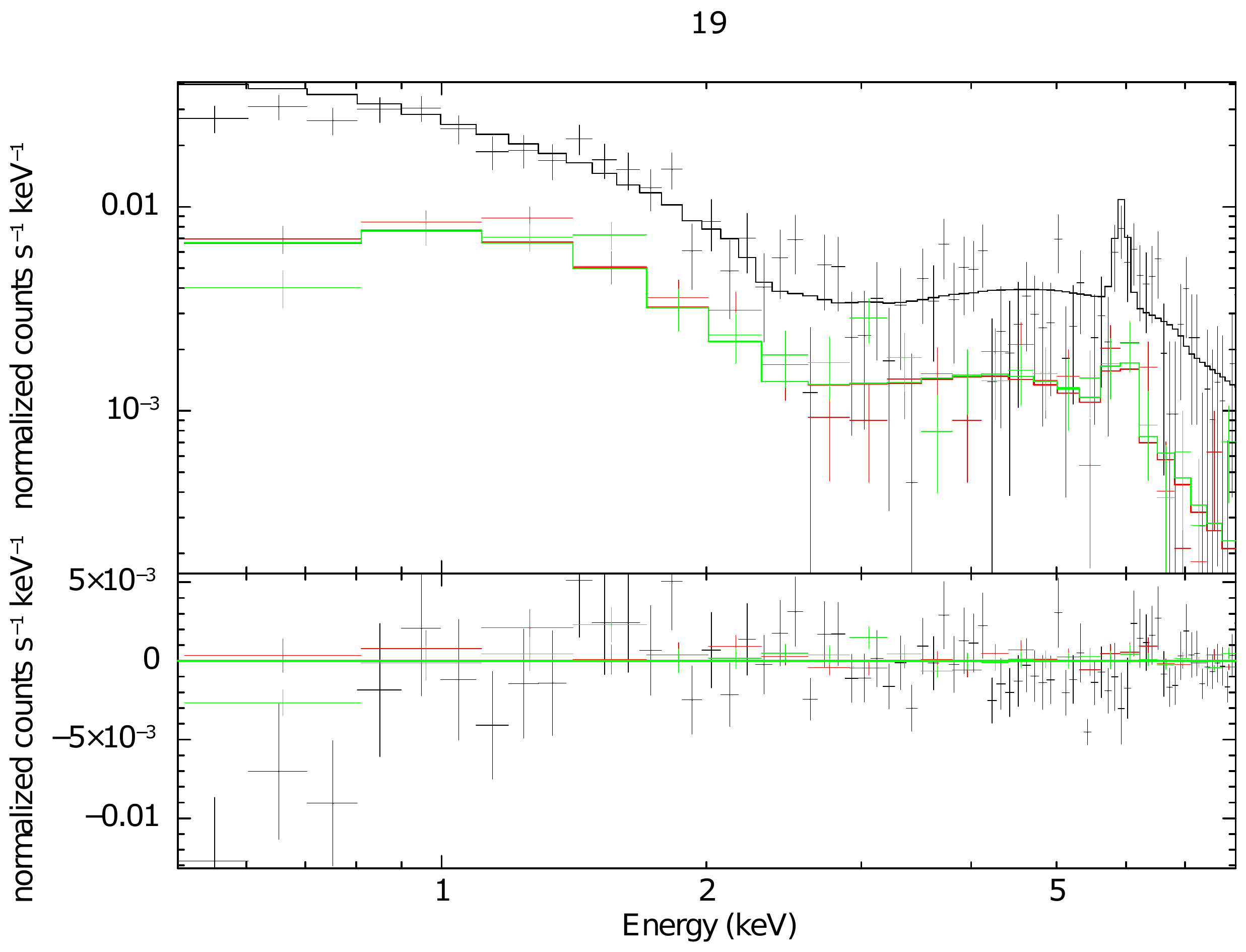} 
\includegraphics[width=5.8cm]{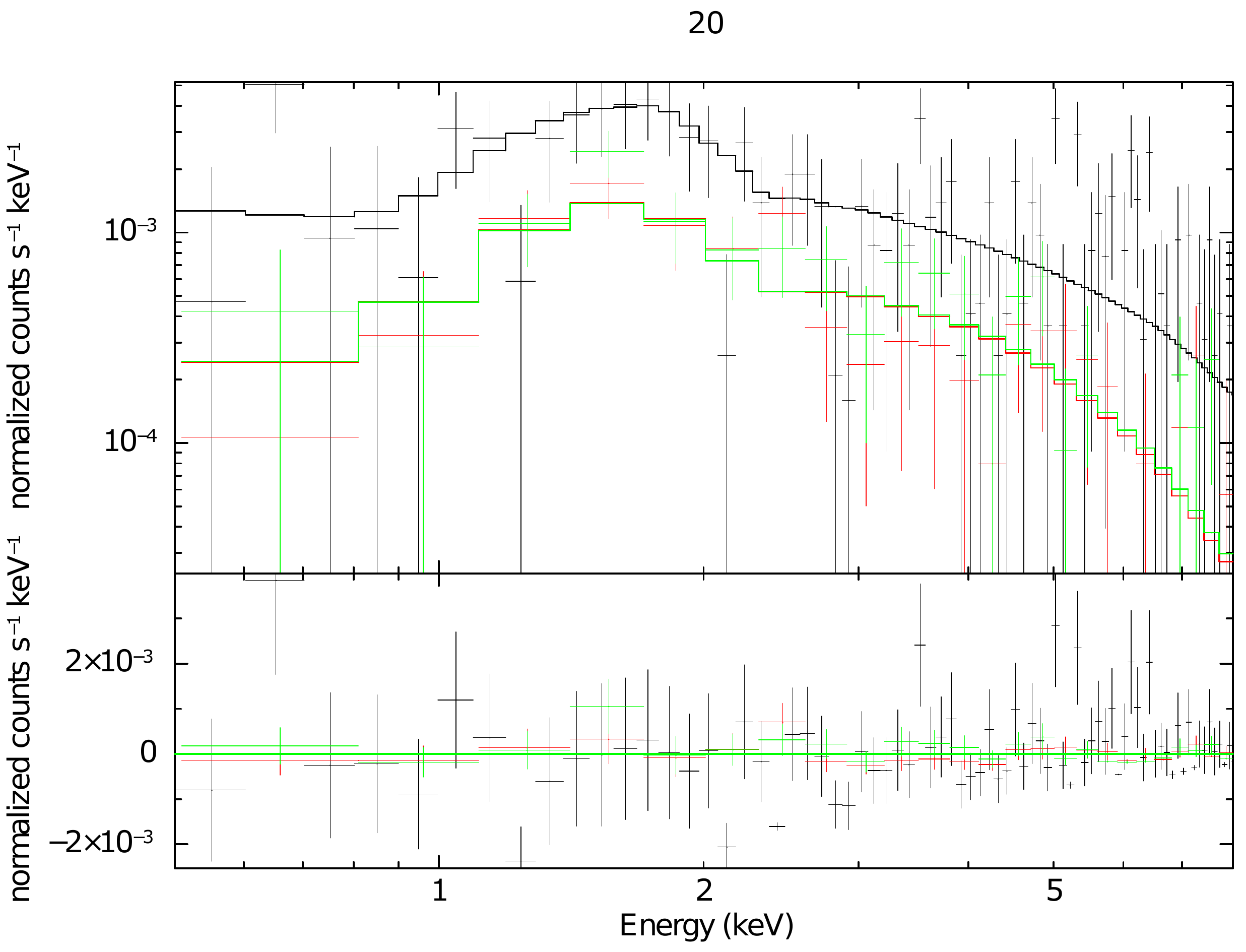} 
\caption{The \emph{XMM--Newton} spectra of the nine sources that show indications of high amounts of obscuration in the X-rays (see text). The best fit models are those described in Tables \ref{double} and \ref{reflection}. In red and green colours we plot the MOS-1 and MOS-2, and in black the pn counts. The residuals of each fit are also shown.}
\label{spectra}
\end{center}
\end{figure*}

Nine sources (marked with a ``$\dagger$" symbol in Table\,\ref{PLFits}) show indications in the X-rays of having a column density higher than $\rm \sim10^{23}$ $\rm cm^{-2}$ as they present a) absorption turn-overs suggestive of column densities higher than $\rm \sim10^{23}\,cm^{-2}$ or b) large EW of the FeK$\alpha$ line and/or flat spectral indices which could be indicative of a reflection dominated spectrum ($\Gamma\sim1.4$ or flatter). For these sources, we have repeated the spectral analysis with more complicated models. We use a two power-law model plus a Gaussian line: {\sc wa*(po+zwa*(po+zga))} in {\sc XSPEC} notation for the new fits, and the results are shown in Table\,\ref{double}. The spectral indices of both power-laws have been fixed to $\Gamma=1.8$ \citep[e.g.][]{Dadina2008}, while the energy of the FeK$\alpha$ line has been fixed to a rest-frame energy of 6.4\,keV. We detect significant absorbing columns ($N_{\rm H}>10^{23}\,{\rm cm^{2}}$) for eight out of the nine sources (source \#\,1 has an upper limit of $\rm 5.6\times10^{23}\,cm^{-2}$, still being consistent with heavy obscuration), while for one (\#\,4) there is direct evidence for Compton-thick absorption with $N_{\rm H}=1.2\times10^{24}\,{\rm cm^{2}}$. The \emph{XMM--Newton} X-ray spectra of the nine sources for which we find indications of heavy absorption are presented in Figure\,\ref{spectra}. For each object the upper panel shows the X-ray spectrum along with the model presented in Tables\,\ref{double} and \ref{reflection}  while the lower panel shows the residuals. For illustration purposes only, the spectra are re-binned every 40 channels (using ``setplot rebin 40 40" in {\sc XSPEC}).

The five sources which present a flat spectral index (\#\,4,6,12,18,19) are similar to the flat-spectrum Compton thick candidates of \citet{Georgantopoulos2013} in the CDFS and \citet{Lanzuisi2013} in the COSMOS field. A flat spectrum alone cannot constitute a Compton-thick source; more evidence is needed in the form of a high equivalent width Fe\,K$\alpha$ line \citep[$\rm EW\gtrsim500\,eV$; see][]{George1991}. According to Table\,\ref{double}, only one of the flat sources (\#\,19) has a relatively strong line, with $\rm EW=480^{+210}_{-160}\,eV$, making it the second possible CT AGN of our sample, based on the X-ray spectra. Moreover, for the five flat sources we use an alternative fit  with a reflection component model \citep{Magdziarz1995} plus a Gaussian line  ({\sc wa*(pexrav+zga)} in {\sc XSPEC} notation).  We fix the incident power law component to $\Gamma=1.8$ and the cosine of the inclination angle of the reflecting slab to 0.45. Again the energy of the FeK$\alpha$ line has been fixed to a rest-frame energy of 6.4\,keV. For three of the five sources the reflection model is excluded at over the 99.9 confidence level, the result of the fits for the rest are given in Table\,\ref{reflection}, where no significant detection is made of an Fe\,K$\alpha$ line.

\begin{table}
\centering
\caption{The X-ray spectral fits using two power-law components and an FeK$\alpha$ line.}
\label{double}
\begin{tabular}{cccccc}
\hline\hline
Number & $z$   & $N_{\rm H}$           & EW                     & c-stat    \\
 (1)   & (2)   & (3)                   & (4)                    & (5)       \\
 \hline
 1     & 1.973 & $<56$                 & $0.74^{+0.66}_{-0.44}$ & 2071/2498 \\ 
 4     & 0.526 & $120^{+66}_{-44}$     & $0.20^{+0.33}_{-0.20}$ & 2729/2498 \\
 6     & 0.039 & $87^{+39}_{-27}$      & $0.19^{+0.14}_{-0.13}$ & 2219/2498 \\
 9     & 0.210 & $81^{+27}_{-22}$      & $0.50^{+0.30}_{-0.22}$ & 1476/2498 \\
12     & 0.027 & $71^{+13}_{-11}$      & $<0.20$                & 1851/2498 \\
13     & 0.094 & $40^{+51}_{-19}$      & $0.41^{+0.6}_{-0.27}$  & 2239/2498 \\
18     & 0.184 & $45^{+300}_{-38}$     & $<0.60$                & 2722/2498 \\
19     & 0.077 & $29^{+12}_{-9}$       & $0.48^{+0.21}_{-0.16}$ & 2613/2498 \\
20     & 2.560 & $31^{+11}_{-8.5}$     & $<0.20$                & 1440/2498 \\
\hline\hline
\end{tabular}
\begin{list}{}{}
\item The columns are: (1) Source number;
                       (2) Redshift;
                       (3) Intrinsic hydrogen column density in units of $\rm 10^{22}\,cm^{-2}$;
                       (3) Equivalent width of the FeK$\alpha$ line in units of keV;
                       (4) c-statistic / degrees of freedom. Both power-law spectral slopes are fixed to $\Gamma=1.8$.
\end{list}
\end{table}

\begin{table}
\centering
\caption{The X-ray spectral fits using a reflection model}
\label{reflection}
\begin{tabular}{cccc}
\hline\hline
Number & $z$   & EW                     & c-stat    \\
(1)    & (2)   & (3)                    & (4)       \\
\hline
 4     & 0.53  & $<0.41$                & 2727/2500 \\
18     & 0.184 & $<0.85$                & 2727/2500 \\
 \hline\hline
\end{tabular}
\begin{list}{}{}
\item The columns are: (1) Source number;
                       (2) Redshift;
                       (3) Equivalent width of the FeK$\alpha$ line in units of keV;
                       (4) c-statistic / degrees of freedom.
\end{list}
\end{table}

\subsection{Optical spectra}
\begin{figure}
\resizebox{\hsize}{!}{\includegraphics{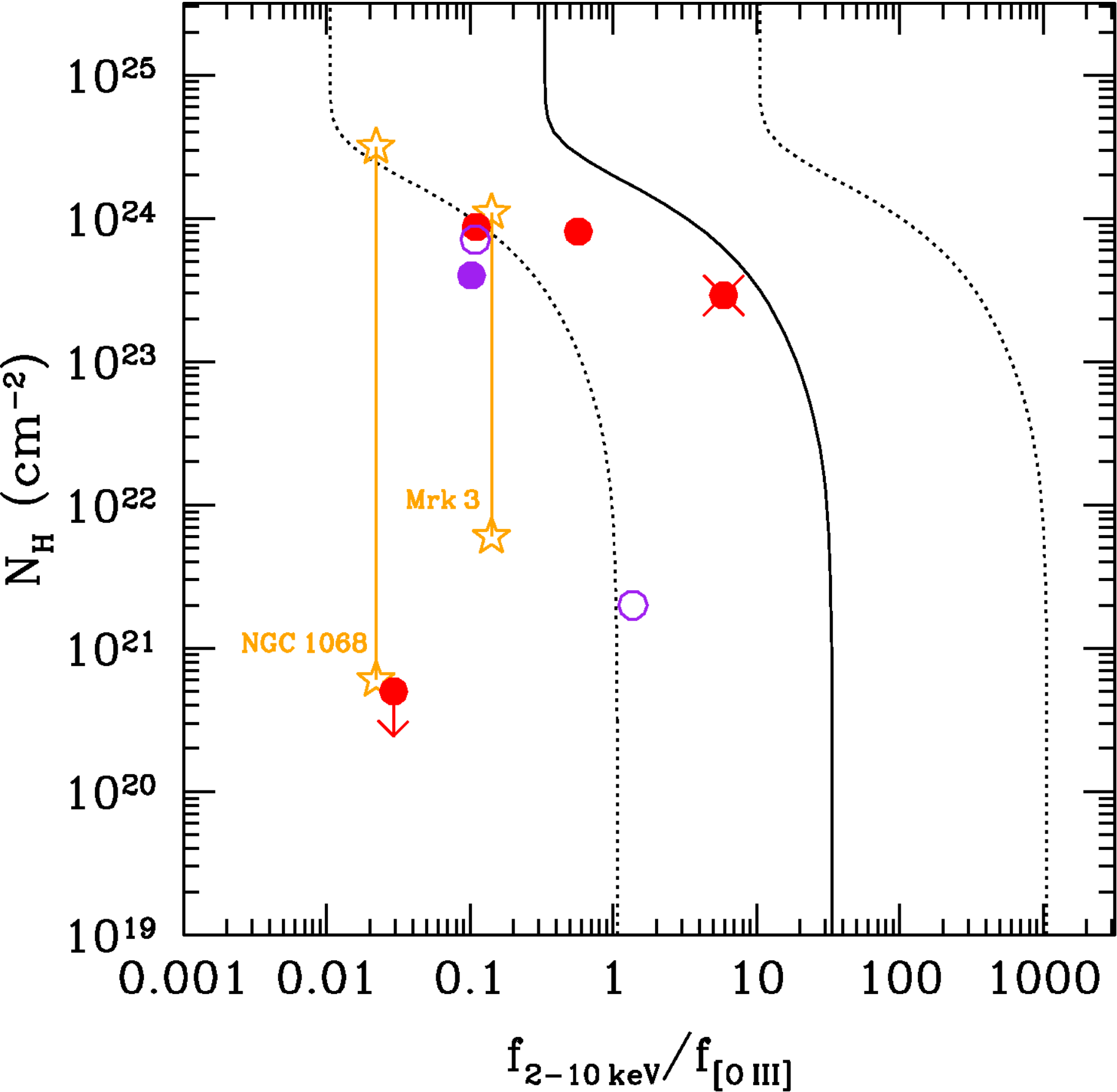}}
  \caption{Hydrogen column density versus $f_{\rm x}/f_{\rm [OIII]}$ ratio for all heavily obscured AGN candidates with [OIII] flux measurements in the SDSS. The X-ray fluxes come from the spectral fits whenever possible, otherwise measured in \citet{Georgakakis2011} and the [OIII] fluxes are corrected for extinction using the Balmer decrement. The $N_{\rm H}$ values are the ones using a single power-law (Table\,\ref{PLFits}), except for sources also listed in Table\,\ref{double}, where two power-law components are fit. The colour coding used is the same as in Figure\,\ref{Lx-z}, with the open symbols representing non X-ray under-luminous sources. The red cross marks a possibly Compton-thick source (\#\,19). The curves are taken from \citet{Akylas2009} and represent the correlation expected for a power-law X-ray spectrum with $\Gamma=1.8$ and a 1\% reflection/scattered component, normalised at $N_{\rm H}=0$ at the mean for Seyfert-1 sources, $\pm3\sigma$. The orange symbols show the difference in measured $N_{\rm H}$ values for two nearby Compton-thick sources, when observed with \emph{Einstein} (low-quality data, low-energies, low $N_{\rm H}$) and \emph{BeppoSAX} (higher-quality data, higher-energies, higher $N_{\rm H}$)}
  \label{fxfOIII}
\end{figure}

In this section we try to investigate the effectiveness of the selection methods used by invoking another parameter frequently used to select heavily obscured AGN, the hard X-ray to [OIII]\,5007 line flux ratio. In Figure\,\ref{fxfOIII} we plot it against the hydrogen column density as calculated using the X-ray spectral fits in Tables\,\ref{PLFits} and \ref{double}. The [OIII]\,5007 line is thought to arise from the narrow-line region and its flux is well correlated with the hard X-ray ($2-10$\,keV) flux for both Seyfert-1 and Seyfert-2 sources, independent tom the inclination angle \citep*[e.g.][]{AlonsoHerrero1997}. It is therefore considered an isotropic indicator of the intrinsic AGN flux. The hard X-ray to [OIII]\,5007 line flux ratio has been used as a diagnostic for X-ray absorption in a number of studies \citep*[e.g.][]{Maiolino1998,Bassani1999,Cappi2006,Vignali2006,Akylas2009,Goulding2011}. In Figure\,\ref{fxfOIII} we use the updated X-ray fluxes from the spectral fits where available and keep the same colour coding as in Figure\,\ref{Lx-z}, plotting sources which are no longer X-ray under luminous with respect to the mid-infrared with open symbols. The [OIII] fluxes are taken from the SDSS database and are corrected for extinction using the relative intensity of the Balmer lines ($f_{\rm [OIII],cor}=f_{\rm [OIII],obs}[{\rm (H\alpha/H\beta)/(H\alpha/H\beta)_{0}}]^{2.94}$) and assuming and intrinsic Balmer decrement of ${\rm (H\alpha/H\beta)_{0}}=2.76$ \citep[see][]{Brocklehurst1971}. The solid line represents the expected correlation of the two quantities assuming an X-ray spectrum in the form of an absorbed power-law with $\Gamma=1.8$ and a reflection/scattered component of 1\% of the intrinsic flux. The normalisation is taken from \citet{Akylas2009} and the dotted lines represent the $3\,\sigma$ limits of the correlation of the Seyfert-1 sample of \citet{Akylas2009}. Note that the limits used by \citet{Maiolino1998} and \citet{Cappi2006} are within those boundaries.

In Figure\,\ref{fxfOIII} we keep the colour notation of Figure\,\ref{Lx-z} and plot the more robustly detected AGN SEDs in red colour. We can see that all the candidate sources for which we have a measurement of the [OIII] flux in the SDSS lie on the left of the mean $f_{\rm x}/f_{\rm [OIII]}$ found in \citet{Akylas2009}, implying some degree of obscuration. It is interesting that source \#\,10, for which there is no solid evidence in the X-rays for the presence of a heavily obscured or Compton-thick nucleus is leftward of the dotted line. For this source the presence of a heavily obscured nucleus is implied by both a low X-ray to mid-infrared luminosity ratio and a low X-ray to [OIII] flux ratio, but it cannot be seen in the broad-band X-ray spectrum. One possibility is that the X-ray flux is variable, as the X-ray, optical, and infrared observations are not taken at the same epoch. Another possibility is that we do not find any evidence in the X-rays because of the poor quality of the X-ray spectra, especially in the hard band. In Figure\,\ref{fxfOIII} we plot in orange symbols the values for two well-known nearby Compton-thick AGN, NGC\,1068 and Mrk\,3. The low-$N_{\rm H}$ stars are measurements of the hydrogen column density based on early spectra taken with the \emph{Einstein Observatory} \citep*{Kruper1990} and the high-$N_{\rm H}$ stars are based on spectra taken with \emph{BeppoSAX}, which extend to very high energies \citep{Cappi1999,Guainazzi1999}. The $f_{\rm x}/f_{\rm [OIII]}$ values are taken from \citet{Bassani1999}. Both sources have a prominent soft-excess attributed to a scattered component, which could be mistaken overall for a steep power-law if the high-energy measurements were missing, or if the quality was poor. We therefore assume that part, or all of the sources plotted in Figures\,\ref{Lx-z} and \ref{fxfOIII} could be heavily obscured despite the fact that there is no apparent evidence in their \emph{XMM--Newton} spectra.

\subsection{Normal galaxies}
\label{contamination}

Galaxies that do not host an AGN can also produce X-rays because of their star-formation activity \citep[see e.g.][]{Ranalli2003,Rovilos2009,Ranalli2012}. According to the hard X-ray to infrared relation of \citet{Ranalli2003}: $\log(L_{\rm 2-10\,keV})=\log(L_{\rm FIR})-(3.62\pm0.29)$. If we use the 105 SED templates of \citet{Chary2001} to calculate the ratio between the integrated FIR luminosity and the monochromatic luminosity at $\rm 12\,\mu m$ we find: $\log(L_{\rm FIR})=\log(\nu L_{\nu}(12\,\mu m))+(0.9\pm0.2)$ for the star formation component. The integrated to monochromatic luminosity relations are best described by a broken power-law \citep[see][]{Chary2001}, however for the purpose of this study we keep the simplistic approach of a linear relation. From these relations we expect that for a normal galaxy, the X-ray to mid-infrared ratio will be $\log(L_{\rm 2-10\,keV}/\nu L_{\nu}(\rm 12\,\mu m))=-2.72\pm0.35$. Moreover, normal galaxies generally have soft X-ray spectra \citep[see e.g.][]{Ranalli2003,Lehmer2008}. The mid-infrared luminosity we measure here comes from the torus component, after having performed a decomposition of the SED. In previous works looking for low X-ray--IR AGN \citep[e.g.][]{Georgakakis2010} the contamination from normal galaxies is coming predominantly from sources that show no evidence for an AGN in their mid-infrared spectra \citep[see also][]{Pope2008}. Therefore, we are not expecting a significant number of normal galaxies in the low X-ray to mid-infrared luminosity sample. Indeed, there are only three sources with $L_{\rm 2-10\,keV}/\nu L_{\nu}(\rm 12\,\mu m)<2.37$ and $HR<-0.35$, and for one of them the torus component is not statistically significant, which demonstrates the significance of the SED decomposition performed.

\section{Discussion}

In this paper we select a number of heavily obscured AGN candidates based on the ratio between their X-ray and mid-infrared luminosities. The AGN produce high energy emission close to the black hole, which heats the dust surrounding the nucleus and is re-emitted in the infrared. Thermal equilibrium of the hot dust in combination with the clumpiness of the medium \citep[see][]{Gandhi2009,Honig2010a,Honig2010b} causes the infrared emission to peak in relatively short wavelengths ($\rm\sim10\,\mu m$). We use this property of the AGN infrared emission to separate it from the infrared emission from circumnuclear, or host galaxy star-formation, whose spectral energy distribution peaks at $\rm\sim100\,\mu m$. Our selection of highly obscured AGN candidates is based on the assumed suppression of the X-ray luminosity with respect to the infrared.

There are 20 heavily obscured AGN candidates, and we look into their X-ray, optical, and infrared properties in greater detail. A more detailed SED decomposition using optical to far-infrared photometry and a three-component fit finds a statistically significant contribution from a warm dust (torus) component in all 20 candidates, however half of them are still X-ray under-luminous if we take into account their infrared flux lower limits. We call those sources the ``robust'' sample. Looking into the X-ray properties of all 20 candidates, we find evidence of high X-ray obscuration in {nine} sources (see Table\,\ref{PLFits}; Figure\,\ref{spectra}), and we call this the ``X-ray obscured'' sample. We also investigate the [OIII] line luminosities in the SDSS optical spectra and comparing them with the observed X-ray luminosities we find evidence for obscuration for seven sources with optical spectra.

\subsection{Comparison with other selection methods}

In this section we compare the results of our method with others used in the literature.

\subsubsection{$\rm 25\,\mu m$ selection}

\begin{figure}
\resizebox{\hsize}{!}{\includegraphics{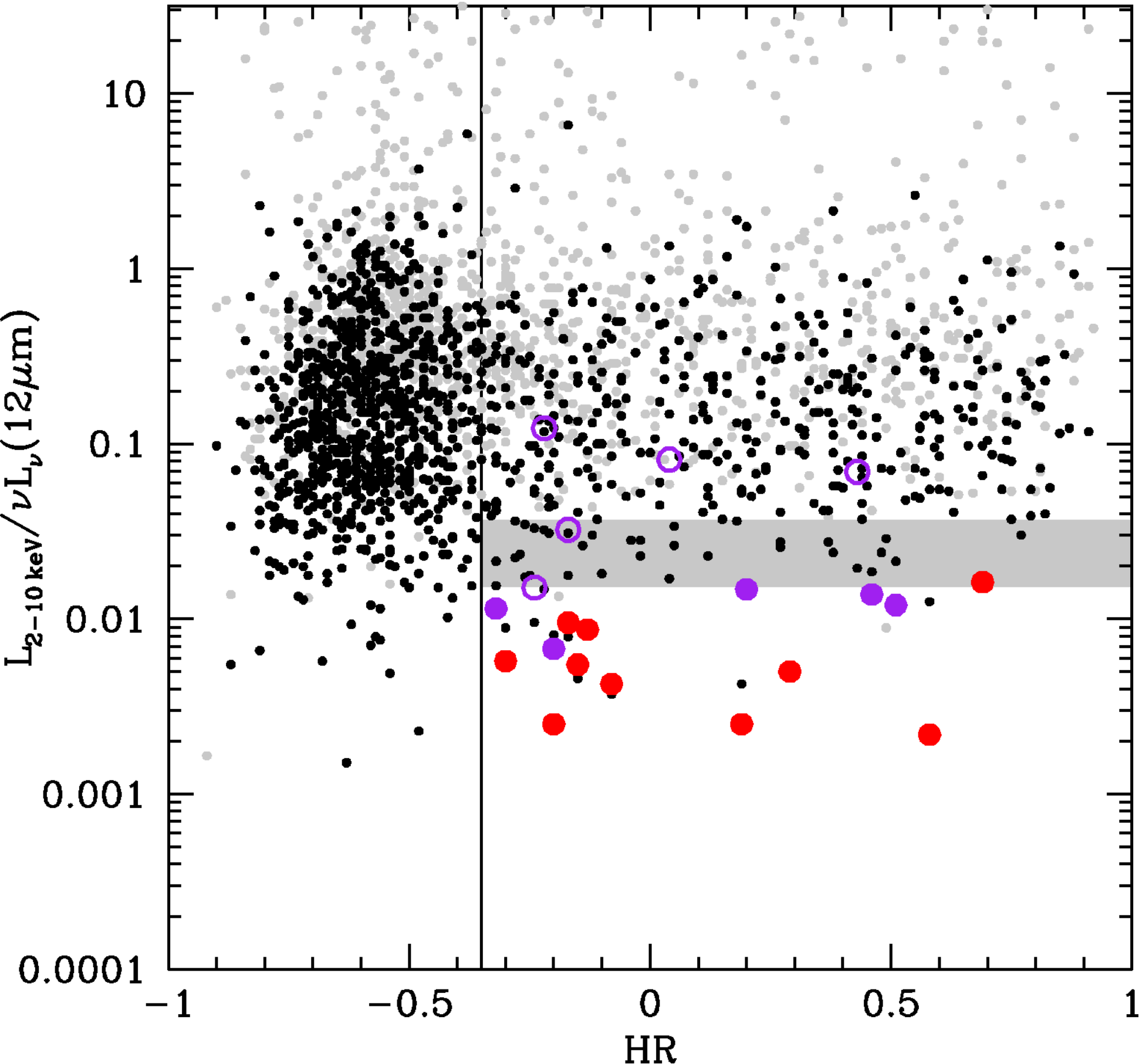}}
  \caption{The X-ray to mid-infrared luminosity ratio plotted against the X-ray hardness ratio for all the X-ray -- mid-infrared sample. We plot the sources with robustly detected ($>2\sigma$) infrared AGN components with black circles, and the rest in grey. The vertical line marks the $HR=-0.35$, which we use to differentiate between soft and hard X-ray sources. The grey area marks the $L_{\rm 2-10\,keV}/\nu L_{\nu}(\rm 12\,\mu m)$ threshold marked with the solid line in Figure\,\ref{lxl12} with the limiting luminosities set to $L_{\rm 2-10\,keV}^{\rm min}=10^{41.5}\,{\rm erg\,s^{-1}}$ and $L_{\rm 2-10\,keV}^{\rm max}=10^{45.0}\,{\rm erg\,s^{-1}}$. We keep the same symbols for the sources in our sample as in Figure\,\ref{fxfOIII}}.
  \label{lxl12_hr}
\end{figure}

Recently, \citet{Severgnini2012} used a similar technique with the one used here to detect nearby Compton-thick sources, comparing the $\rm 2-12\,keV$ X-ray flux and the total $\rm 25\,\mu m$ infrared flux, and using the hardness ratio between the $\rm 2-4.5\,keV$ and $\rm 4.5-12\,keV$ bands of \emph{XMM--Newton} ($HR4$), they found that $\sim84\%$ of AGN with $f_{\rm 2-12\,keV}/\nu f_{\nu}{\rm(25\,\mu m)}<0.02$ and $HR4>-0.2$ show CT characteristics in their X-ray spectra \citep[see Fig.\,1 of][]{Severgnini2012}. In Figure\,\ref{lxl12_hr} we plot the equivalent plot for the \emph{XMM}--\emph{WISE} sample, using the $L_{\rm 2-10\,keV}/\nu L_{\nu}{\rm(12\,\mu m)}$ and $HR$ thresholds described in \S\,\ref{CTselection}, and the $L_{\rm 2-10\,keV}$ values derived from the spectral fits. We plot the sources with robustly detected ($>2\sigma$) infrared AGN components with black circles, and the rest in grey. We use luminosities instead of fluxes here in order to take into account the different k-corrections for each source, since our sample expands to relatively high redshifts; the sample of \citet{Severgnini2012} on the other hand consists of local sources, where the k-correction is negligible. The vertical line is at $HR=-0.35$, and the grey area represents the range of $L_{\rm 2-10\,keV}/\nu L_{\nu}{\rm(12\,\mu m)}$ threshold values within the range of $L_{\rm 2-10\,keV}=10^{41.5}-10^{45}\,{\rm erg\,s^{-1}}$, since the relation of \citet{Gandhi2009} is not linear, and a simple $L_{\rm 2-10\,keV}/\nu L_{\nu}{\rm(12\,\mu m)}$ threshold cannot characterise our selection. We use the same symbols as in Figure\,\ref{fxfOIII} for the 20 sources of our sample.

Using the SED decomposition technique described earlier we also have a measure of the AGN monochromatic luminosity in the $\rm 25\,\mu m$ band. If we apply a $L_{\rm 2-10\,keV}/\nu L_{\nu}{\rm(25\,\mu m)}$ limit of 0.02, as in \citet{Severgnini2012}, and our $HR$ limit of --0.35, we find the same 17 of the 20 sources of our sample (and all the ``robust'' ones), plus four sources that could be CT candidates and are not in the low $L_{\rm 2-10\,keV}/\nu L_{\nu}{\rm(12\,\mu m)}$ sample. However, the $\rm 25\,\mu m$ luminosity measurement is an extrapolation from the best-fitting SED, expect for sources with a $\rm 22\,\mu m$ flux measurement at $z>1.14$. For the four sources, the $L_{\rm 2-10\,keV}/\nu L_{\nu}{\rm(12\,\mu m)}$ values are very close to the limit used in this paper, so they are considered statistical fluctuations. Therefore we conclude that selecting CT candidates using the $\rm 25\,\mu m$ instead of the $\rm 12\,\mu m$ luminosity would not yield a significantly different sample, and therefore our selection is very similar to the one used by \citet{Severgnini2012}.

\subsubsection{WISE colour--colour selection}

The \emph{WISE} catalogue provides photometry in four different mid-infrared bands, so different colour--colour techniques have been proposed to select AGN, similar to the techniques used for \emph{Spitzer}--IRAC sources \citep[e.g.][]{Lacy2004,Stern2005,Donley2012}. A useful selection is the 3-band colour--colour diagram using the 3.4, 4.6, and $\rm 12\,\mu m$ bands, where optically and radio selected QSOs occupy a distinct region \citep[e.g.][]{Yan2013}. Seyfert galaxies (spectroscopically identified) in this study however lie in a similar region to that occupied by starbursts. Similarly, \citet{Mateos2012} defined a region in the $[3.4]-[4.6]$ vs. $[4.6]-[12]$ diagram where the bulk of the X-ray luminous ($L_{\rm 2-10\,keV}>10^{44}\,{\rm erg\,s^{-1}}$) AGN are located, which lies around the line defined by a power-law ($f_{\nu}\propto\nu^{\alpha}$) with $\alpha<-0.3$ \citep[see also][]{AlonsoHerrero2006,Donley2007}. \citet{Stern2012} argue that a similar result is reached when the selection is made using only the $[3.4]-[4.6]$ colour, with $[3.4]-[4.6]>0.8$, and find that the technique has similar properties to the \emph{Spitzer}--IRAC selection of \citet{Stern2005}. However, there is a dependency of both the reliability and the efficiency of the method with the flux limits used, as previously claimed \citep[e.g.][]{Barmby2006,Brusa2009,Rovilos2011,Mendez2013}.

\begin{figure*}
\resizebox{\hsize}{!}{\includegraphics{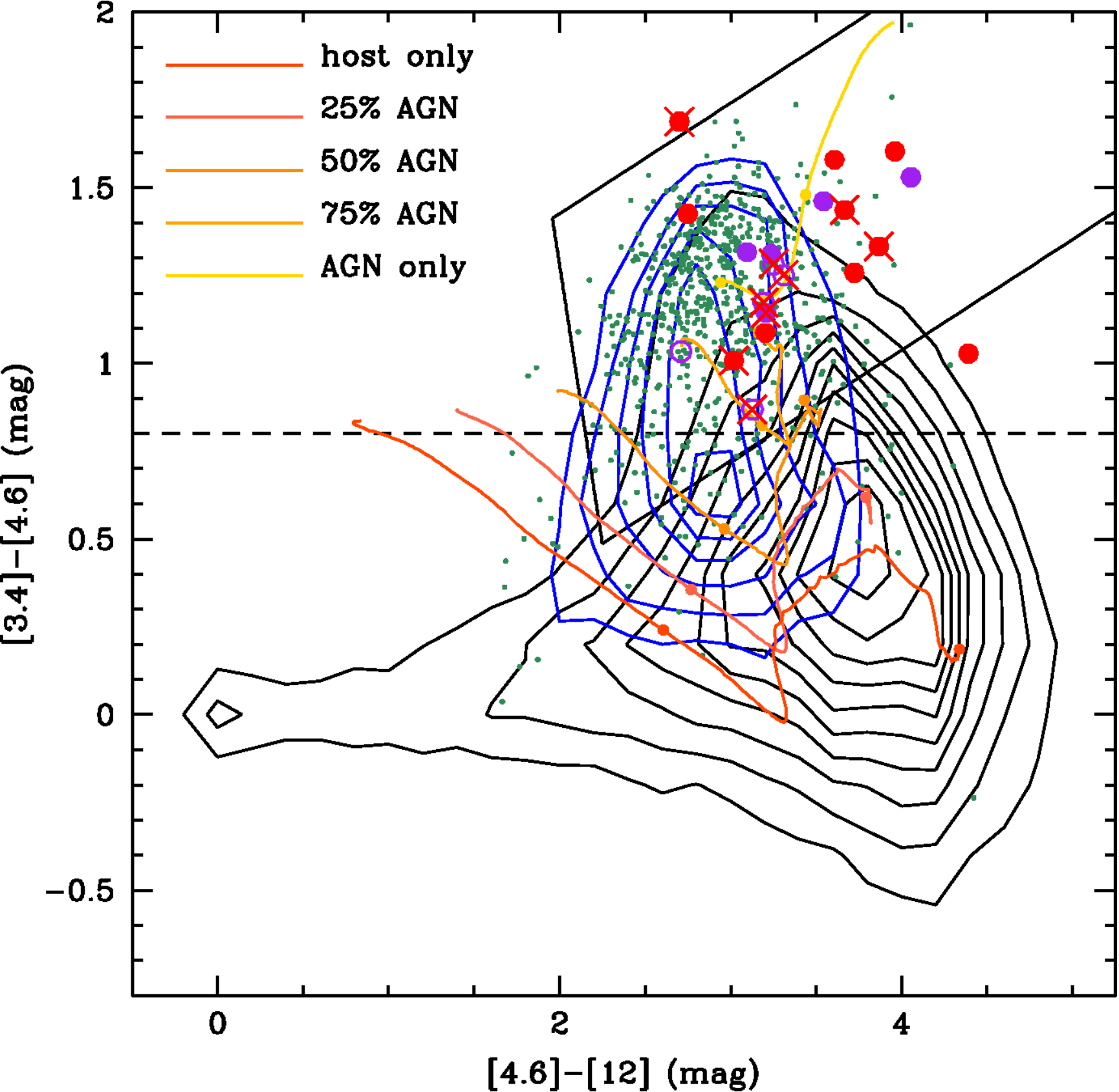}}
  \caption{The \emph{WISE} colour--colour diagram. The black contours represent all the infrared sources, while the blue contours represent the X-ray sources. The bulk of the X-ray sources is offset from the bulk of the infrared sources, but not in the areas where most of the luminous X-ray sources reside, according to \citet{Mateos2012} (solid lines) and \citet{Stern2012} (dashed line). The AGN with $L_{\rm 2-10\,keV}>10^{44}\,{\rm erg\,s^{-1}}$ are plotted with green dots, and the majority of them are in the aforementioned areas. The heavily obscured candidates are plotted in red and purple symbols, following Figues\,\ref{fxfOIII} and \ref{lxl12_hr}. Red crosses mark sources that have an indication of heavy obscuration in their X-ray spectra. The orange to yellow contours are the tracks of a pure host SED gradually contaminated with an AGN (torus) SED.}
  \label{wise_colours}
\end{figure*}

In Figure\,\ref{wise_colours} we plot the $[3.4]-[4.6]$ vs. the $[4.6]-[12]$ colours for all the sources in our sample. The black contours represent the density in the $[3.4]-[4.6]$ -- $[4.6]-[12]$ parameter space of all of the \emph{WISE} sources that lie within a 1\,arcmin radius of an X-ray source, therefore an infrared-selected sample, or a ``background'' sample, since the vast majority of the infrared sources out to this radius are chance matches to the X-ray sources. The contours represent regions with point densities of 100--1\,000 sources per $\rm0.2\times0.2\,mag^{2}$ pixel, with a step of 100. The region where the 2\,844 X-ray sources with fitted infrared SEDs lie is represented with the blue contours, with 15--90 sources per $\rm0.2\times0.2\,mag^{2}$ pixel, with a step of 15. This is the region where the moderate X-ray luminosity AGN lie (2\,181/2\,844 have $L_{\rm 2-10\,keV}<10^{44}\,{\rm erg\,s^{-1}}$) and it is different from the background, although the highest density point is not in the ``QSO region'' of \citet{Mateos2012}, marked with the solid lines, or above the ``QSO limit'' of \citet{Stern2012}, marked with the dashed line. These selection criteria are reliable for selecting the 661 sources with $L_{\rm 2-10\,keV}\geq10^{44}\,{\rm erg\,s^{-1}}$, plotted with green circles. This is indicative of the contamination of the mid-infrared colours of AGNs by the host galaxy; \citet{delMoro2013} find that the fraction of AGN detected in the mid-infrared using SED decomposition increases with increasing X-ray luminosity. The \emph{WISE} colours blue contour and the green circles regions are in good agreement with the \emph{WISE} colours of the [OIII]-selected and hard-X-ray selected AGN respectively of \citet{Mateos2013}, which again reflects on the X-ray luminosity distributions of those samples.

The red and purple points in Figure\,\ref{wise_colours} mark the colours of the 20 highly obscured AGN candidates, the red points representing the most robust sample, and the open open purple circles representing the sources which do not meet the initial $L_{\rm x}/L_{\rm ir}$ criteria if the X-ray luminosities are measured from the spectra. 8/10 and 10/10 of the red points, and 18/20 and 20/20 of all points comply with the criteria defined by \citet{Mateos2012} and \citet{Stern2012}, respectively. It is interesting that for the points in Figure\,\ref{wise_colours} the fraction that comply with the ``QSO selection'' criteria and have $L_{\rm 2-10\,keV}>10^{44}\,{\rm erg\,s^{-1}}$ is significantly smaller than that of the general AGN population: 4/18 (22.2\%) and 5/20 (20.0\%) of the red sources that comply with the \citet{Mateos2012} and \citet{Stern2012} criteria respectively have high X-ray luminosities, compared to 594/1483 (40.0\%) and 577/1258 (45.9\%) respectively for the general X-ray source population. We keep the original X-ray luminosities for this discussion, because an a priori of the luminosities from the X-ray spectra is not possible. We attribute the difference to the obscuration of the X-rays. If we use the intrinsic X-ray luminosities calculated from the \citet{Gandhi2009} relation (column 5 of Table\,\ref{candidates}), the fractions become 12/18 (66.7\%) and 13/20 (65.0\%), even higher than that of the overall X-ray population, showing that the ``infrared-selected" sample is not clean \citep[see also][]{Georgantopoulos2011}. \citet{Georgantopoulos2011} find that the X-ray to mid-infrared luminosity selection appears to be complete in the local universe and incomplete at higher redshifts; here we find signs of incompleteness in the local universe (all four sources have $z<0.04$), which is potentially even higher, since all our candidates were initially selected using a mid-infrared criterion.

The mid-infrared colours can be used as an indication of the contamination of the mid-infrared flux by the host galaxy \citep[see][]{Rovilos2011}. This is also evident if we plot in Figure\,\ref{wise_colours} the track of a pure starburst galaxy form the \citet{Chary2001} library used for the SED decomposition, and gradually add a torus template from the \citet{Silva2004} sample; this is seen going from orange to yellow tracks in Figure\,\ref{wise_colours}. The 25\% step refers to the fraction of $\rm 12\,\mu m$ monochromatic flux coming from the torus template, and the redshift range is $0\leq z\leq 2$, with a circle marking $z=0$ and $z=1$. In this case the high percentage of X-ray luminous sources in the ``QSO selection'' regions shows that the highest X-ray luminosity AGN are less contaminated by the host galaxy in the mid-infrared wavelengths \citep[see also][]{delMoro2013}. This trend is not as strong for the Compton-thick candidates, which is a direct implication of their selection criterion (low $L_{\rm x}/L_{\rm ir}$), meaning that the obscuration is affecting the X-rays more than the infrared wavelengths. It is also interesting that 6/20 and 5/20 of the infrared-selected candidates would not have been selected if we used a selection scheme based on a combination between the \citet{Mateos2012} or \citet{Stern2012} criterion respectively, and a low observed X-ray luminosity threshold (e.g. $L_{\rm 2-10keV}<10^{44}\,{\rm erg\,s^{-1}}$). The reason for this is that the SED decomposition is more accurate in identifying the AGN component from the mid-infrared photometry than a simple colour--colour selection, as it uses more information (more infrared bands), and the statistics are easier to handle and interpret.

\subsection{Number of heavily obscured AGN}
\label{CTnumber}

We use three indicators for a heavily obscured AGN, the X-ray spectrum, the X-ray to mid-infrared luminosity ratio, and the X-ray to [OIII] line flux ratio. For the latter, we assume a ratio of $f_{\rm 2-10\,keV}/f_{\rm[OIII]}=1.0$, on the dotted line of Figure\,\ref{fxfOIII} for low column densities. In Table\,\ref{obstests} we present the results of the three indicators: we use a ``$\checked$'' symbol for a positive indication, a ``$\times$'' symbol for negative, and a ``-" symbol for no indication, or an inconclusive indication. For nine of the 20 candidates (\#\,2,6,9,10,11,12,13,19,20) there are two or more indicators for the presence of a heavily obscured AGN, and another three (\#\,1,4,18) are heavily obscured according to their X-ray spectra, which is the most reliable indicator. Four sources (\#\,3,8,14,16) are heavily obscured candidates from their mid-infrared luminosities with no information from either their X-ray spectra or the [OIII] flux, and three (\#\,7,15,17) have low X-ray to mid-infrared luminosities, while their X-ray spectra argue against them being heavily obscured; they could be either variable sources, or their obscuration might be missed in the X-rays because of their low X-ray fluxes. Those three sources (along with source \#\,10 which has no obscuration indication in the X-rays but is heavily obscured according to both its infrared and its [OIII] luminosity) are the faintest sources with fitted spectra in our candidate sample and their spectra are expected to be of low quality, so that an obscured nucleus can easily be missed. Finally, source \#\,5 has no indications of obscuration. The Compton-thick sources are even fewer: we have direct indications from the X-ray spectra for two sources (\#\,4,19), and another two (\#\,2,11) are X-ray under-luminous with respect to both the mid-infrared and the [OIII] luminosities, while they are too faint in the X-rays for a useful spectrum to be extracted. Finally, source \#\,10 is in the robust mid-infrared sample, with its X-ray to [OIII] flux ratio indicating that it is Compton-thick; there is no indication in its X-ray spectrum, but it has the lowest number of counts of all the sources with fitted spectra. Summing up, there are 12--19 heavily obscured AGN in the sample, and 2--5 of them are Compton-thick.

\begin{table}
\centering
\caption{Obscuration tests: ``$\checked$'' indicates an obscured AGN, ``$\times$'' an unobscured AGN, and ``-'' inconclusive or no indication.}
\label{obstests}
\begin{tabular}{cccc}
\hline\hline
Number & mid-IR & X-rays & [OIII] \\
 (1) & (2) & (3) & (4) \\
\hline
 1 & $\times$            & $\checked$ & -          \\
 2 & $\checked\checked$  & -          & $\checked$ \\
 3 & $\checked$          & -          & -          \\
 4 & $\times$            & $\checked$ & -          \\
 5 & $\times$            & $\times$   & $\times$   \\
 6 & $\checked\checked$  & $\checked$ & $\checked$ \\
 7 & $\checked$          & $\times$   & -          \\
 8 & $\checked\checked$  & -          & -          \\
 9 & $\checked\checked$  & $\checked$ & $\checked$ \\
10 & $\checked\checked$  & $\times$   & $\checked$ \\
11 & $\checked\checked$  & -          & $\checked$ \\
12 & $\times$            & $\checked$ & $\checked$ \\
13 & $\checked$          & $\checked$ & $\checked$ \\
14 & $\checked$          & -          & -          \\
15 & $\checked$          & $\times$   & -          \\
16 & $\checked\checked$  & -          & -          \\
17 & $\checked\checked$  & $\times$   & -          \\
18 & $\times$            & $\checked$ & -          \\
19 & $\checked\checked$  & $\checked$ & $\times$   \\
20 & $\checked\checked$  & $\checked$ & -          \\
\hline
Total & 15(10)/20 & 9/14 & 7/9 \\
\hline\hline
\end{tabular}
\begin{list}{}{}
\item The columns are: (1) Source number;
                       (2) $L_{\rm 2-10\,keV}/\nu L_{\nu}{\rm(12\,\mu m)}$ being significantly lower than the \citet{Gandhi2009} relation. A double ``$\checked$'' mark notes that the $2\sigma$ lower limit is still below the relation;
                       (3) Detection of an Fe\,K$\alpha$ line or a flat X-ray spectrum;
                       (4) $f_{\rm 2-10\,keV}/f_{\rm[OIII]}<1.0$
\end{list}
\end{table}

We compare the above numbers to the predictions of X-ray background synthesis models for the number of heavily obscured
($N_{\rm H}\sim 10^{23-24}\,{\rm cm^{-2}}$) and Compton-thick AGN expected in the current XMM/WISE survey. X-ray background synthesis models predict the number of Compton-thick AGN using as constraints the spectrum of the X-ray background and/or  the number of Compton-thick AGN observed in very hard ($>10\,{\rm keV}$) X-ray surveys performed with \emph{Swift} and \emph{INTEGRAL} at bright fluxes. There have been several X-ray background synthesis models publicly available online \citep*[e.g.][]{Gilli2007,Draper2009,Treister2009,Akylas2012}. We choose to work with the model of \citet{Akylas2012}, as this model uses the most updated constraints on the number of Compton-thick AGN observed in the local Universe with Swift \citep{Burlon2011}. We use the area curve of our \emph{XMM--Newton} survey as given in \citet{Georgakakis2011} and assume an intrinsic Compton-thick fraction of 15\%, which is the best fit model in \citet{Akylas2012}. We find a fraction of heavily obscured and Compton-thick AGN of 12\% and 0.8\% respectively. Given that we fit the SEDs of 2\,844 sources in our sample (those with a hard band detection, either spectroscopic or photometric redshift, and three or more infrared photometry data-points), we expect 23 and 342 Compton-thick and heavily obscured AGN respectively. These numbers are well above the numbers found in the previous paragraph (2--5 and 12--19). This discrepancy can be explained if there are highly obscured sources that do not present a low X-ray to mid-infrared luminosity ratio, and therefore they are not selected by our selection procedure. This is the case for two Compton-thick AGN in the \emph{XMM}--CDFS observations of \citet{Comastri2011}. \citet{Georgantopoulos2011} discuss that these have a high X-ray to mid-infrared luminosity ratio typical of unobscured AGN. Moreover, we have included here an initial selection of sources that have a redshift determination in the SDSS, which means that we are not including a large number of optically faint sources, which are on average more obscured in the X-rays than the general X-ray population \citep*[see e.g.][]{Civano2005,Rovilos2010}.

Another factor which we have to include when assessing the number of heavily obscured AGN found with the low X-ray to mid-infrared luminosity method is the scatter in the X-ray to mid-infrared flux relation. Detailed analysis of the nuclear regions of local Seyferts done by \citet{Gandhi2009} finds a relation with a scatter of 0.23\,dex for the well resolved sample and 0.36\,dex for the full sample. Including more sources and low-luminosity AGN, \citet{Asmus2011} find a similar relation with a similar scatter (0.35\,dex). \citet{Georgakakis2010} find that the host contribution to the infrared flux can affect the X-ray to infrared ratio, and in this work we remove the host contribution from the infrared flux using SED decomposition. We find a scatter in the $L_{\rm x}-L_{\rm MIR}$ relation of 0.5\,dex, larger that found both by cited{Gandhi2009} and \citet{Asmus2011}. Part of our scatter can be introduced by X-ray absorption, which is not corrected for, and another part is a result of the $\Gamma=1.4$ assumption made in calculating the X-ray luminosities. Comparing the X-ray luminosities in Tables\,\ref{candidates} and \ref{PLFits} we can see that in some cases they can differ up to one order of magnitude. Removing this source of scatted would require fitting the X-ray spectra of all the 2\,844 sources, which is not feasible. We have removed it for 14 sources of our sample and indeed five of them are no longer low X-ray to mid-infrared candidates, but we do not have information on how many sources would make it into the mid-infrared selected sample if we corrected all the sources.

In summary, the X-ray to mid-infrared luminosity ratio technique is somewhat reliable for finding obscured sources i.e. most low X-ray to IR luminosity sources have large amounts of obscuration. However, the technique suffers from incompleteness, because of the selection made in optical--infrared wavelengths and because of scatter of the $L_{\rm x}-L_{\rm MIR}$ relation, introduced by observational constrains.

\section{Conclusions}

In this work we combine the \emph{XMM}--SDSS survey with the all-sky mid-infrared survey of \emph{WISE} to select highly obscured AGN. We use only sources detected in the hard X-ray band (2--10\,keV) in order to have an initial estimate of the hardness ratio and therefore avoid normal galaxies, whose X-ray emission does not originate from an AGN and have similar X-ray to infrared luminosity ratios to heavily obscured AGN, and other unobscured sources. We also use only X-ray sources that have an SDSS counterpart bright enough to provide a spectroscopic or photometric redshift, and with a detection in at a least three bands with \emph{WISE} and 2MASS to be able to perform SED fitting. These selections limit the initial number of sources from 39\,830 to 2\,844. Out of those sources, we select 20 heavily obscured AGN candidates on the basis of their low X-ray to mid-infrared luminosity ratios and relatively hard X-ray spectra ($HR>-0.35; \Gamma<1.4$). We then investigate further their optical to far-infrared SEDs using a three-component fit, their optical properties (spectra), and their X-ray spectra from \emph{XMM--Newton}. Our results are summarised as follows:

\begin{itemize}
\item Detailed SED decomposition using photometry from the near-ultraviolet to the far-infrared and three components (stellar, AGN, star formation) finds robust evidence for the presence of a luminous AGN component in the infrared for all 20 sources.
\item The broad-band X-ray spectra indicate the presence of a heavily obscured AGN in nine out of the 14 sources for which X-ray spectra could be fitted. The X-ray criteria used are either: i) the presence of a high equivalent width FeK$\alpha$ line, ii) a flat spectral index, 3) a reflection component, or a combination of the three.
\item We have an [OIII]\,5007 line measurement for nine out of 20 heavily obscured candidates, and assuming a threshold of $f_{\rm2-10\,keV}/f_{\rm[OIII]}=1.0$, seven of them are likely to host an obscured AGN.
\item Taking all the above criteria into account, we deduce that the number of heavily obscured sources selected with our method is 12--19, with 2--5 of them being Compton-thick. This number is a factor of 20 less than what would be expected from X-ray background population synthesis models for heavily obscured AGN, and a factor of five for Compton-thick AGN. We attribute those differences on the initial selection method been based on a relatively high mid-infrared luminosity and an optical detection, characteristics not shared by many obscured AGN, and the scatter of the $L_{\rm x}-L_{\rm MIR}$ relation, mostly introduced by observational constraints. This shows the limitations of this method for selecting large numbers of heavily obscured AGN in wide-shallow surveys.
\item We test popular obscured AGN selection methods based on mid-infrared colours, and find that the probability of an AGN to be selected by its mid-infrared colours increases with the intrinsic X-ray luminosity, while the (observed) X-ray luminosities of heavily obscured AGN are relatively low ($L_{\rm 2-10\,keV}<10^{44}\,{\rm erg\,s^{-1}}$). In fact, a selection scheme based on a relatively low X-ray luminosity and QSO mid-infrared colours would not select a quarter of the heavily obscured AGN of our sample.
\end{itemize}

\section*{Acknowledgements}

The research leading to these results has received funding from the People Programme (Marie Curie Actions) of the European Union's Seventh Framework Programme (FP7/2007--2013) under REA grant agreement number 298480.
D.M.A., A.D.M. and J.R.M. thank the Science and Technology Facilities Council (STFC) and the Leverhulme Trust for support.
A.C. acknowledges financial contribution from the agreement ASI--INAF I/009/10/0 and INAF--PRIN 2011.
P.G. acknowledges support from STFC grant reference ST/J00369711.
This publication makes use of data products from the \emph{Wide-field Infrared Survey Explorer}, which is a joint project of the University of California, Los Angeles, and the Jet Propulsion Laboratory/California Institute of Technology, funded by the National Aeronautics and Space Administration.

\clearpage

\appendix

\section[]{SED fitting details}
\label{decomposition_details}

The method used for the SED decomposition is maximum likelihood. We assume a gaussian profile for the photometry in the different filters, and for a given combination and normalisation, we calculate the likelihood:
\begin{equation}
\label{basic_equation}
L_{m}=\prod_{i}\frac{1}{\sigma_{i}\sqrt{2\pi}}\exp\left[-\frac{\left(f_{i}-f_{i}^{m}\right)^{2}}{2\sigma_{i}^{2}}\right]
\end{equation}
where ``$i$'' refers to the photometric data-point and ``$m$'' refers to the model, i.e. each combination and normalisation.

In the general case of three component fitting (including synthetic stellar templates), we have three sets of templates, codenamed ``stellar'', ``SB'', and ``AGN''. For the stellar templates, we use the \citet{Bruzual2003} stellar population models, with solar metallicity and a range of star-formation histories and ages; in total we use 75 templates. We then redden each template using a \citet{Calzetti2000} dust extinction law, with a range of $E(B-V)$ from 0 to 2 in steps of 0.1. This gives us a library of 1\,500 synthetic stellar templates. For the SB templates we use the library of \citet{Chary2001} (105 templates), or the library of \citet{Mullaney2011} (5 templates), which we crop at rest-frame wavelengths below $\rm4.5\,\mu m$ in order not to duplicate the stellar population, since it is the main contributor in the optical--near-infrared wavelengths in the \citet{Chary2001} templates. For the AGN templates we use the library of \citet*{Silva2004}, which includes five templates: four of them are ``torus'' templates with varying extinction ranging form $N_{\rm H}=0$ to $N_{\rm H}=10^{24}\,{\rm cm^{-2}}$ and the fifth is the $N_{\rm H}=0$ template with a blue bump.

We use a Monte-Carlo Markov Chain (MCMC) sampling method for the SED fitting, taking into consideration all the possible combinations of templates ($1500\times105\times5$ or $1500\times5\times5$). For each combination, we calculate the minimum and maximum contribution of each component, we reduce the minimum contribution by a factor of 100, and use this range to vary the contribution of each template with the MCMC sampling. For each ``test combination'' we calculate the likelihood using Equation\,\ref{basic_equation}. As the best-fit we keep the combination and normalisation that gives the maximum likelihood. We keep the likelihood values of all trial fits and we plot their natural logarithm (log-likelihood) against key values corresponding to the given combination and normalisations. Examples of such plots are given in Figure\,\ref{SED_figures} -- left panel.

In the general case we do not use any prior information in our SED fits, and if we use the natural logarithm of Equation\,\ref{basic_equation}, it becomes:
\begin{equation}
\ln L_{m}=\sum_{i}\ln\left[\frac{1}{\sigma_{i}\sqrt{2\pi}}\right]-\sum_{i}\left[\frac{\left(f_{i}-f_{i}^{m}\right)^{2}}{2\sigma_{i}^{2}}\right]=C(D)-\frac{\chi^{2}}{2}
\end{equation}
where $C(D)$ is a constant depending only on the data, and $\chi^{2}$ is the chi-square statistic of each trial fit. Translating this to the $\chi^{2}$ difference between the best-fitting model (which also has the lowest $\chi^{2}$) and a trial fit, we get:
\begin{equation}
\Delta\chi^{2}=-2\Delta(\ln L)
\end{equation}
so we can use the log-likelihood differences between the trial fits and the best fit to estimate the 1-$\sigma$, 2-$\sigma$ and 3-$\sigma$ confidence intervals of a value in question using $\Delta\chi^{2}=-2.30,-6.17,-11.8$ respectively, which corresponds to $\Delta(\ln L)=1.15,3.09,5.9$. These are the horizontal dashed lines in Figure\,\ref{SED_figures} -- left panel.

For some cases where we are detecting a point-source in the SDSS images, we have a prior information, that the bulk of the flux in some filter comes form the AGN. We quantify the probability density of this information by assuming a Gaussian distribution of the calculated flux density of the AGN component with a $\sigma$ equal to 10\% of the value. We then multiply each trial likelihood with this prior and get:
\begin{equation}
L'_{m}=\prod_{i}\frac{1}{2\pi\sigma_{i}\sigma_{i,AGN}^{m}}\exp\left[-\frac{\left(f_{i}-f_{i}^{m}\right)^{2}}{2\sigma_{i}^{2}}\right]\exp\left[-\frac{\left(f_{i}-f_{i,AGN}^{m}\right)^{2}}{2\left(\sigma_{i,AGN}^{m}\right)^2}\right]
\end{equation}
where $L'_{m}$ is now the Bayesian likelihood. In this case we use the template library of \citet{Polletta2007} which better samples the blue bump of the AGN SEDs, instead of the ones of \citet{Silva2004}.

The resulting SED fits for the 31 candidate heavily obscured AGN are shown in Figure\,\ref{SEDs}.

\begin{figure*}
\begin{center}
\includegraphics[width=5.8cm]{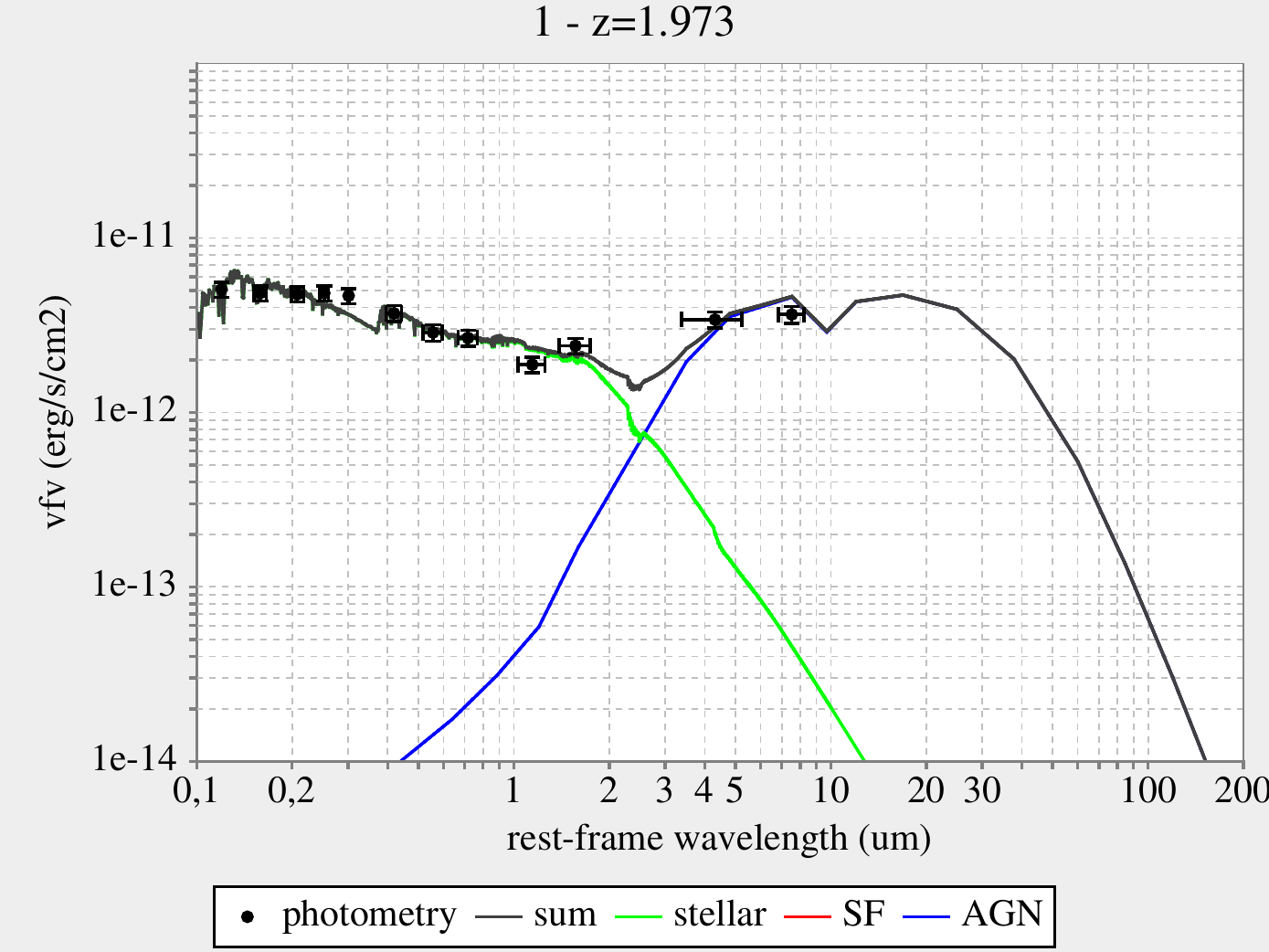} 
\includegraphics[width=5.8cm]{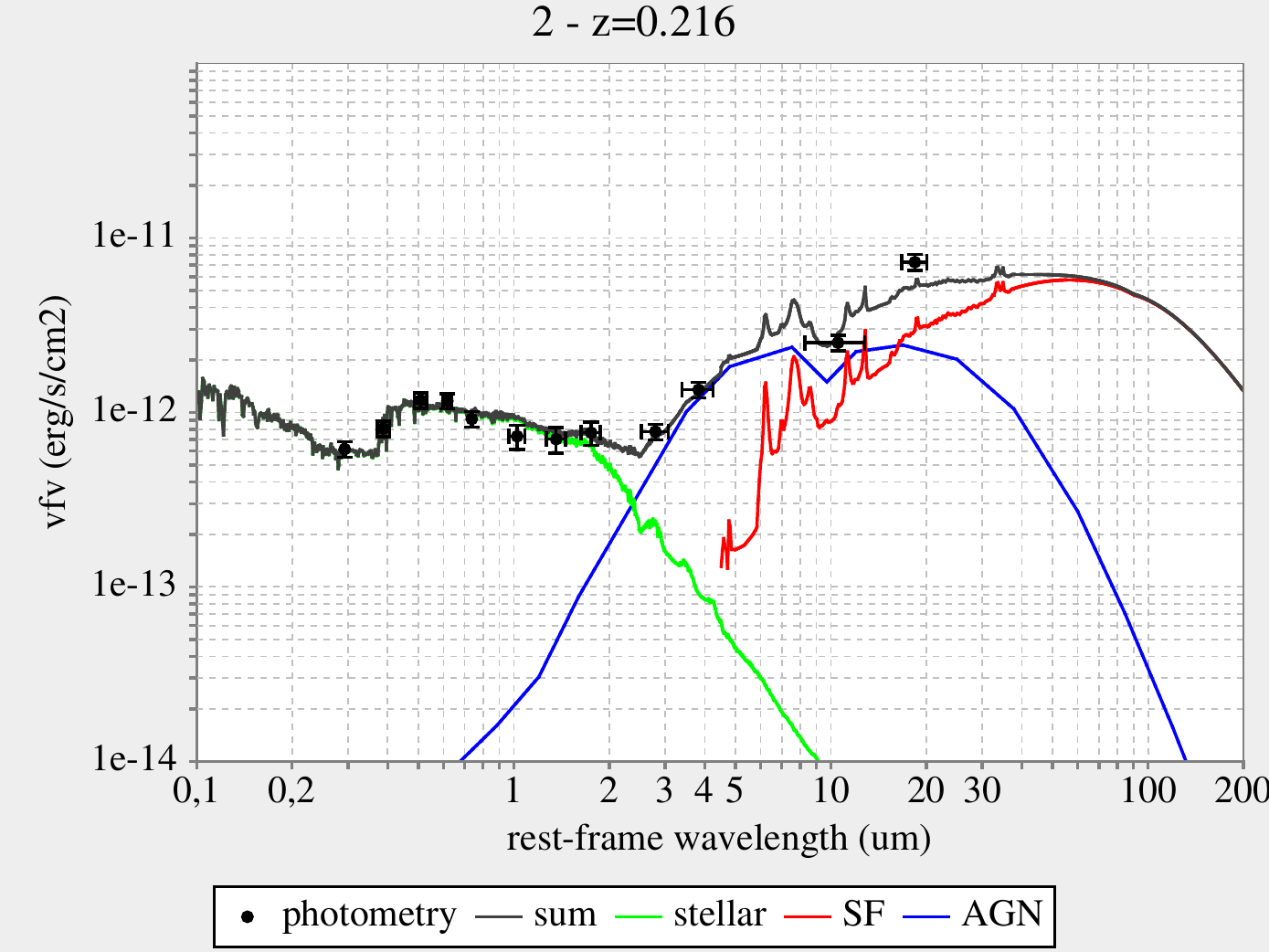} 
\includegraphics[width=5.8cm]{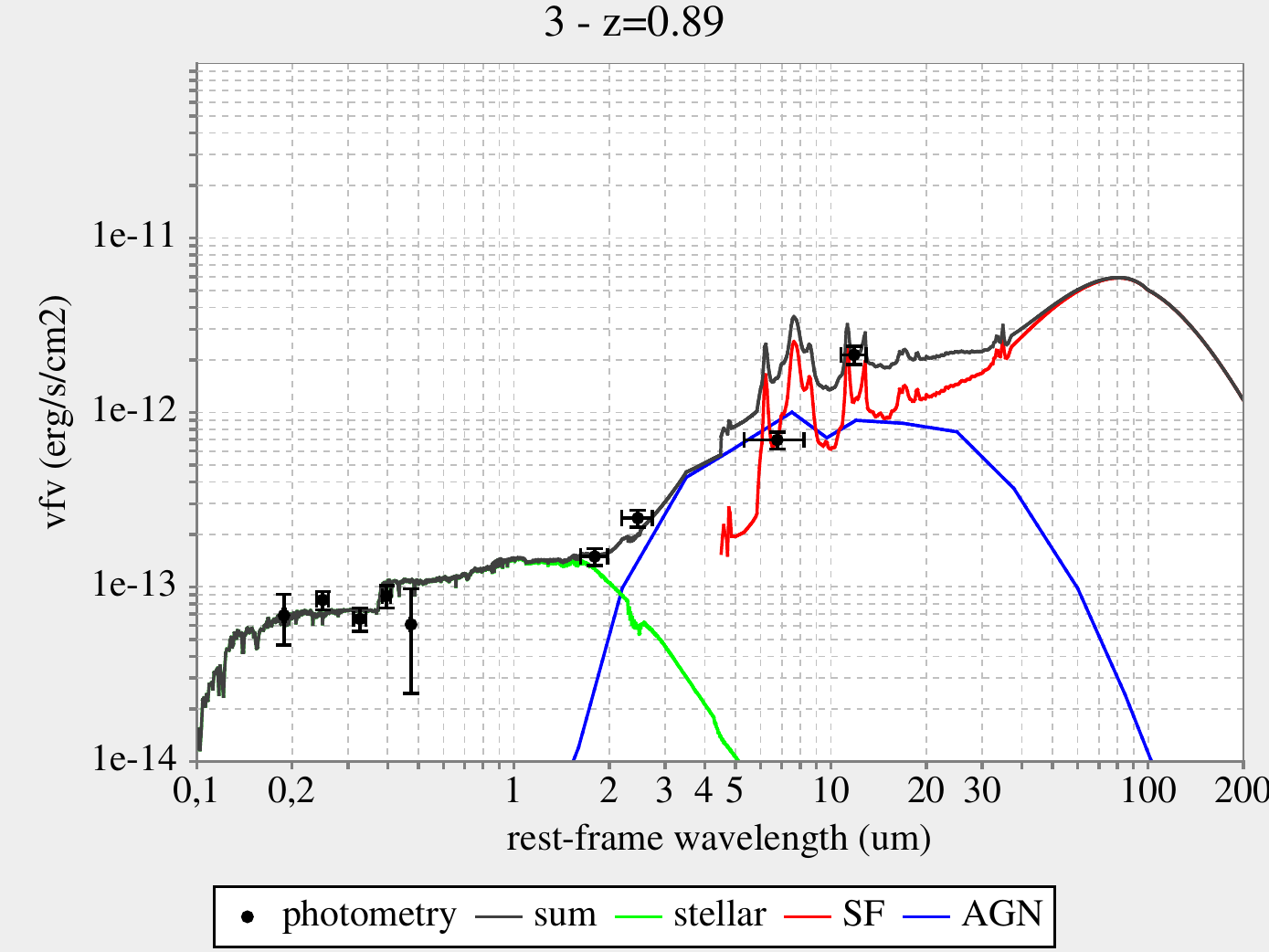} 
\includegraphics[width=5.8cm]{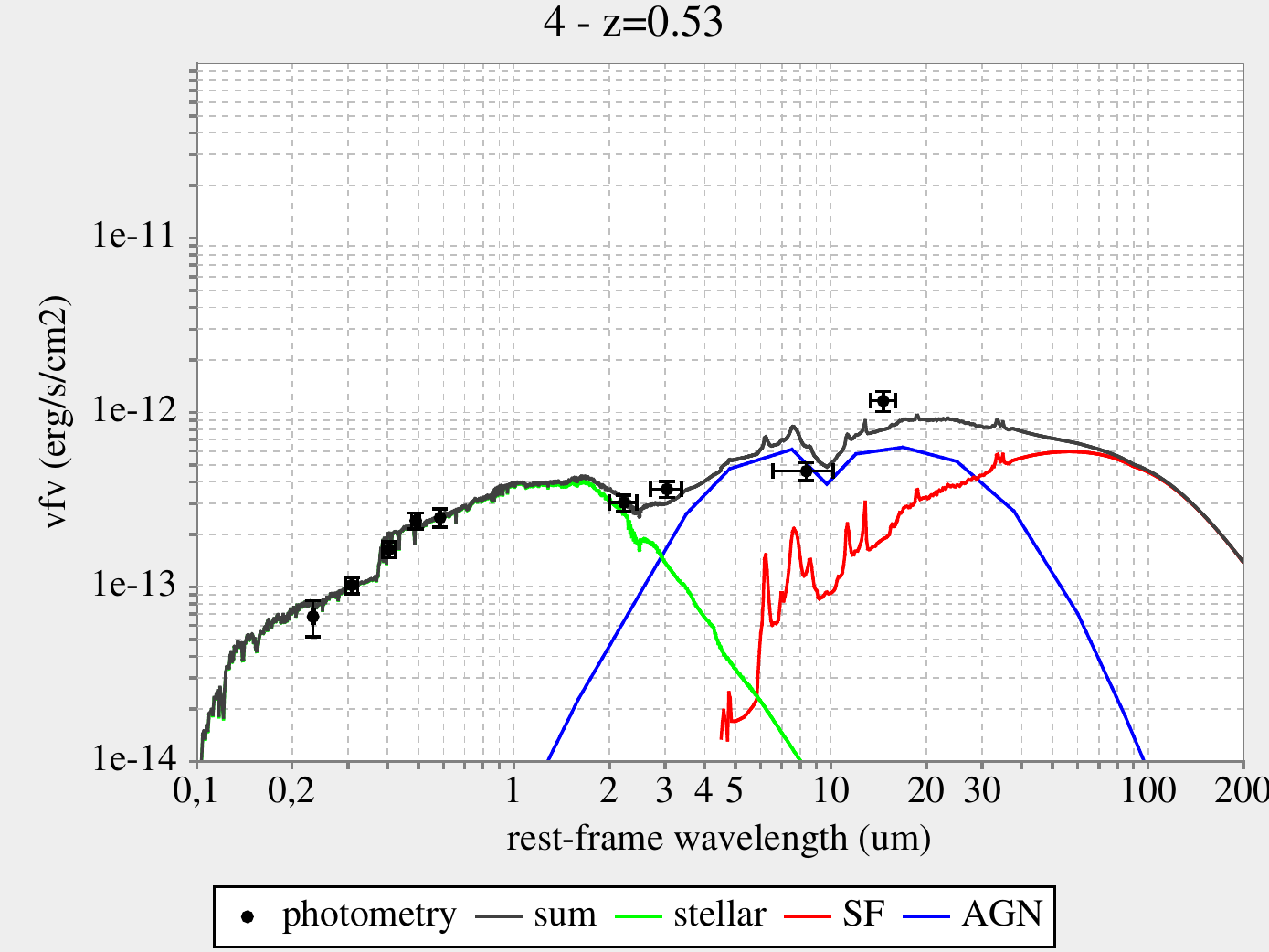} 
\includegraphics[width=5.8cm]{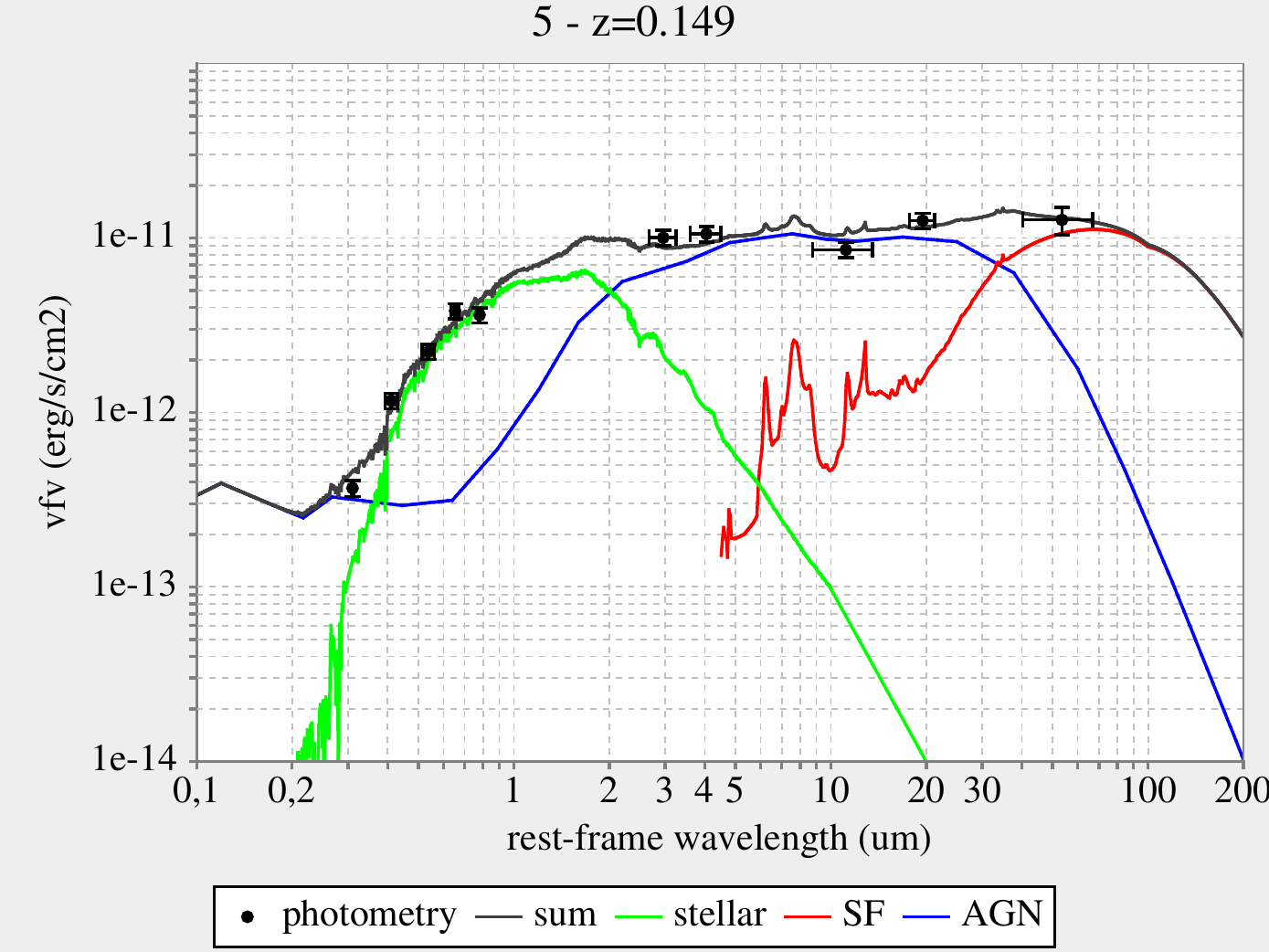} 
\includegraphics[width=5.8cm]{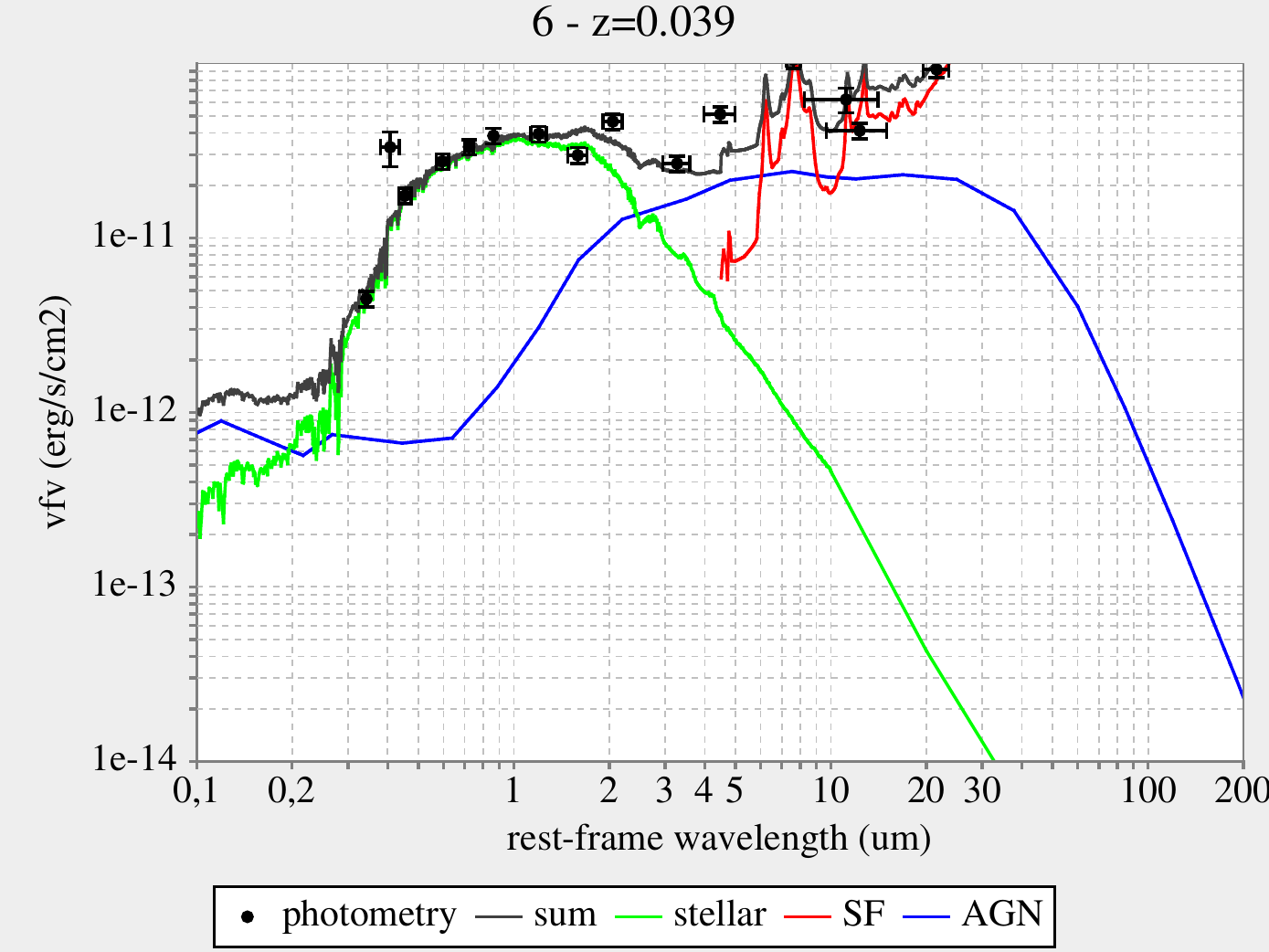} 
\includegraphics[width=5.8cm]{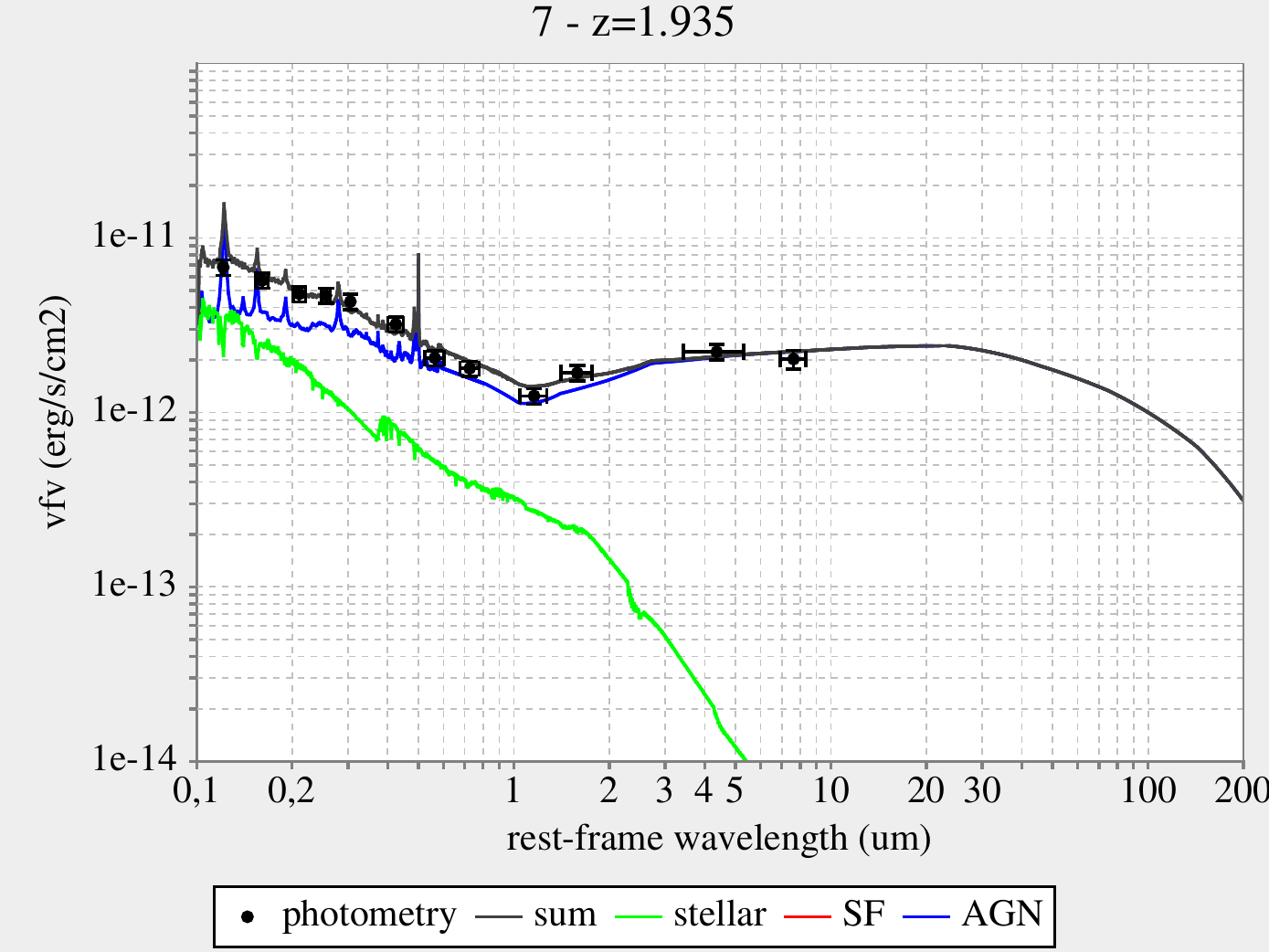} 
\includegraphics[width=5.8cm]{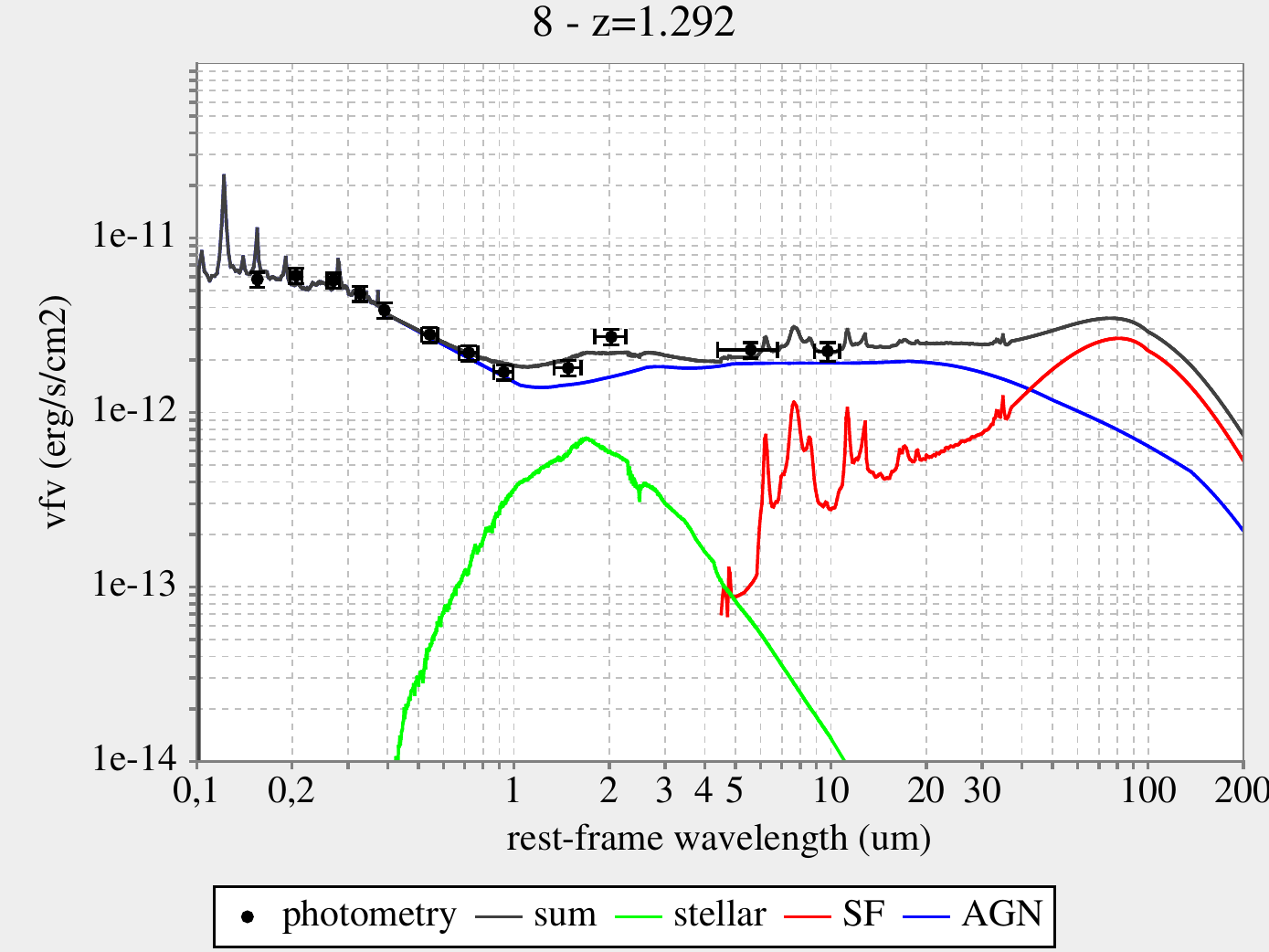}
\includegraphics[width=5.8cm]{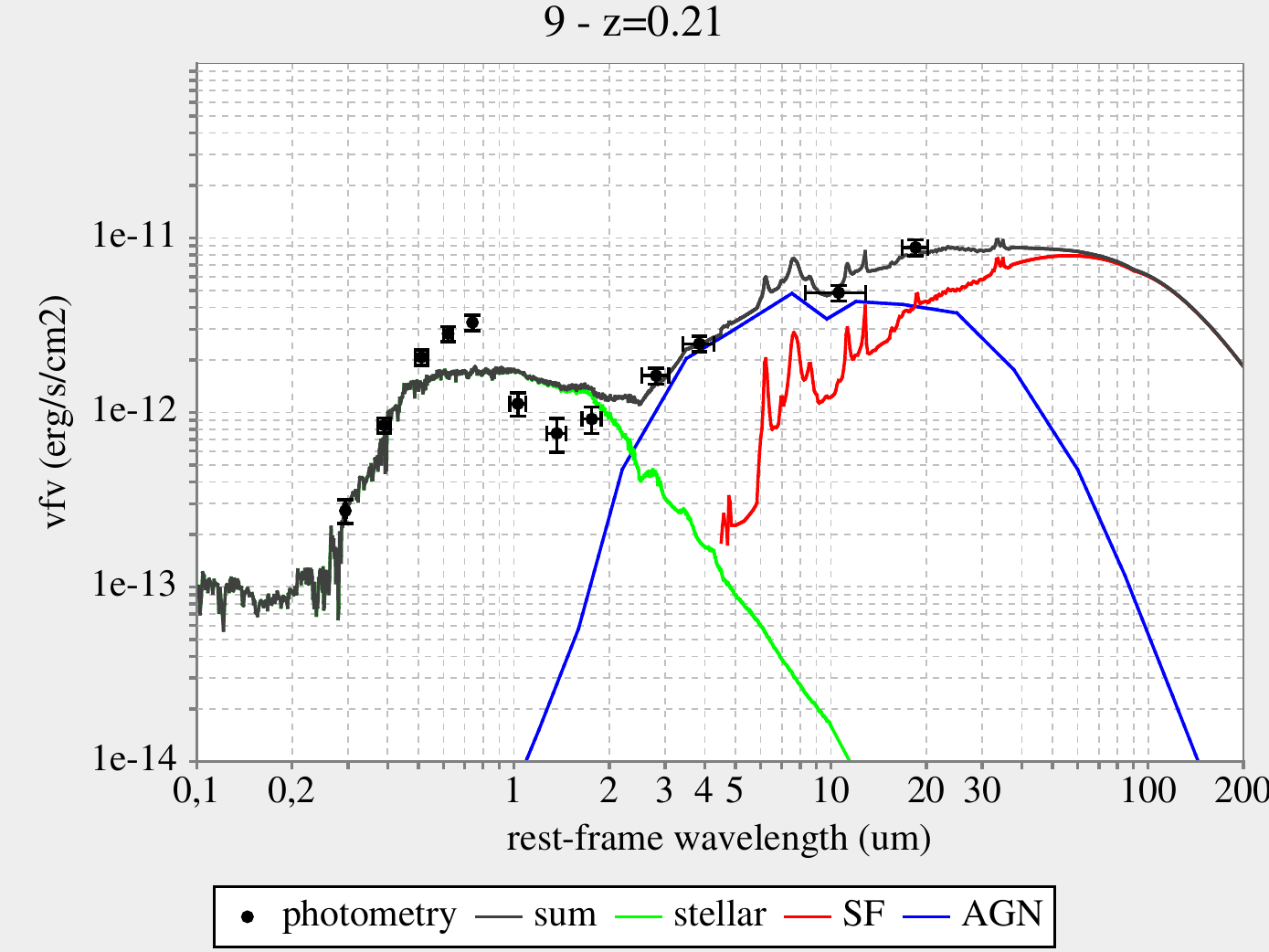} 
\includegraphics[width=5.8cm]{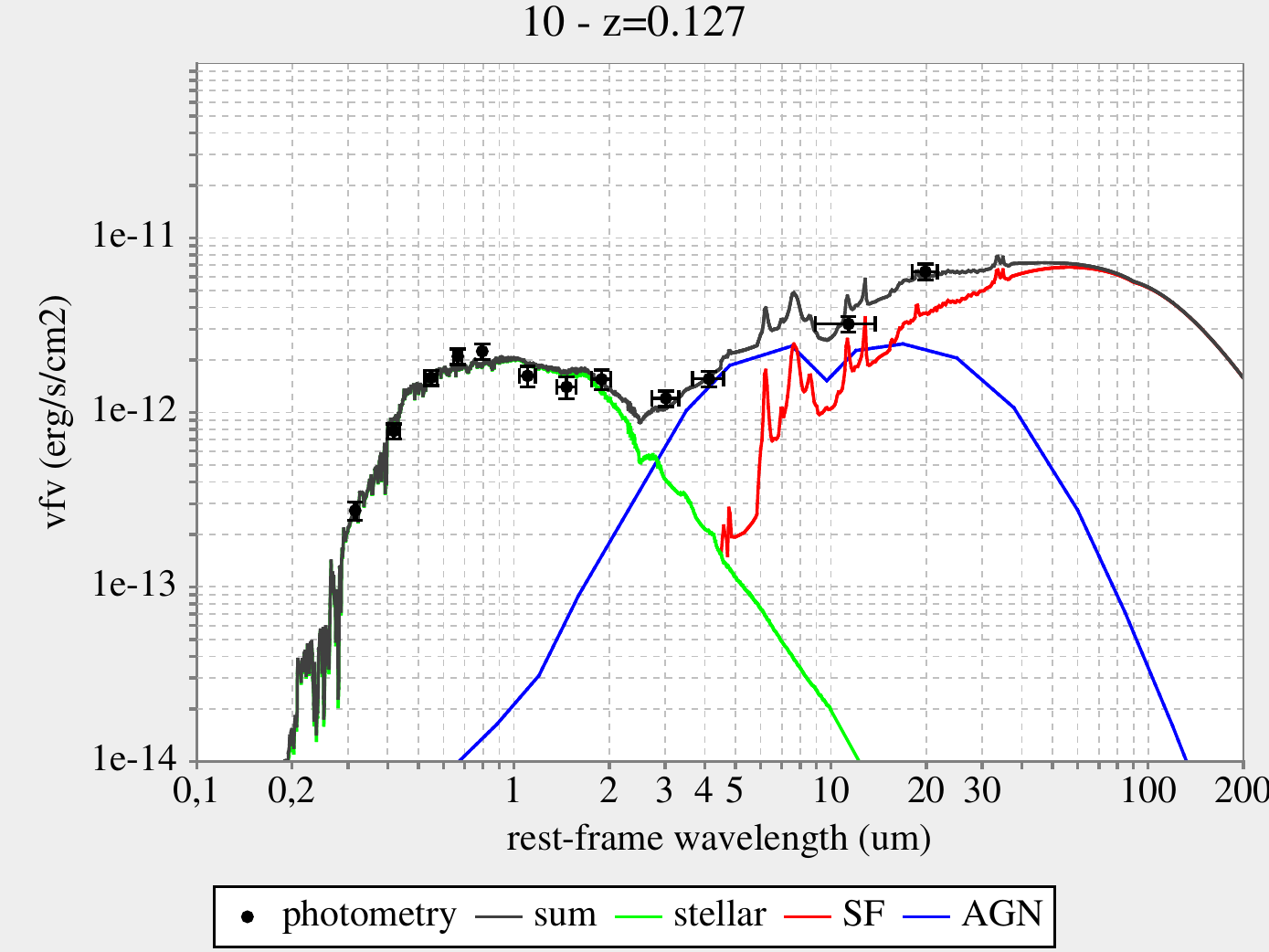} 
\includegraphics[width=5.8cm]{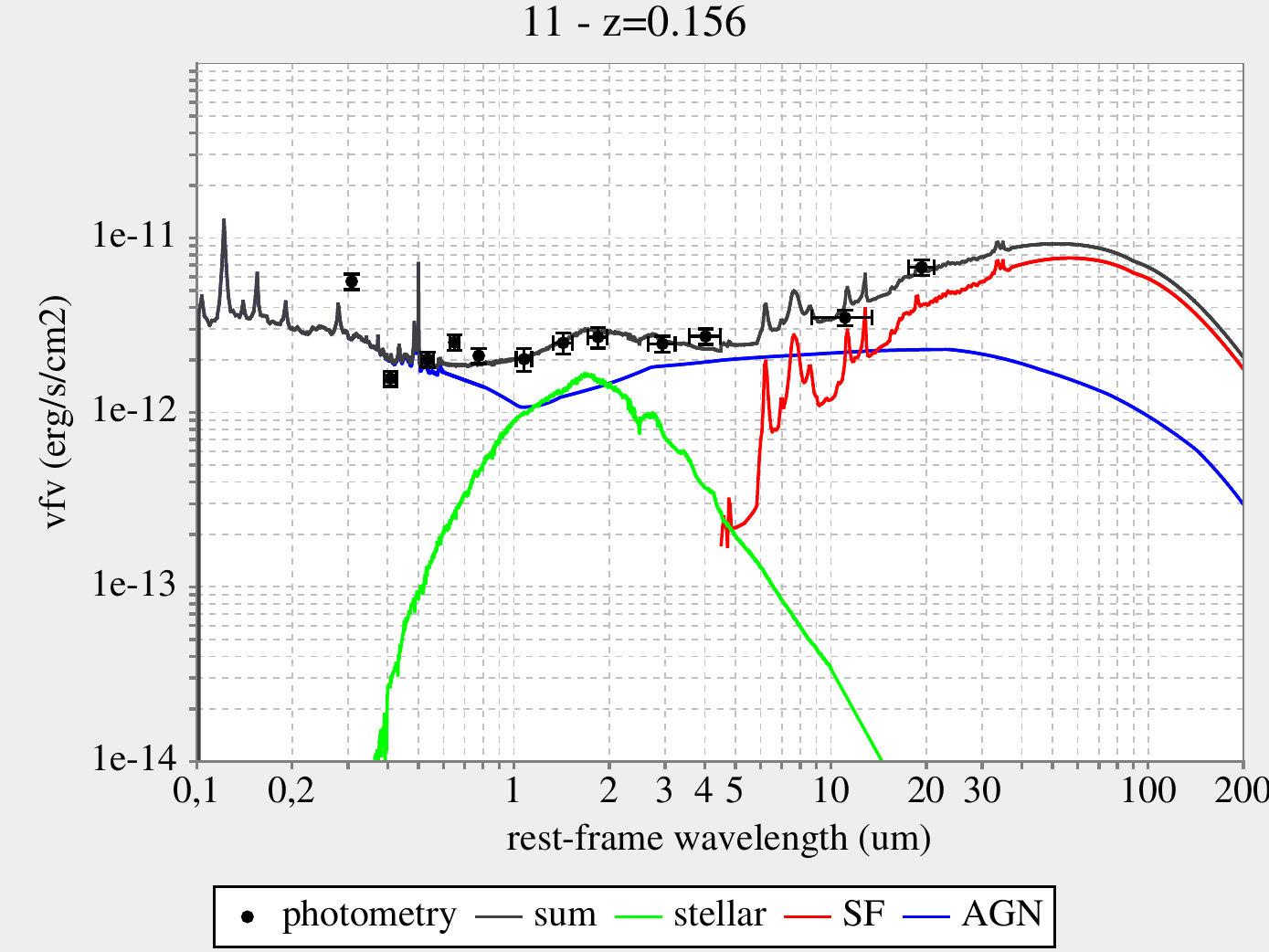} 
\includegraphics[width=5.8cm]{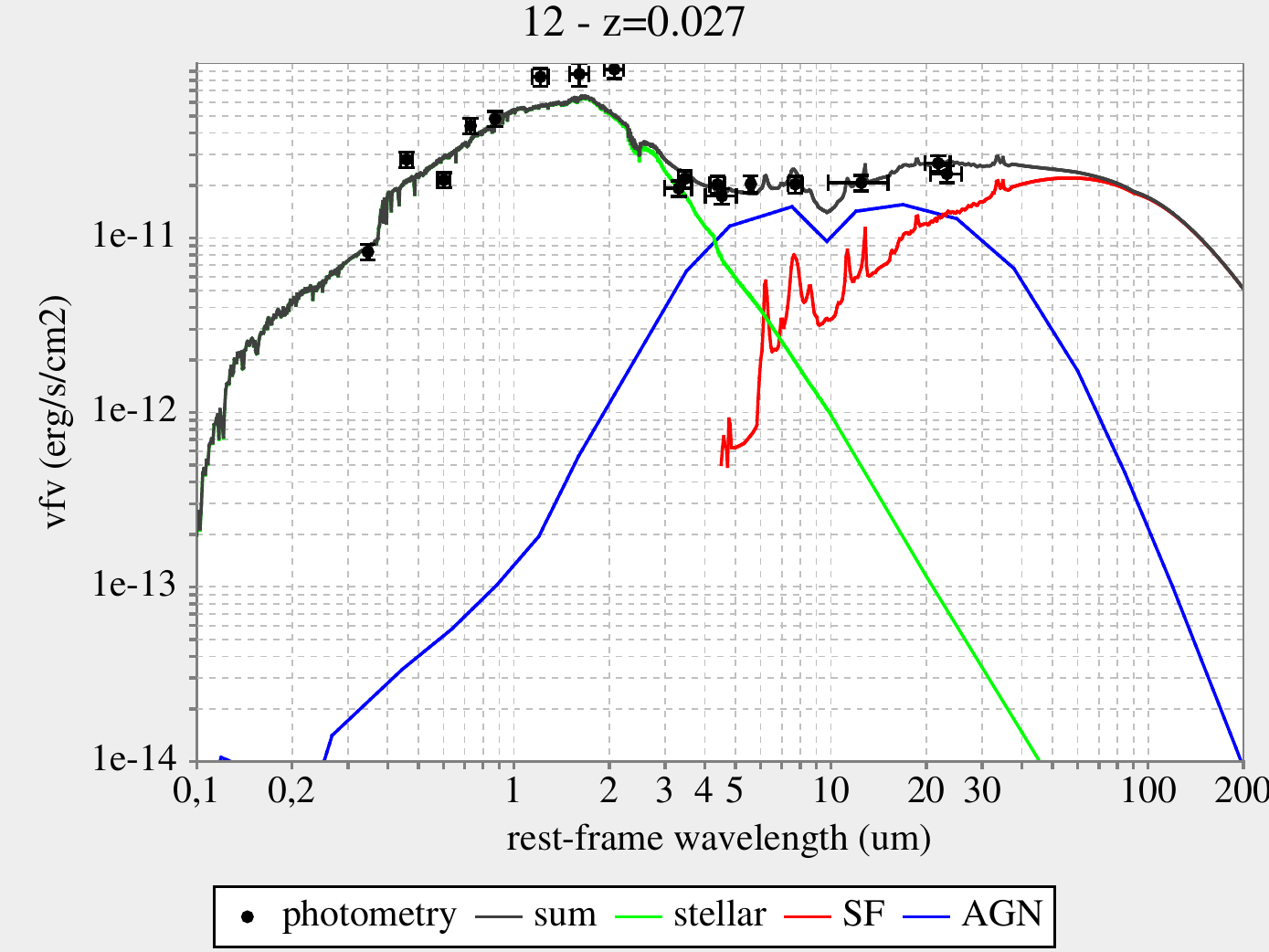} 
\includegraphics[width=5.8cm]{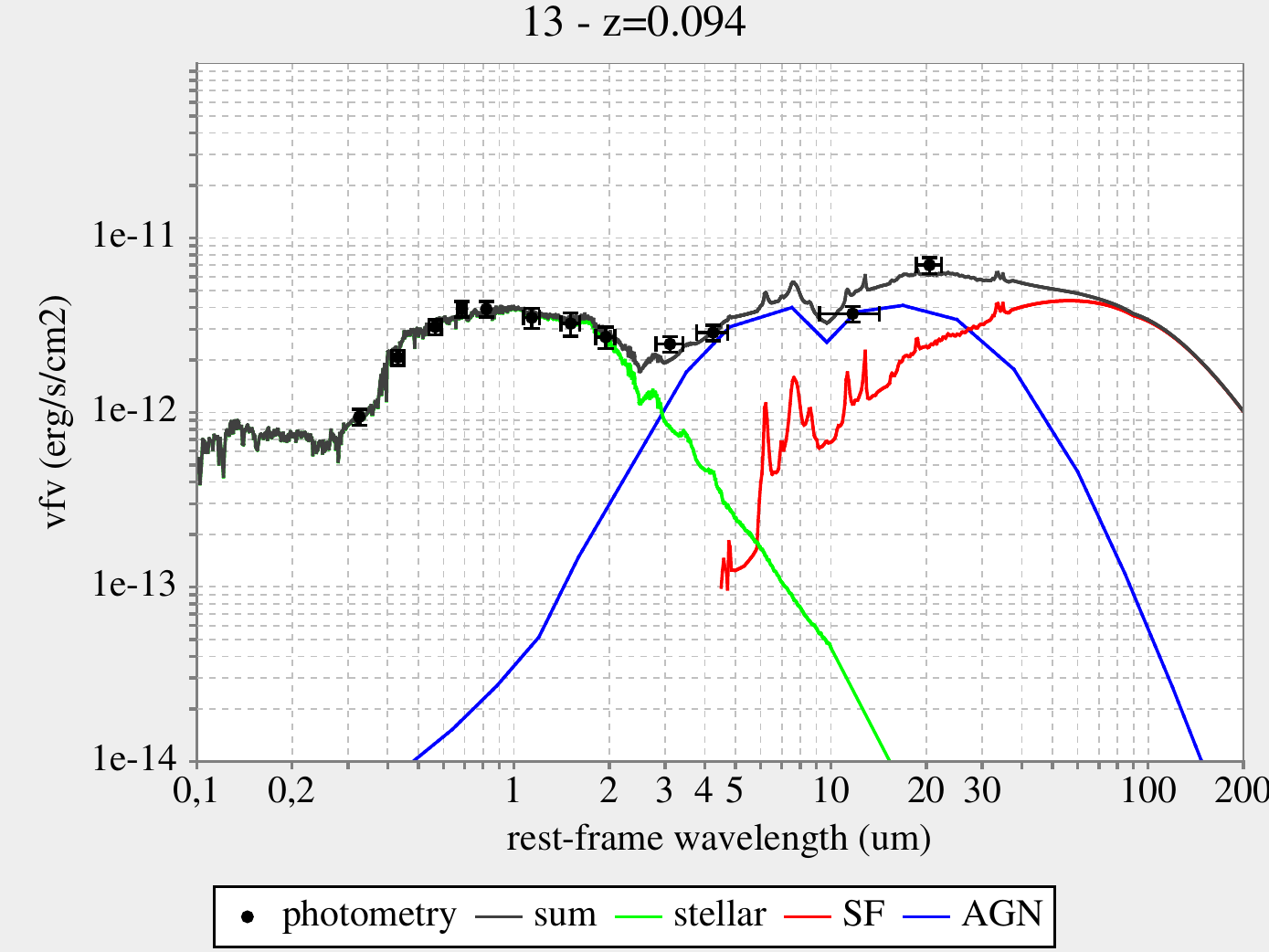} 
\includegraphics[width=5.8cm]{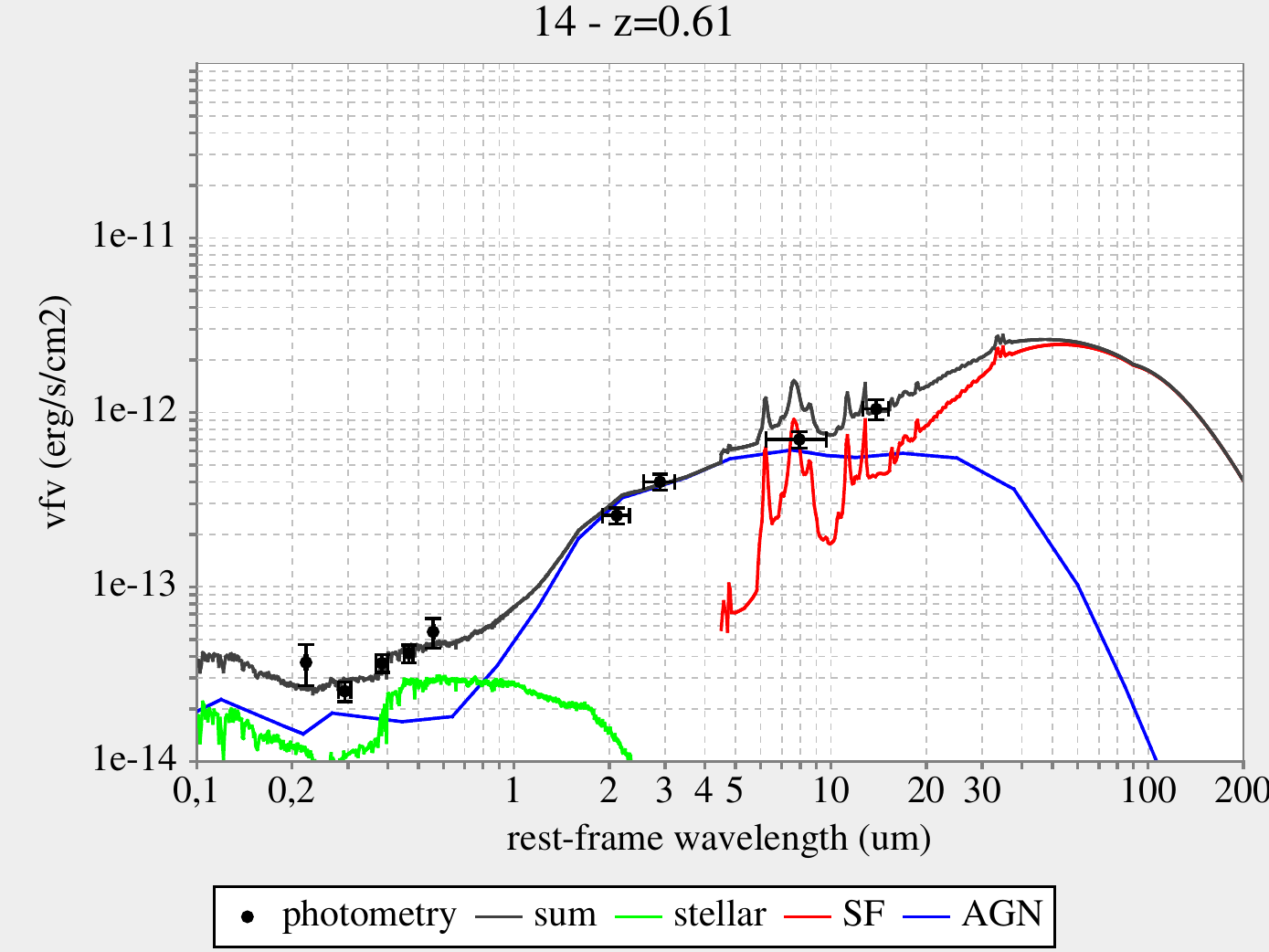} 
\includegraphics[width=5.8cm]{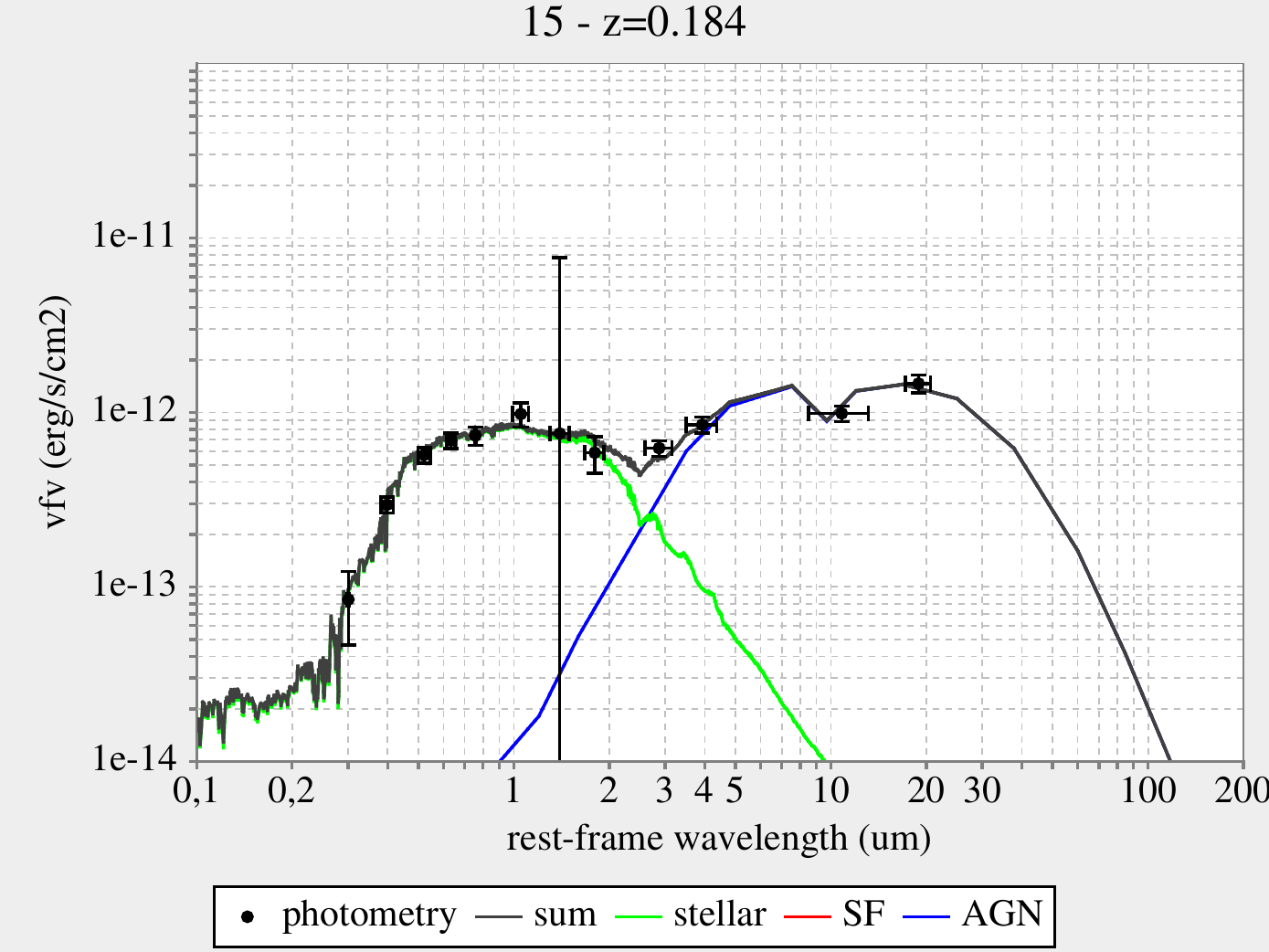} 
\caption{SED decomposition results using three templates (stellar -- AGN -- star formation), shown in the green, blue, and red lines, respectively. The dark grey line is the combination of the three templates and the black points and error bars are the photometry data-points, with the error-bar on the X-axis representing the effective width of the filter.}
\label{SEDs}
\end{center}
\end{figure*}

\setcounter{figure}{0}

\begin{figure*}
\begin{center}
\includegraphics[width=5.8cm]{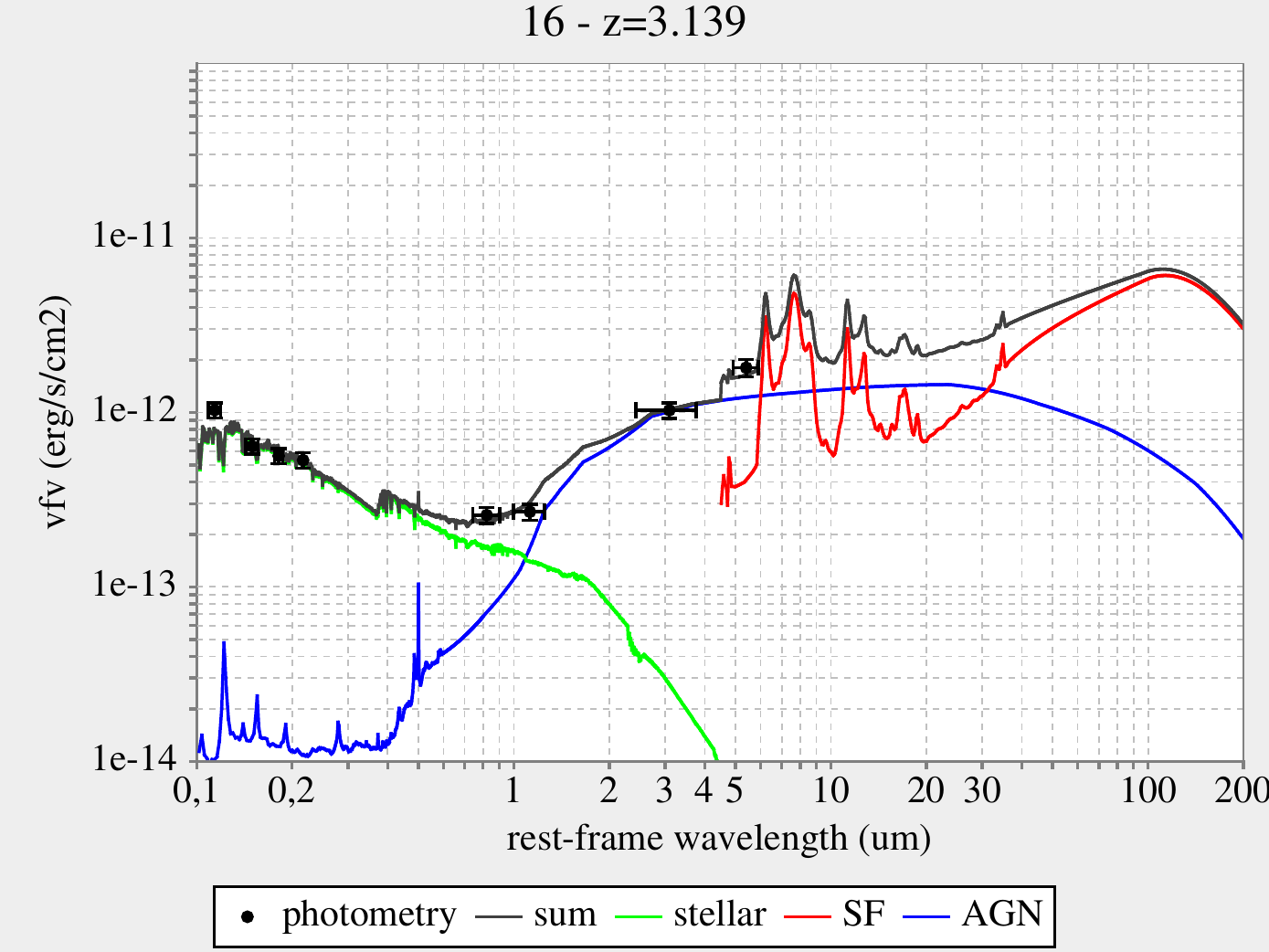} 
\includegraphics[width=5.8cm]{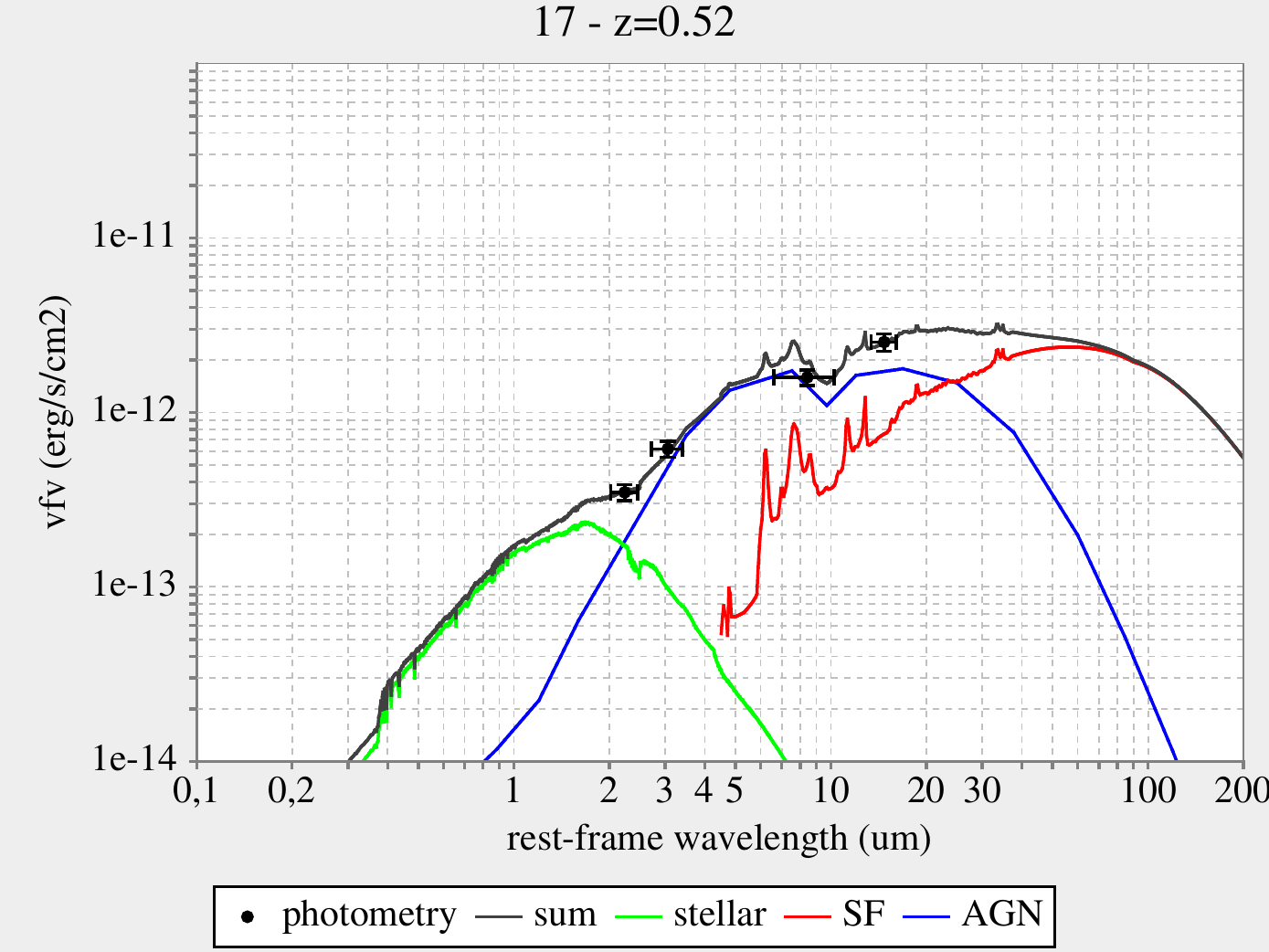} 
\includegraphics[width=5.8cm]{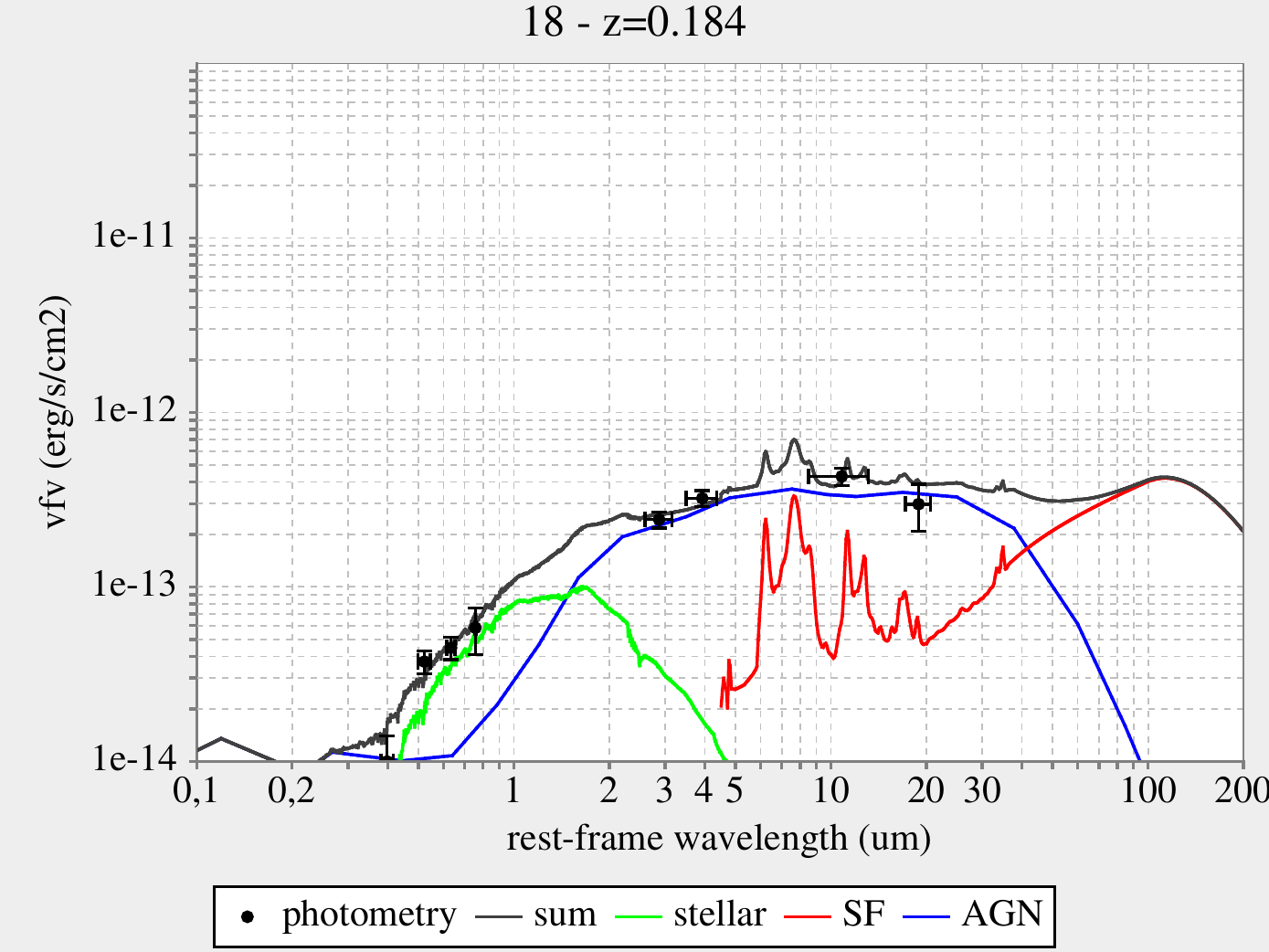} 
\includegraphics[width=5.8cm]{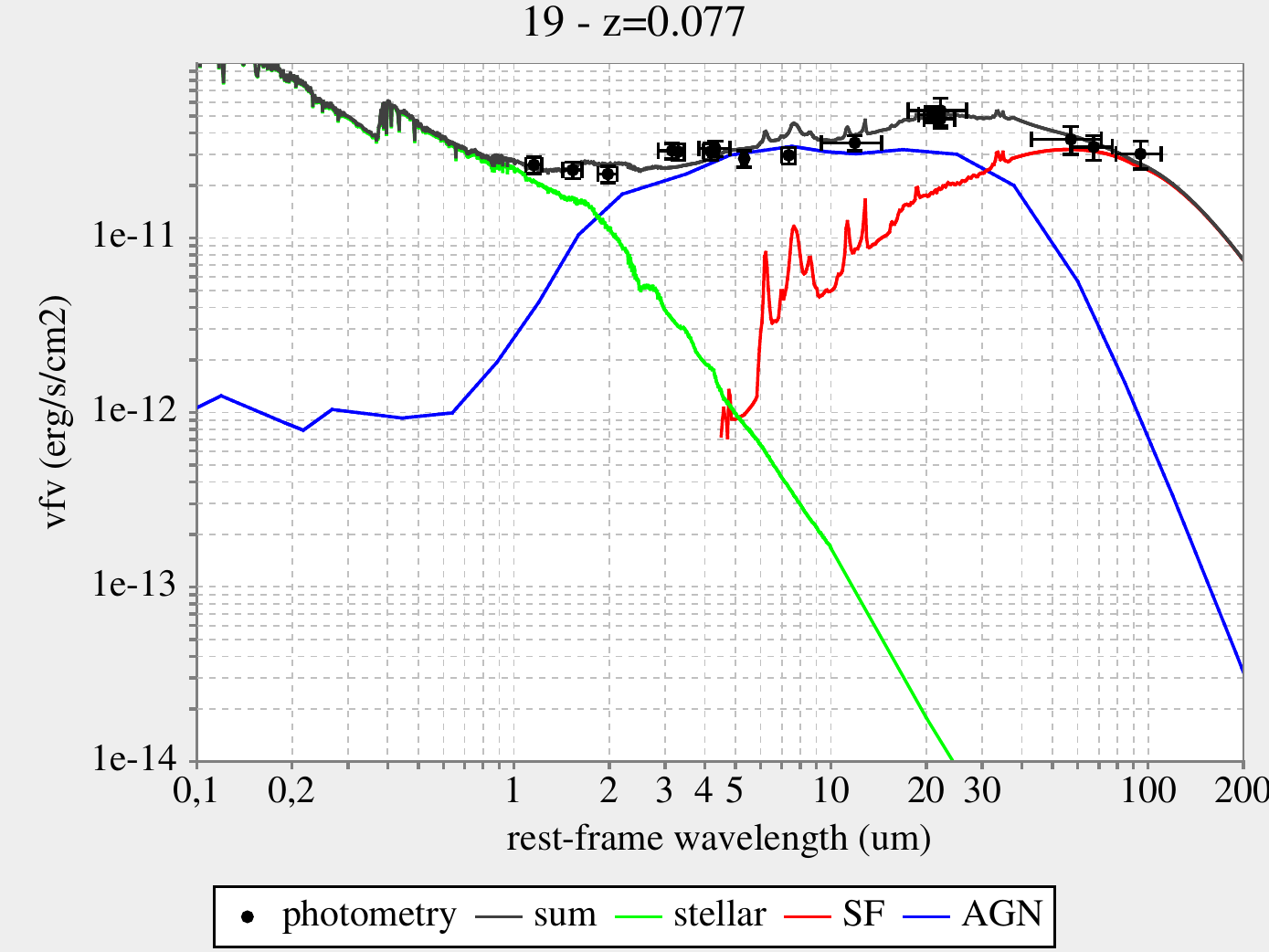} 
\includegraphics[width=5.8cm]{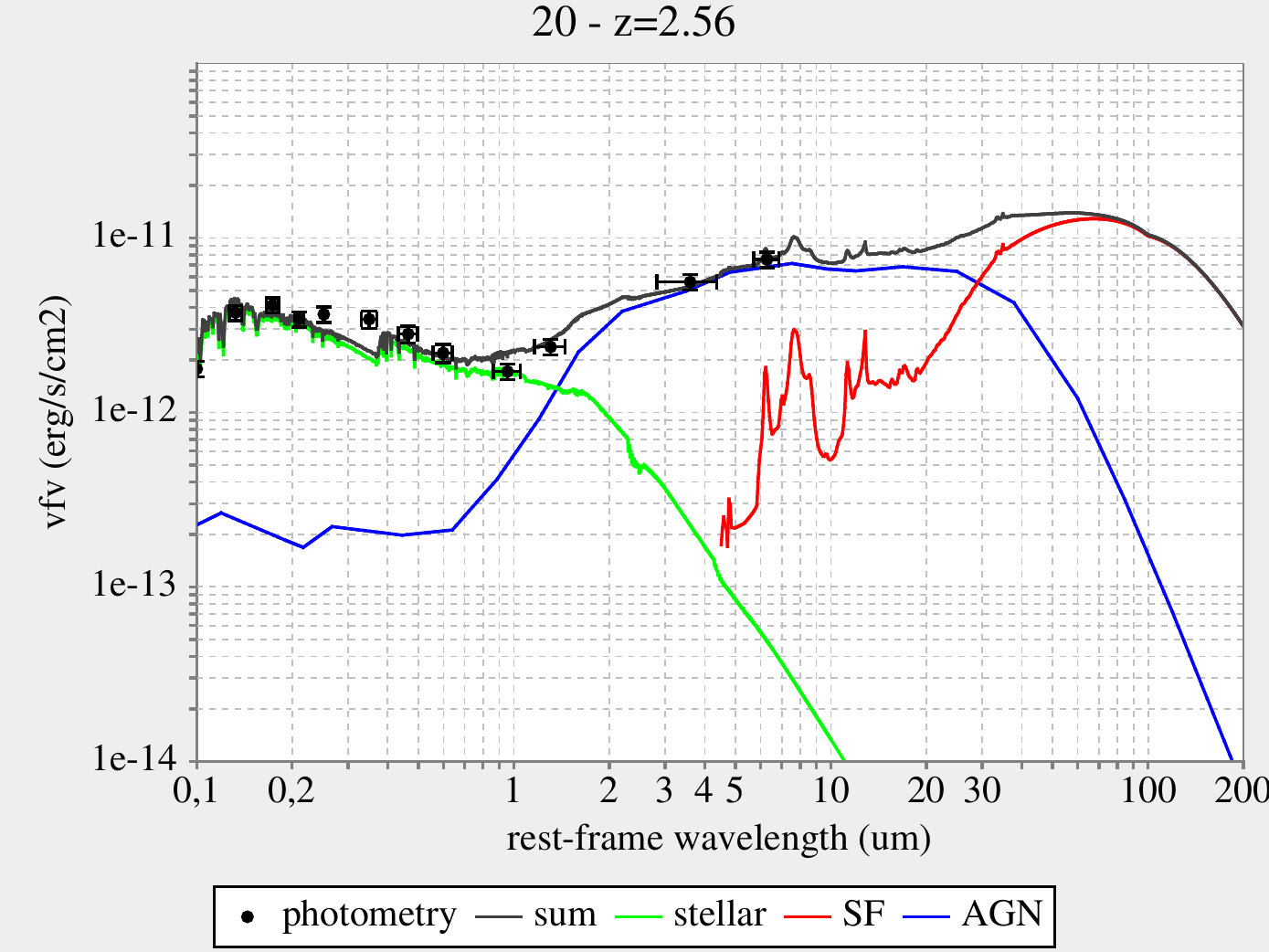} 
\caption{continued}
\end{center}
\end{figure*}

\bsp

\label{lastpage}

\end{document}